\documentclass[preprint,11pt]{elsarticle}

\usepackage{amsmath}
\usepackage{amssymb}
\usepackage{colordvi}
\usepackage{float}

\newcommand{\be}{\begin{equation}}
\newcommand{\ee}{\end{equation}}

\newcommand{\bea}{\begin{eqnarray}}
\newcommand{\eea}{\end{eqnarray}}

\begin{document}

\title{Pattern recognition at different scales: \\ 
a statistical perspective}
\author{Matteo Colangeli$^1$, Francesco Rugiano$^{1}$, Eros Pasero$^1$}
\address{$^1$Dipartimento di Elettronica e Telecomunicazioni, Politecnico di Torino\\Corso Duca degli Abruzzi 24, 10129 Torino, Italy}

\ead{matteo.colangeli@polito.it, francesco.rugiano@polito.it, eros.pasero@polito.it}

\begin{abstract}
In this paper we borrow concepts from Information Theory and Statistical Mechanics to perform a pattern recognition procedure on a set of x-ray hazelnut images. We identify two relevant statistical scales, whose ratio affects the performance of a machine learning algorithm based on statistical observables, and discuss the dependence of such scales on the image resolution. Finally, by averaging the performance of a Support Vector Machines algorithm over a set of training samples, we numerically verify the predicted onset of an ``optimal'' scale of resolution, at which the pattern recognition is favoured.
\end{abstract}

\maketitle

\section{Introduction}
\label{sec:intro}
The theory of artificial neural networks (ANN) represents an open research field setting the stage for the implementation of a statistical mechanical approach in novel interdisciplinary problems, such as the modeling of the collective behavior of the human brain neurons. An important field of application of ANN is represented by the pattern recognition analysis \cite{Egmont,Wang}, which has received an increasing interest in the literature, witnessed by the extensive application of ANN to tackle complex real-word problems, e.g. in medical diagnosis \cite{Haya,Jiang,Shayea} and in biological sequences analysis \cite{Ding,Qian, Condon,Cart}.
Recent works, in this field, paved also the way to the systematic use of technical tools borrowed from Information Theory and Statistical Mechanics \cite{McKay,Anand,Tkacik}.\\
In this paper, in particular, we adopt information theoretic methods \cite{Kim,Vihn} to classify a sequence of hazelnuts images, and show how our approach allows for improving the performance of pattern recognition procedures performed via ANN algorithms.
From a preliminary statistical analysis on the image histograms, we identify some relevant observables to be used in the implementation of a machine learning algorithm. A special focus of our approach is on the role of \textit{fluctuations} of the histograms around the corresponding \textit{mean} distribution. In particular, by making use of various notions of ``distance'' between histograms, we introduce two statistical scales, whose magnitude affects the performance of a machine learning algorithm in disentangling and extracting the distinctive features of the hazelnuts.\\
The paper is organized as follows.\\
In Sec. \ref{sec:sec1} we introduce the two aforementioned statistical scales and discuss their dependence on a quantity referred to as the ``image resolution''. We comment on the need of a large separation between two such scales to obtain an efficient pattern recognition: the lack of a wide separation between them is due to large histograms fluctuations which blur the distinctive features of the hazelnuts, thus hindering a proper classification of the data. \\
In Sec. \ref{sec:sec2} we test, then, the prediction of our statistical analysis by employing a machine learning algorithm, known as Support Vector Machines (SVM) \cite{Haykin,Webb}. The numerical results we obtained not only confirm the relevance of the aforementioned scale separation, but also show that the predicted onset of an optimal scale of description can be recovered through the use of a SVM algorithm, provided that its performance is \textit{averaged} over a sufficiently large set of training samples. \\
Conclusions are finally drawn in Sec. \ref{sec:conc}.\\
The main results of this work can be summarized as follows:
\begin{itemize}
	\item We introduce two typical statistical scales, whose magnitude critically affects the performance of a pattern recognition algorithm based on statistical variables;
        \item We describe the dependence of such scales on the scale of resolution, thus unveiling the onset of an optimal resolution at which the pattern recognition is favoured;
        \item We numerically recover the results of the statistical analysis by using a SVM algorithm, and also shed light on the role of \textit{averaging} the performance of a SVM over sufficiently many training samples.
\end{itemize}

\section{The original set of hazelnut images: a statistical approach}
\label{sec:sec1}

\begin{figure}[H]
\centering
\includegraphics[width=0.15\textwidth, height=0.15\textwidth]{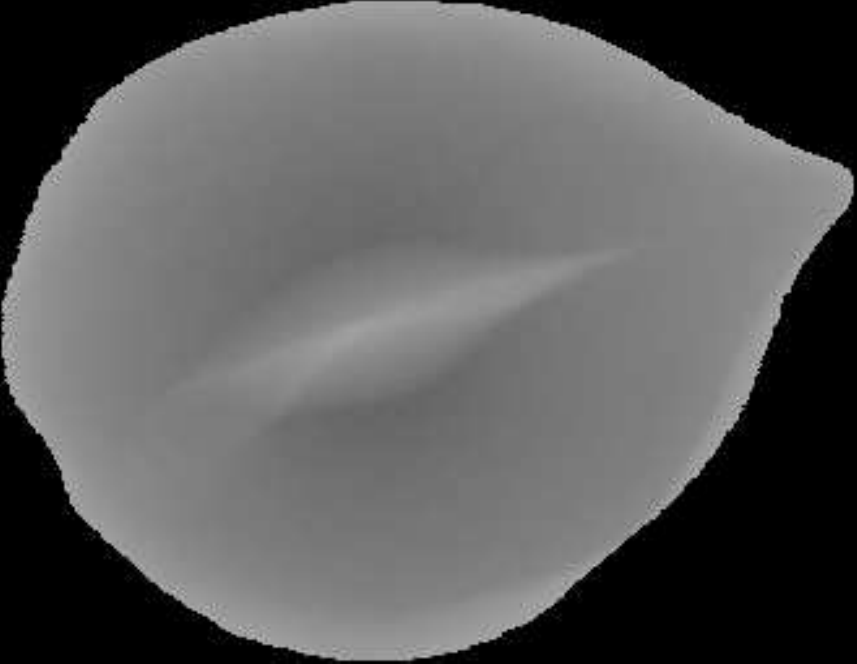}
\hspace{1mm}
\includegraphics[width=0.15\textwidth, height=0.15\textwidth]{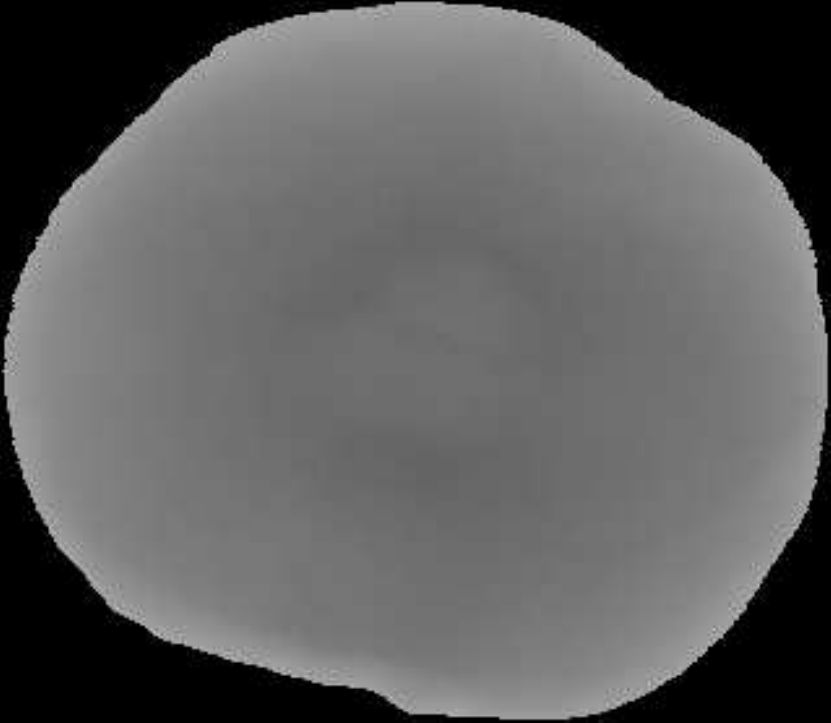}
\hspace{1mm}
\includegraphics[width=0.15\textwidth, height=0.15\textwidth]{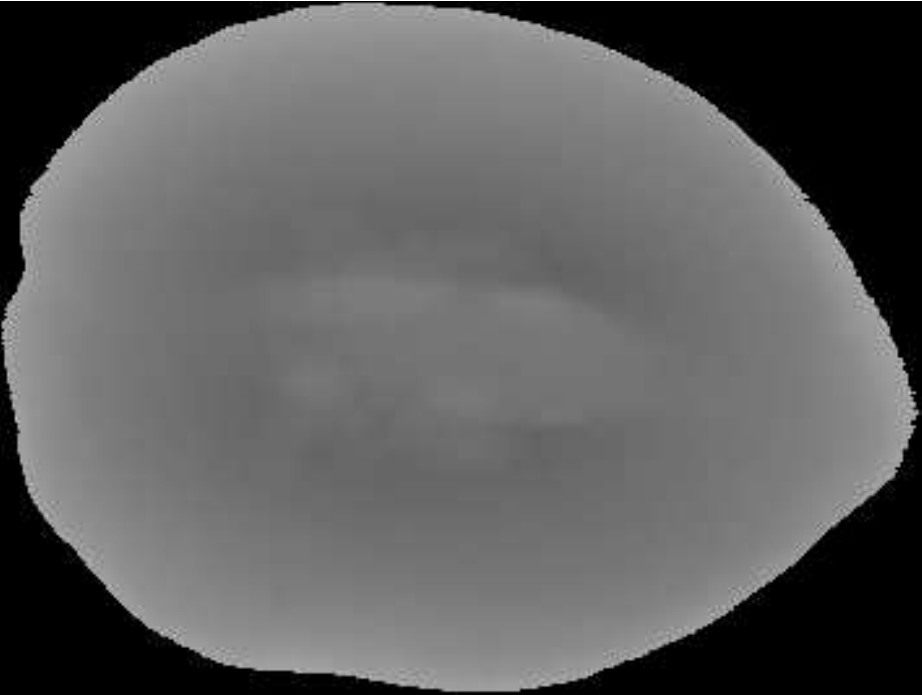}\\
\vspace{2mm}
\includegraphics[width=0.15\textwidth, height=0.15\textwidth]{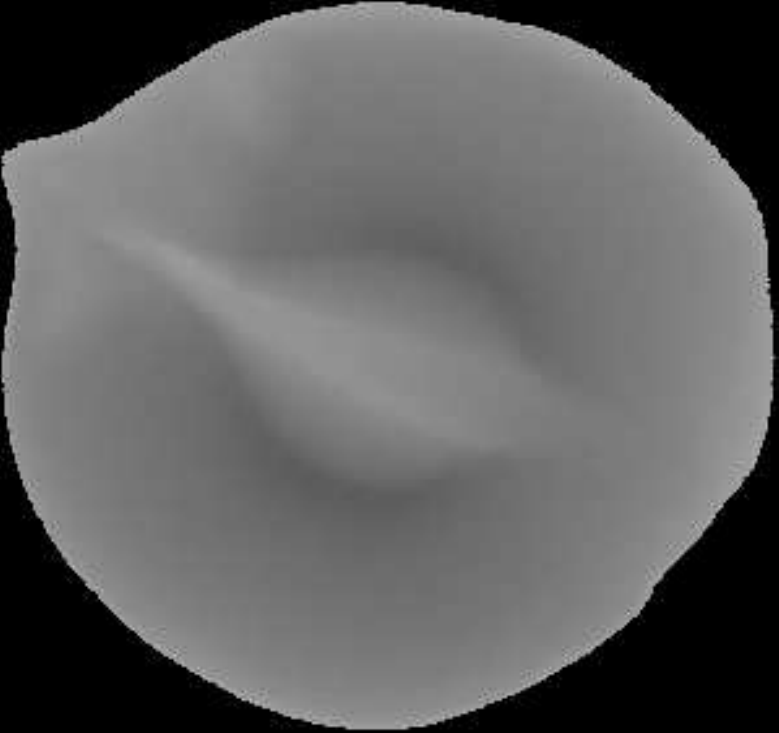}
\hspace{1mm}
\includegraphics[width=0.15\textwidth, height=0.15\textwidth]{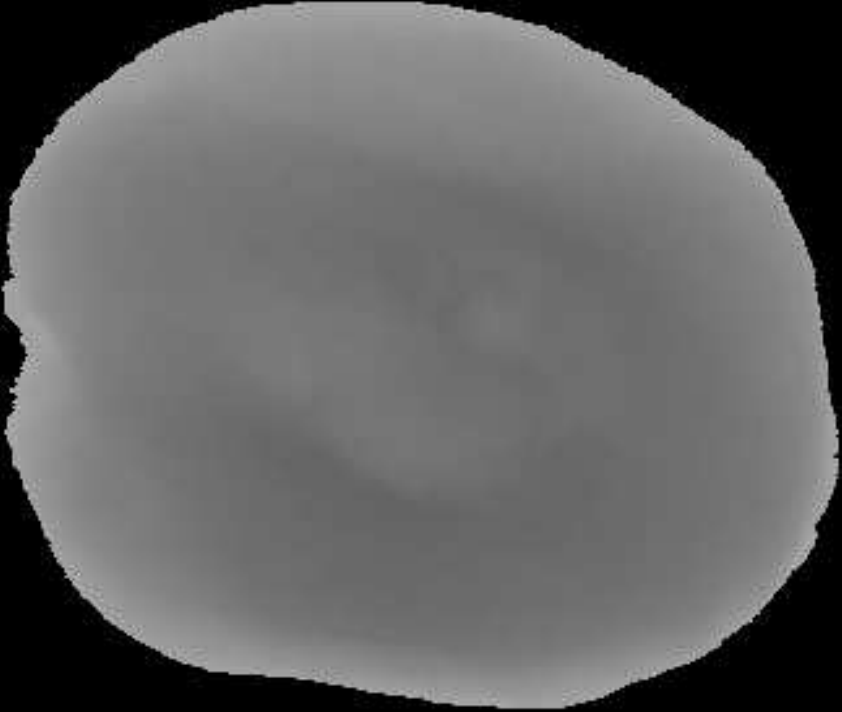}
\hspace{1mm}
\includegraphics[width=0.15\textwidth, height=0.15\textwidth]{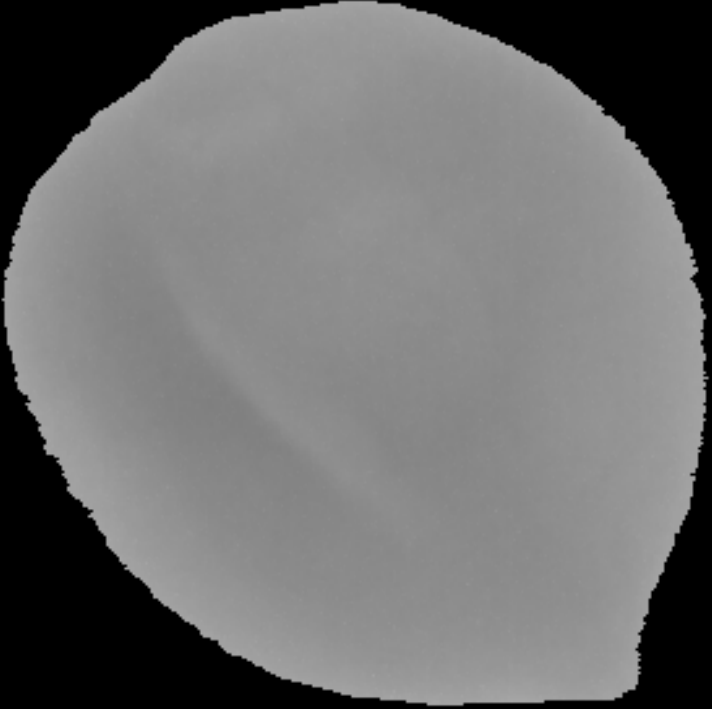}\\
\vspace{2mm}
\includegraphics[width=0.15\textwidth, height=0.15\textwidth]{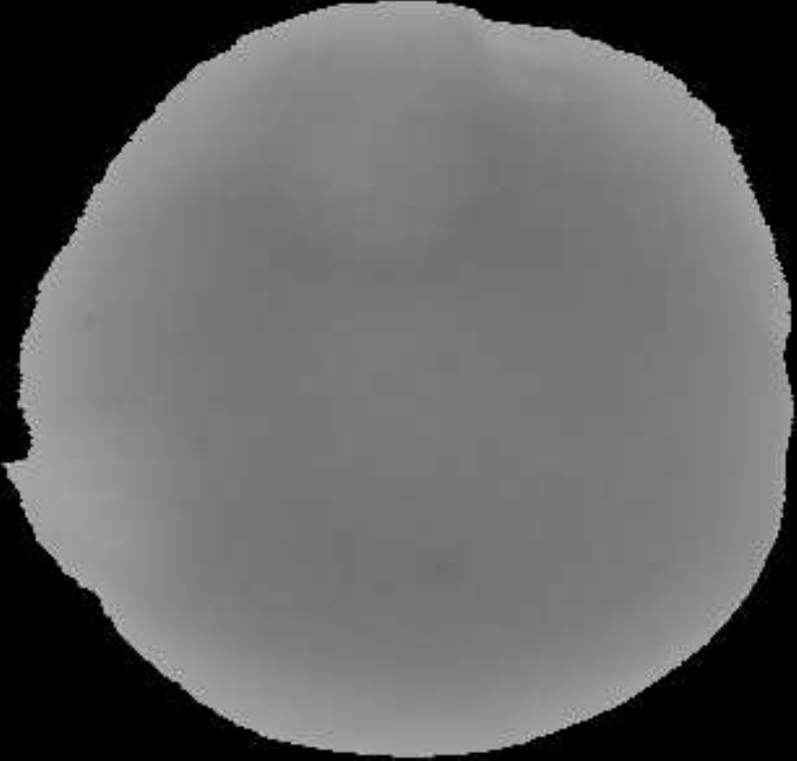}
\hspace{1mm}
\includegraphics[width=0.15\textwidth, height=0.15\textwidth]{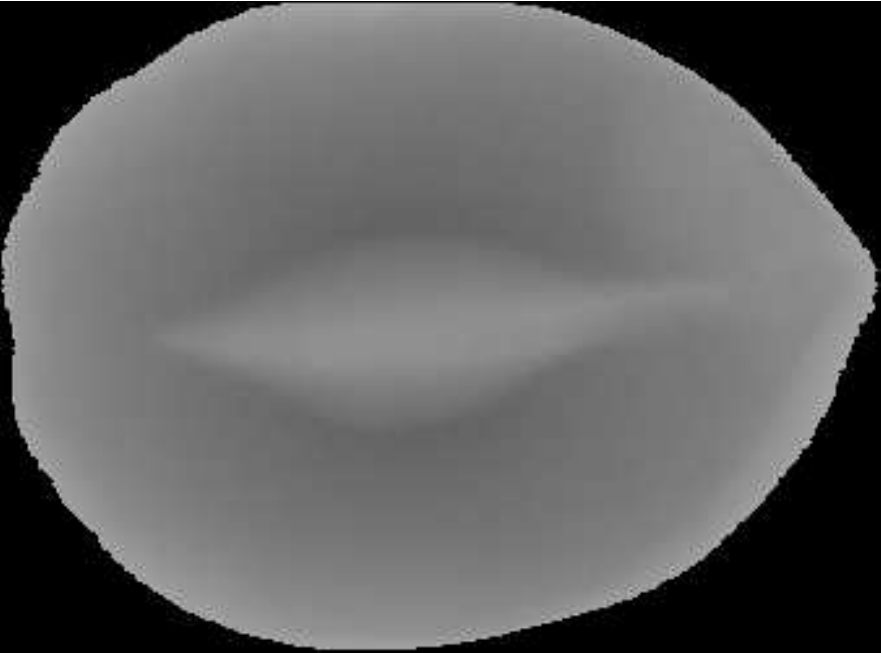}
\hspace{1mm}
\includegraphics[width=0.15\textwidth, height=0.15\textwidth]{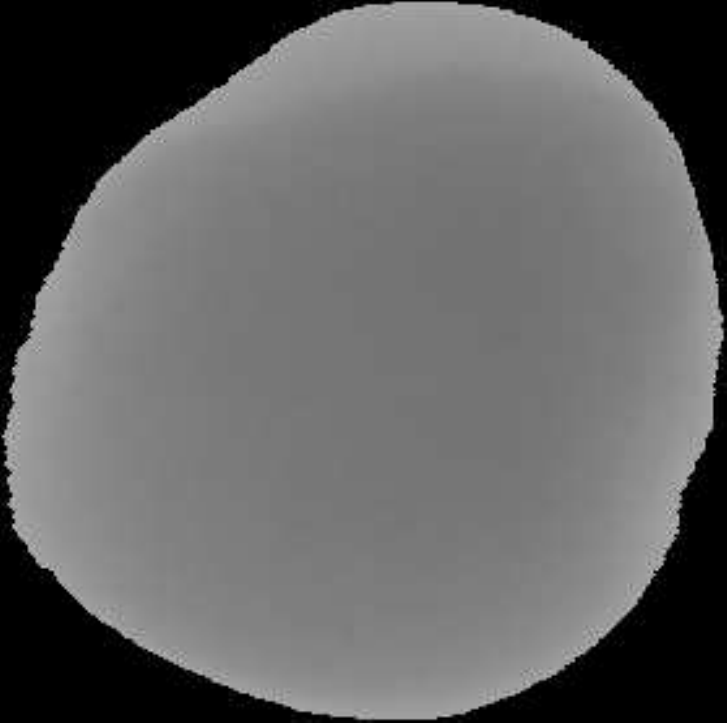}
\caption{X-ray scanned images of good hazelnuts (top row), damaged hazelnuts (middle row) and infected hazelnuts (bottom row).}\label{hazelnuts}
\end{figure}

\begin{figure}[H]
\centering
\includegraphics[width=0.45\textwidth]{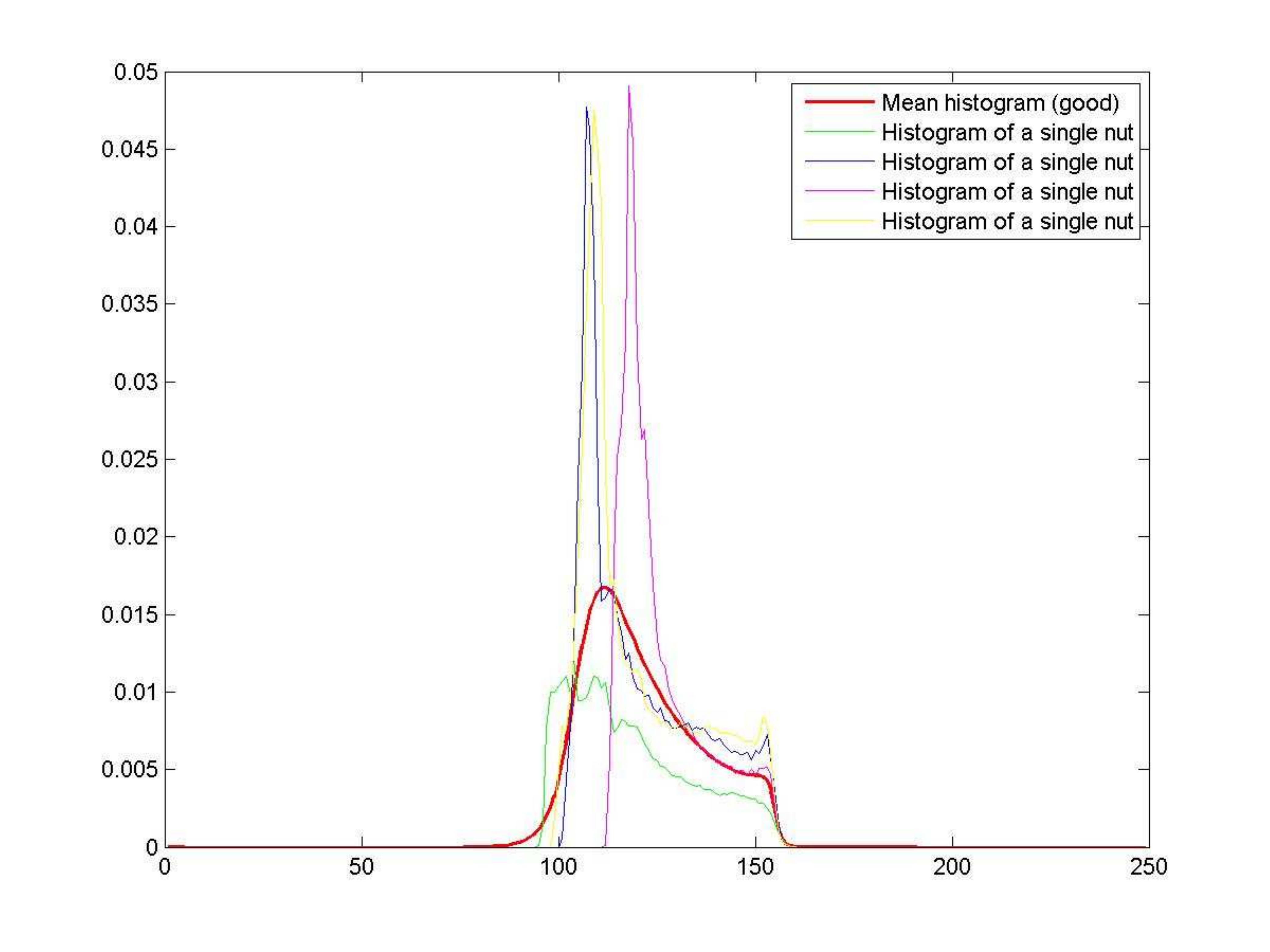}
\includegraphics[width=0.45\textwidth]{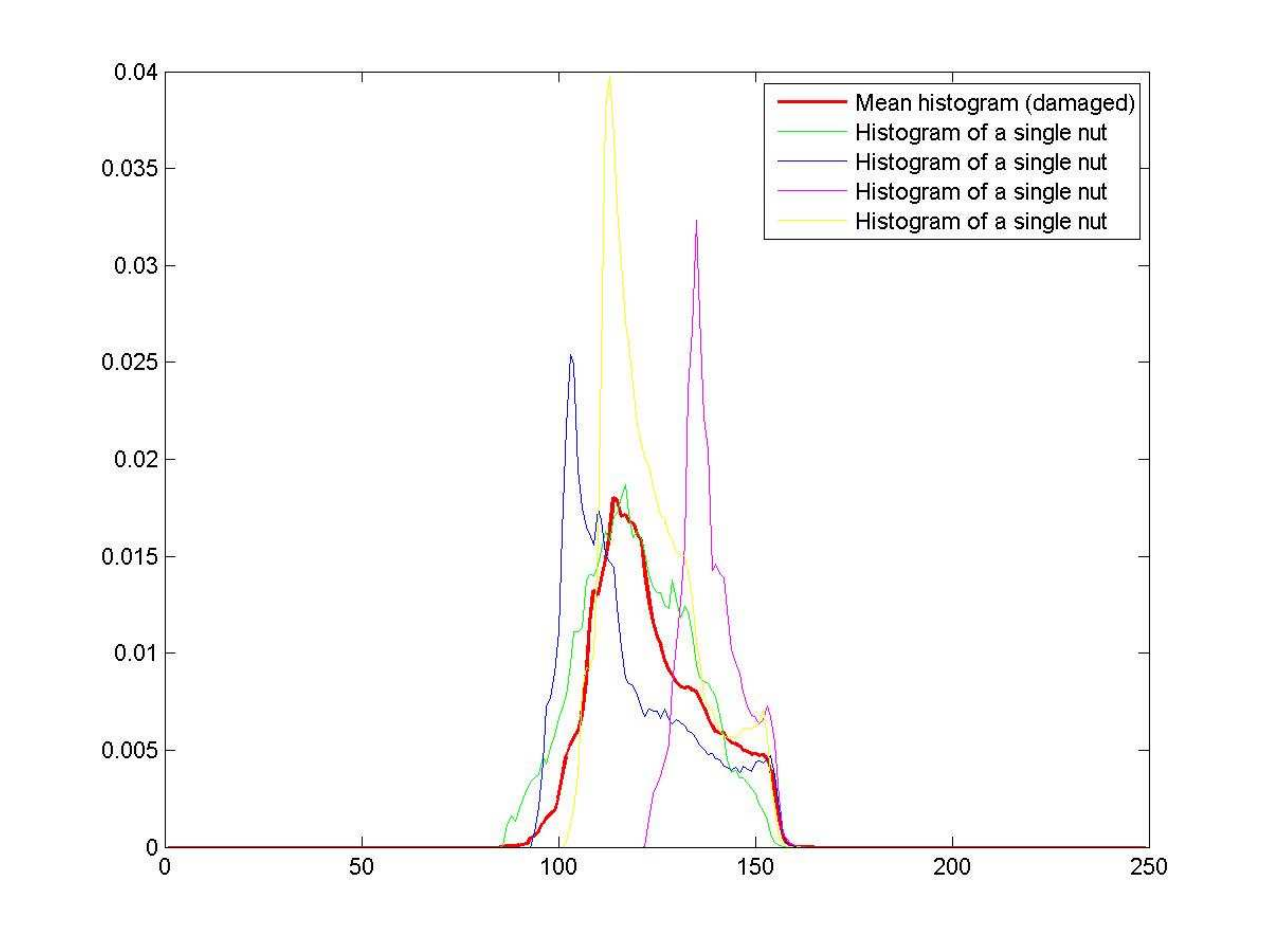}\\
\includegraphics[width=0.45\textwidth]{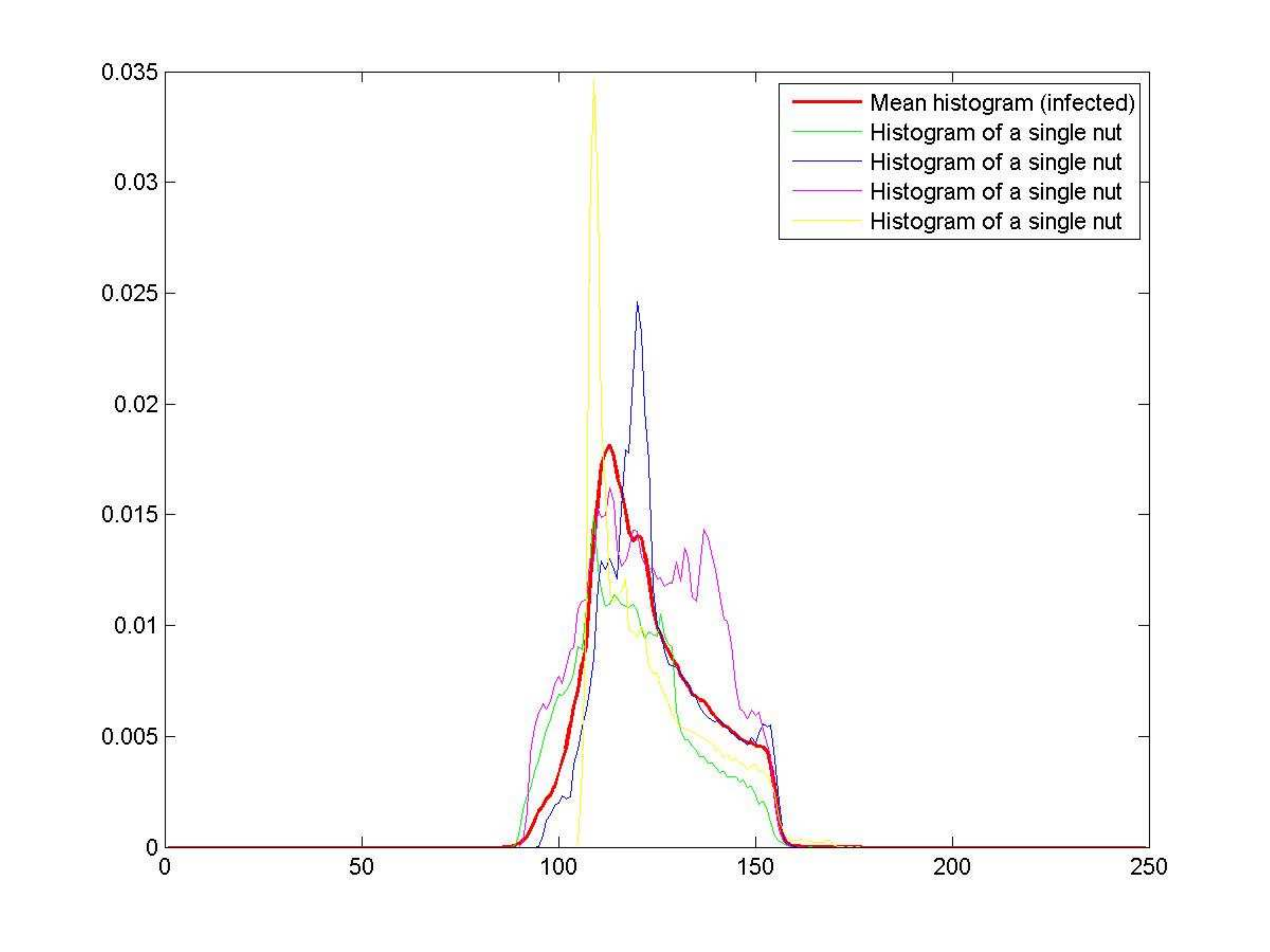}
\includegraphics[width=0.45\textwidth]{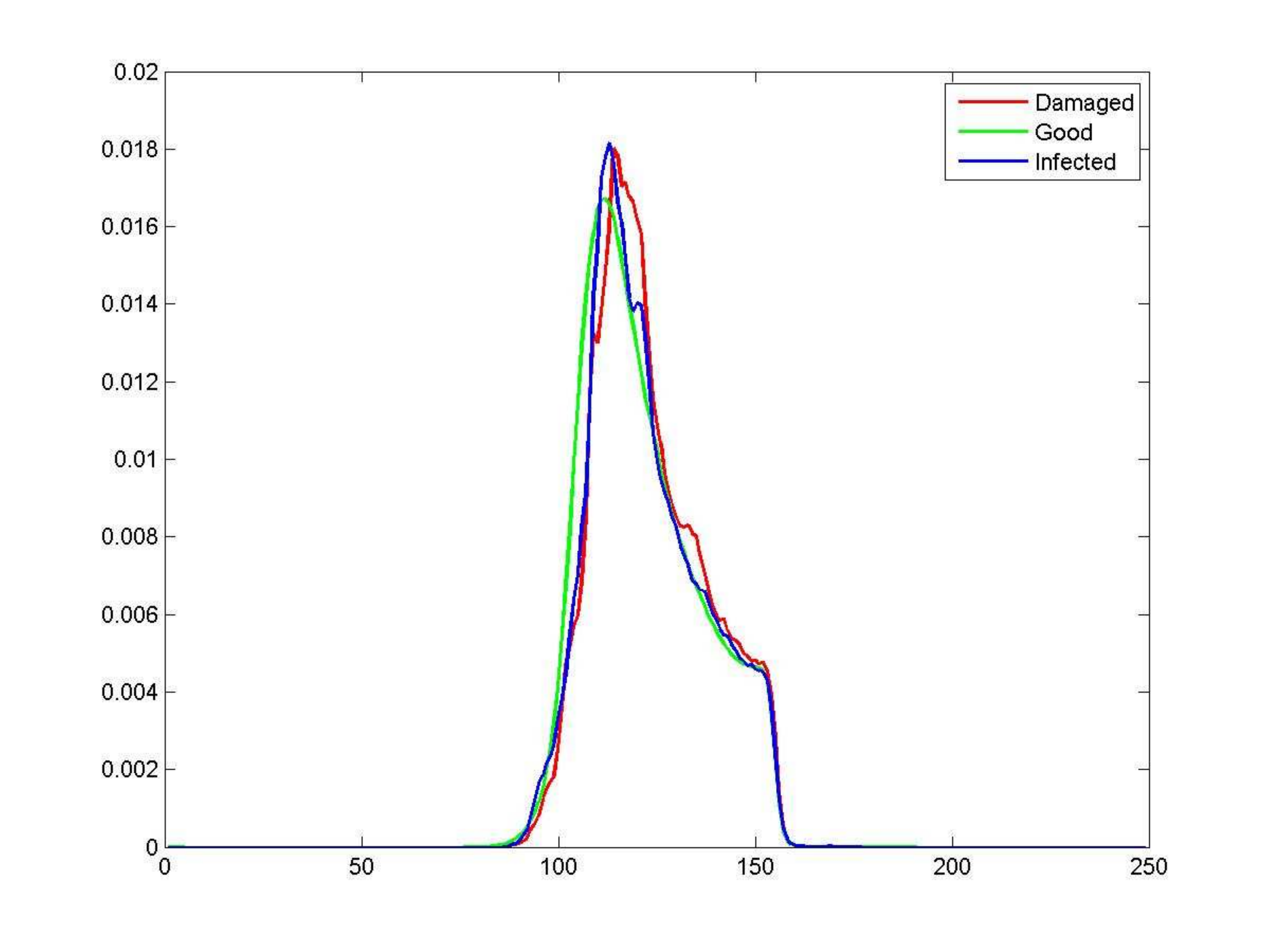}
\caption{Image histograms of good hazelnuts (top left), damaged hazelnuts (top right), infected hazelnuts (bottom left). The horizontal axis displays the shades of gray, conventionally running from $0$ to $255$. For each of the three sets, the figures display the (normalized) histograms of single hazelnuts as well as the (normalized) mean histogram. On the bottom right corner, the mean histograms of the three different sets are compared.}\label{hazelnuts2}
\end{figure}

In this work we consider the problem of pattern recognition applied to a sequence of hazelnut images, to be categorized into three different sets: ``good'' ($G$), ``damaged'' ($D$) and ``infected'' ($I$). In the sequel, we will use the shorthand notation $\mathcal{S}=\{G,D,I\}$, and, for any $A \in \mathcal{S}$, we will also denote $N_A=card(A)$. Our database consists of a set of $800$ x-ray scanned images, cf. Fig. \ref{hazelnuts}, with $N_G=750$, $N_D=25$ and $N_I=25$. The analysis outlined below is meant to provide a guiding strategy to assess, and possibly enhance, the performance of pattern recognition methods based on ANN algorithms. The prominent distinctive features of the three sets $G$, $D$ and $I$ are not detectable from a solely visual inspection of the x-ray images. Hence, in order to extract some valuable information, we relied on the computation of the histograms of the hazelnut images, shown in Fig. \ref{hazelnuts2}.

Therefore, for any $A \in \mathcal{S}$, we computed the number of pixels, in the image pertaining to the $i$-th hazelnut belonging to the set $A$ (with $i= 1,...,N_{A}$), characterized by the shade of gray $j$ (conventionally running from the value $0$ - black - to $255$ - white). After normalizing wrt the total number of pixels forming the same image, we thus obtained the so-called image histogram $p_i^{(A)}(j)$. We could also compute, then, the mean histogram pertaining to $A$, denoted by $\overline{p}_i^{(A)}(j)$, which was obtained by averaging over the $N_A$ histograms $p_i^{(A)}(j)$.

A quantitative characterization of the images can be afforded by introducing various notions of ``distance'' between different histograms \cite{Cha}: we considered, in particular, the norm in $L^1$, in $L^2$ (euclidean), in $L^\infty$, the Squared $\chi^2$ distance and the Jeffrey's divergence \cite{SHC}. 
It is worth briefly recalling some basic aspects concerning the latter two notions of distance, borrowed from probability theory.
The Squared $\chi^2$ distance corresponds to the symmetrized version of the Pearson's $\chi^2$ test \cite{Plackett}, which, given a histogram $p(j)$ and a reference histogram $q(j)$,  defines their relative distance as:
\be
d_{\chi^2}=\sum_j\frac{(p(j)-q(j))^2}{q(j)} \label{chisq} \quad .
\ee
Thus, the quantity $d_{\chi^2}$ in (\ref{chisq}) resembles the standard euclidean distance between the two histograms, except that it introduces a weight corresponding to the inverse of the reference histogram.\\
On the other hand, the Jeffreys' divergence \cite{Jeffreys} belongs to the Shannon entropy family \cite{Beck}, and corresponds to the symmetrized version of the Kullback-Leibler (K-L) divergence (or \textit{relative entropy}) \cite{KL}, defined as:
\be
d_{K-L}(p\|q)=\sum_j \left(p(j)\log\left(\frac{p(j)}{q(j)}\right) \right)=H(p,q)-H(p) \label{KL} \quad ,
\ee
where $H(p,q)$ is the cross entropy of $p$ and $q$, and $H(p)$ is the entropy of $p$ \cite{Kull,Jay}.
More in general, the K-L divergence (\ref{KL}), is a member of the family of the so-called $f$-divergencies \cite{Mori,Ali} and stems as a limiting case of the more general R\'enyi's (or $\alpha$-) divergence \cite{Xu}. It is worth recalling its definition: given any two continuous distributions $p$ and $q$, over a space $\Omega$, with $p$ absolutely continuous wrt $q$, the $f$-divergence of $p$ from $q$ is
\be
d_{f}(p\|q)=\int_\Omega f\left(\frac{dp}{dq}\right)dq \quad ,
\ee
where $f$ is a convex function such that $f(1)=0$.\\
Then, for any $A \in \mathcal{S}$, we considered the distance (or \textit{fluctuation}), defined according to the various notions introduced above, between the histogram $p_i^{(A)}(j)$ and the corresponding mean $\overline{p}_i^{(A)}(j)$. Next, by averaging over the set $A$, one obtains a characteristic ``statistical scale'' (still depending on the chosen notion of distance) characterizing the fluctuations within each set $A$. 
To clarify the meaning of the entries in Tab. \ref{normtot}, let us illustrate, for instance, the procedure to calculate the quantity $\langle d \rangle^{(A)}_2$. To this aim, we introduce the euclidean distance between the histograms $p_i^{(A)}(j)$ and $\overline{p}_i^{(A)}(j)$:
\be
d_{2,i}^{(A)}=\sqrt{\sum_{j=1}^{N_g}|p_i^{(A)}(j)-\overline{p}_i^{(A)}(j)|^2} \label{eucl}
\ee

\begin{table}[bth]
\centering
\begin{tabular}{c|c|c|c|c|c|}
    & $\langle d \rangle^{(A)}_1$ & $\langle d \rangle^{(A)}_2$ & $\langle d \rangle^{(A)}_\infty$ & $\langle d \rangle^{(A)}_{\chi^2}$ & $\langle d \rangle^{(A)}_{J}$\\
\hline
  $A =G$ & $0.2079$ & $0.0372$ & $0.0139$ & $0.0495$ & $0.0369$ \\
\hline
  $A =D$ & $0.2485$ & $0.0488$ & $0.0162$ & $0.0776$ & $0.0477$ \\   
\hline
  $A =I$ & $0.2097$ & $0.0379$ & $0.0145$ & $0.0435$ & $0.0401$ \\
  \hline
\end{tabular}
\caption{Typical fluctuation of the histograms of the hazelnuts from the corresponding mean histogram, within each of the sets $G$, $D$, and $I$. The quantities $\langle d \rangle^{(A)}$ are evaluated by using different notions of distances: norm in $L^1$, in $L^2$ (euclidean), in $L^\infty$, Squared $\chi^2$ distance and Jeffreys divergence.}
          \label{normtot}
          \end{table}

\begin{table}[bth]
\centering
\begin{tabular}{c|c|c|c|c|c|}
    & $\Delta^{(A,B)}_{1}$ & $\Delta^{(A,B)}_{2}$ & $\Delta^{(A,B)}_{\infty}$ & $\Delta^{(A,B)}_{\chi^2}$ & $\Delta^{(A,B)}_{J}$\\
\hline
  $A=G, B=D$ & $0.0923$ & $0.0162$ & $0.0036$ & $0.0089$ & $0.0200$ \\
\hline
  $A=D, B=I$ & $0.0533$ & $0.0090$ & $0.0028$ & $0.0021$ & $0.0030$ \\   
\hline
  $A=G, B=I$ & $0.0526$ & $0.0115$ & $0.0051$ & $0.0044$ & $0.0124$ \\  
  \hline
\end{tabular}
\caption{Average distances between between pairs of mean histograms referring to two different sets $A$ and $B$, evaluated, as in Tab \ref{normtot}, using different notions of distance: norm in $L^1$, in $L^2$ (euclidean), in $L^\infty$, Squared $\chi^2$ distance and Jeffreys divergence.}
          \label{Deltatot}
          \end{table}

\begin{figure}[H]
\centering
\includegraphics[width=0.31\textwidth]{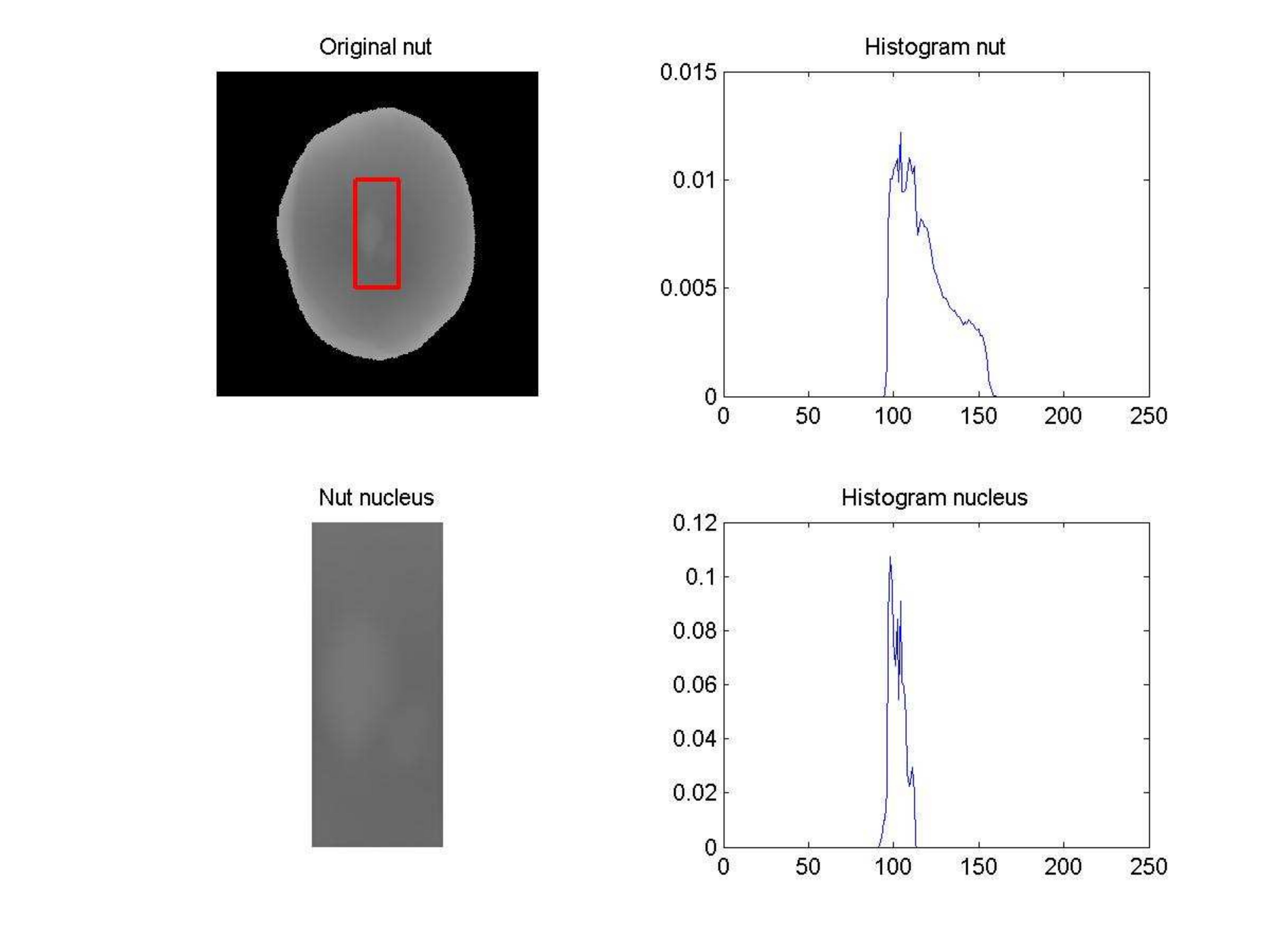}
\includegraphics[width=0.31\textwidth]{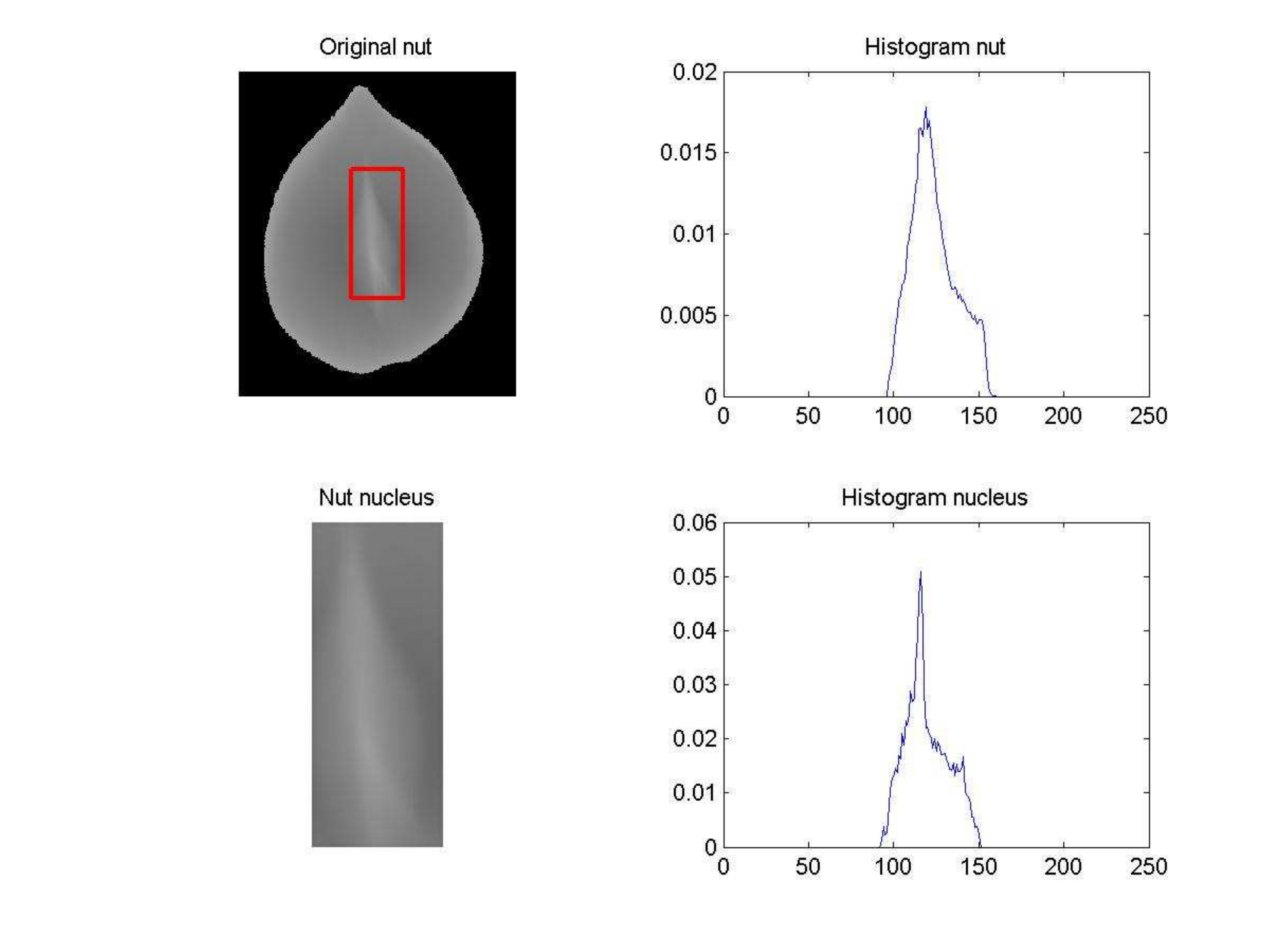}
\includegraphics[width=0.31\textwidth]{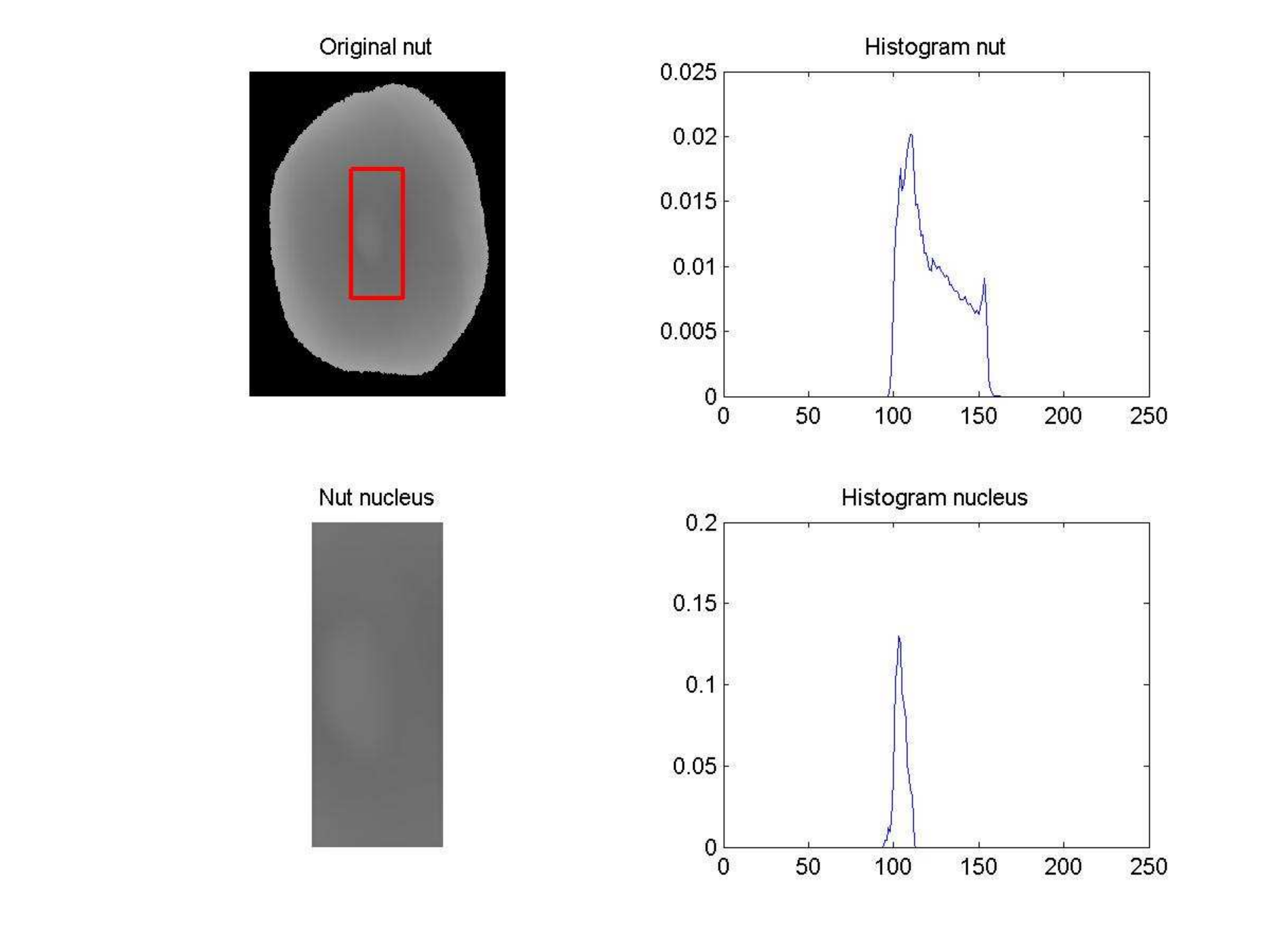}\\
\includegraphics[width=0.31\textwidth]{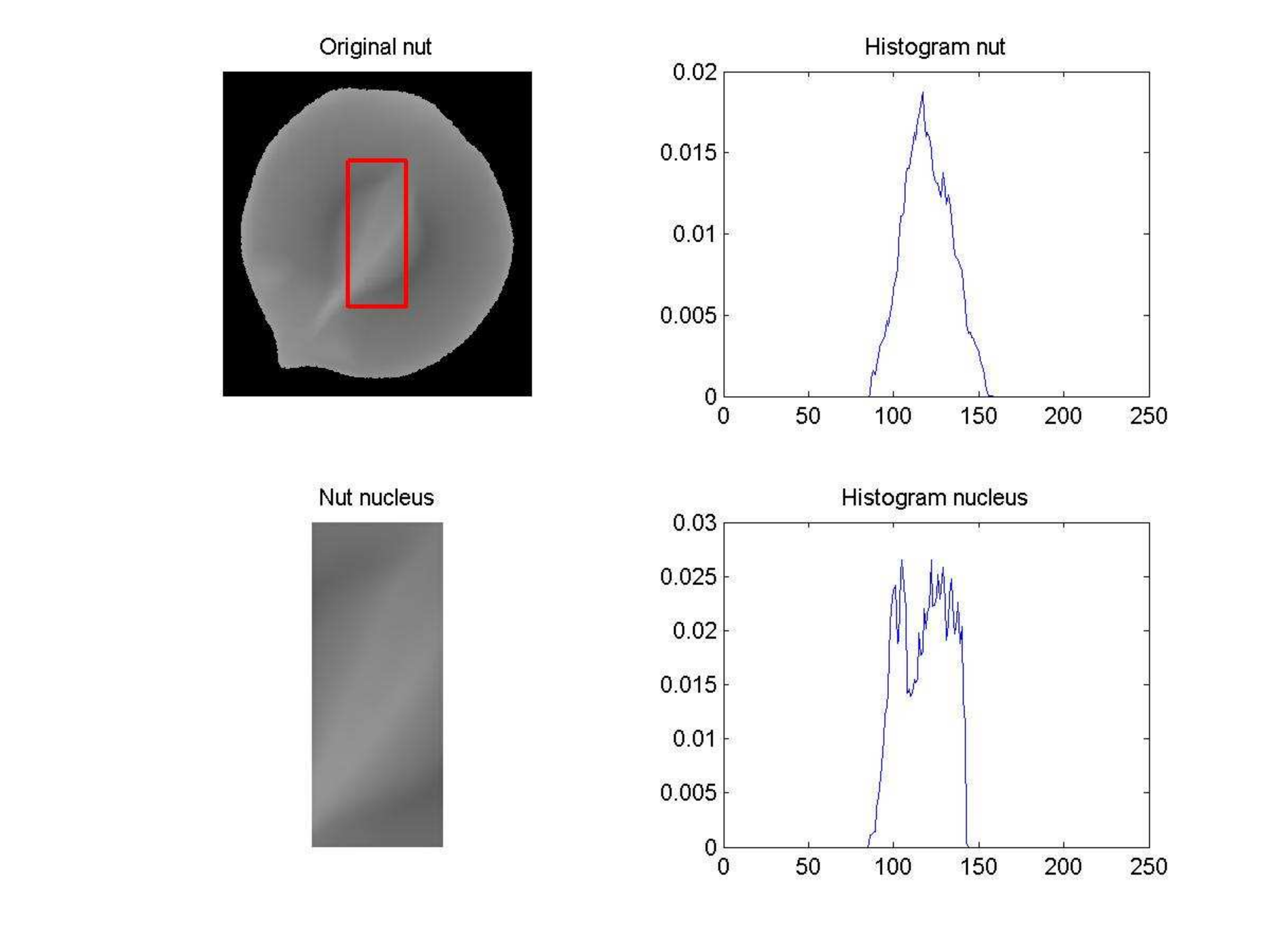}
\includegraphics[width=0.31\textwidth]{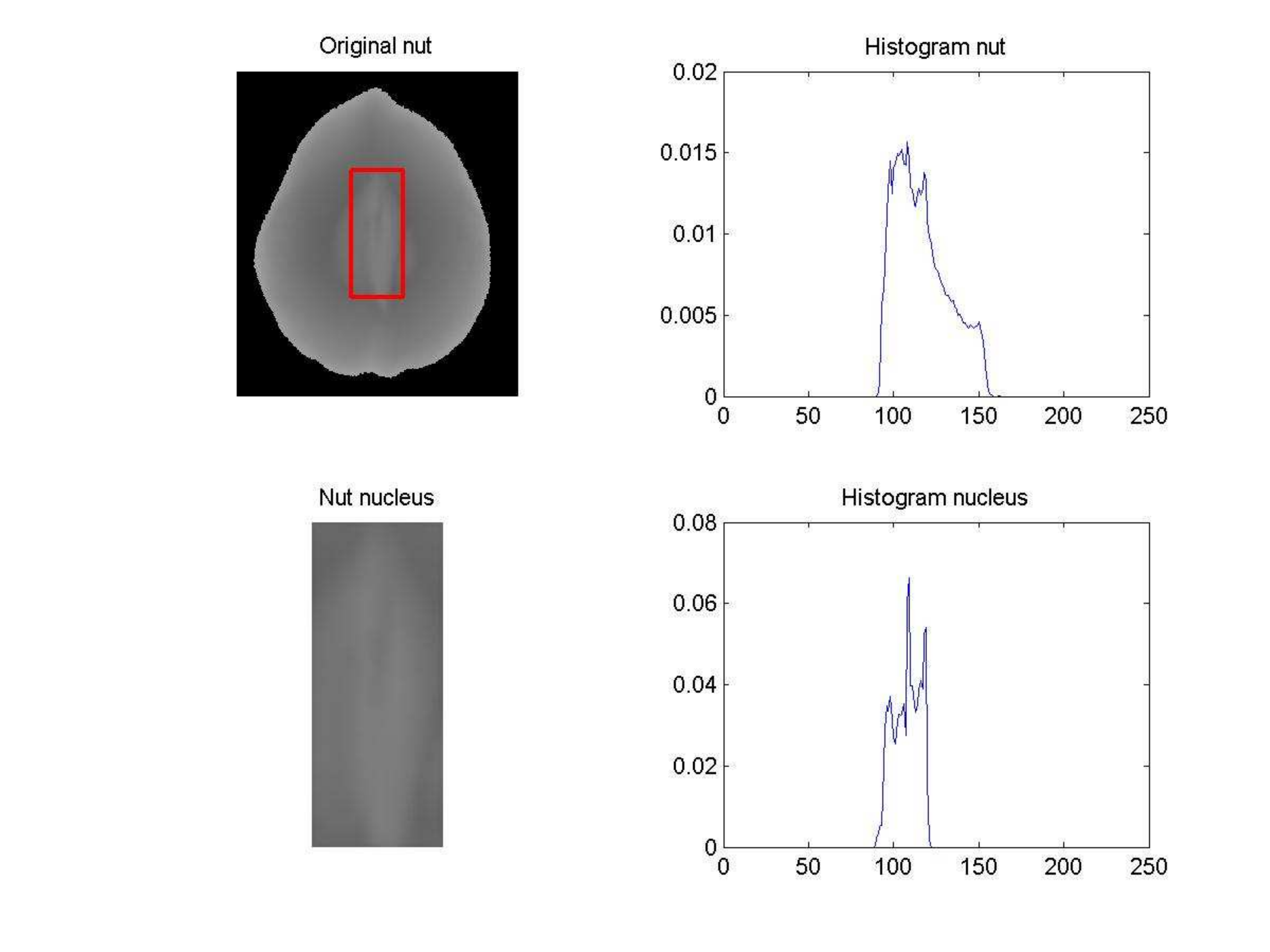}
\includegraphics[width=0.31\textwidth]{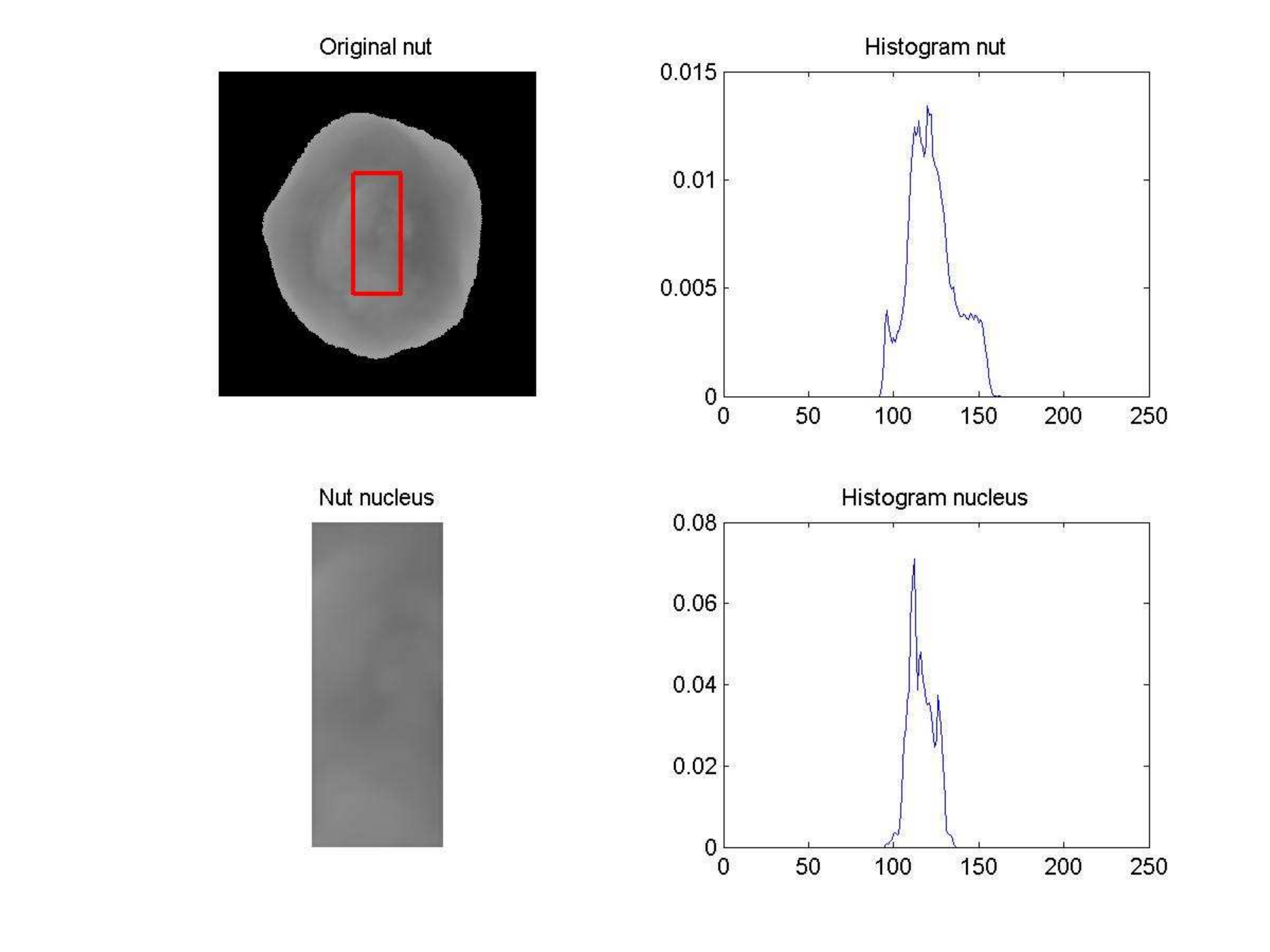}\\
\includegraphics[width=0.31\textwidth]{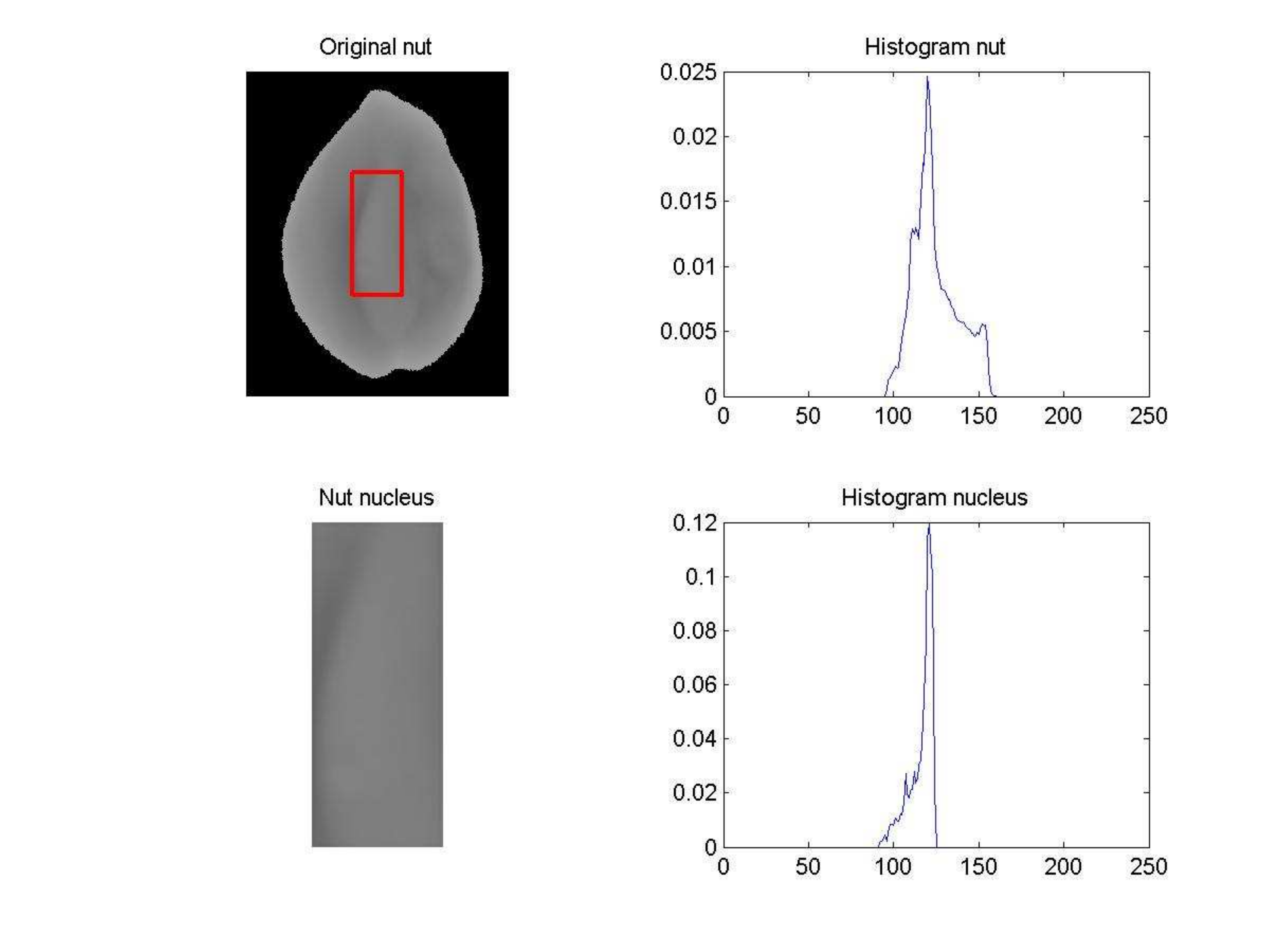}
\includegraphics[width=0.31\textwidth]{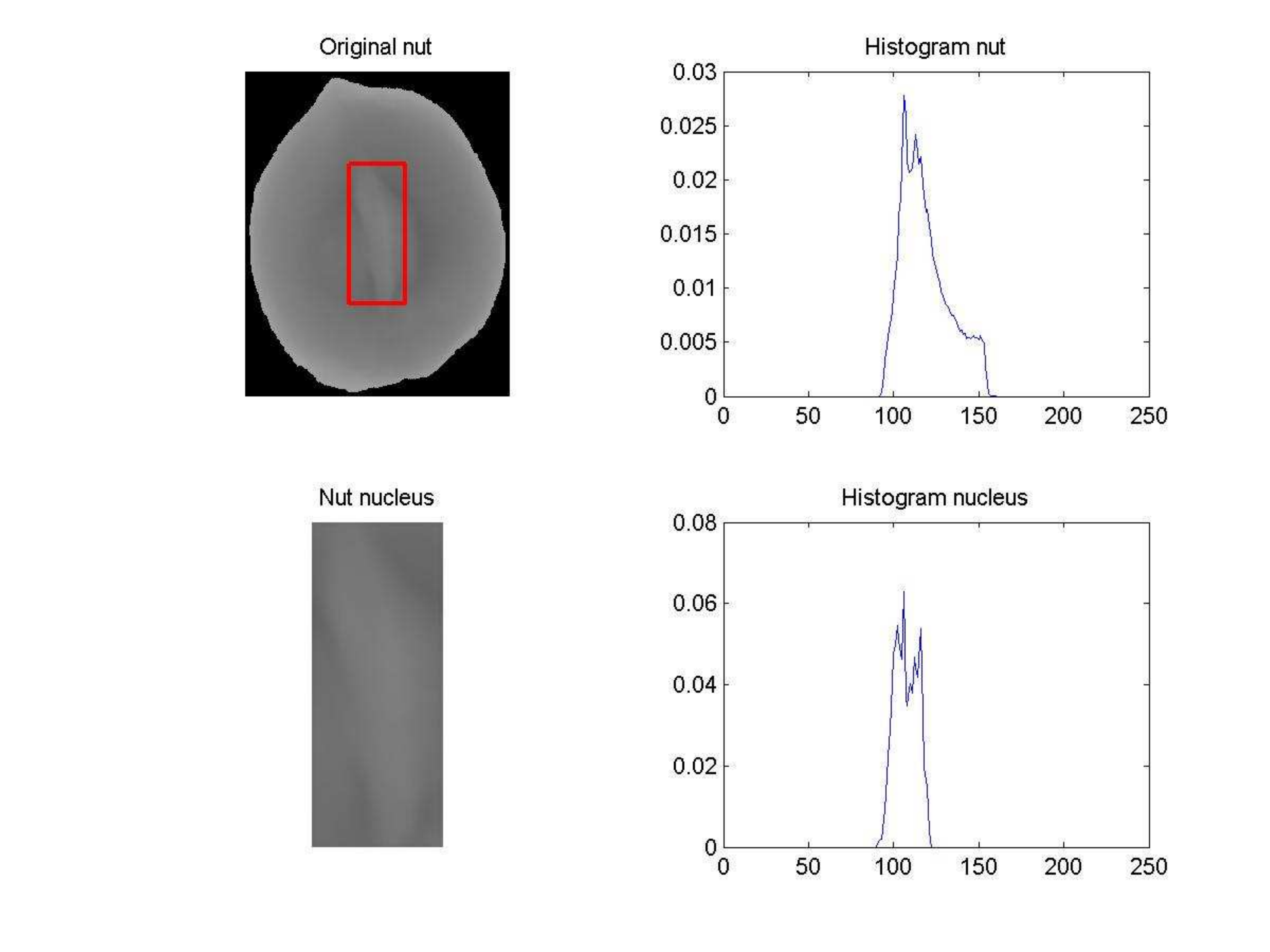}
\includegraphics[width=0.31\textwidth]{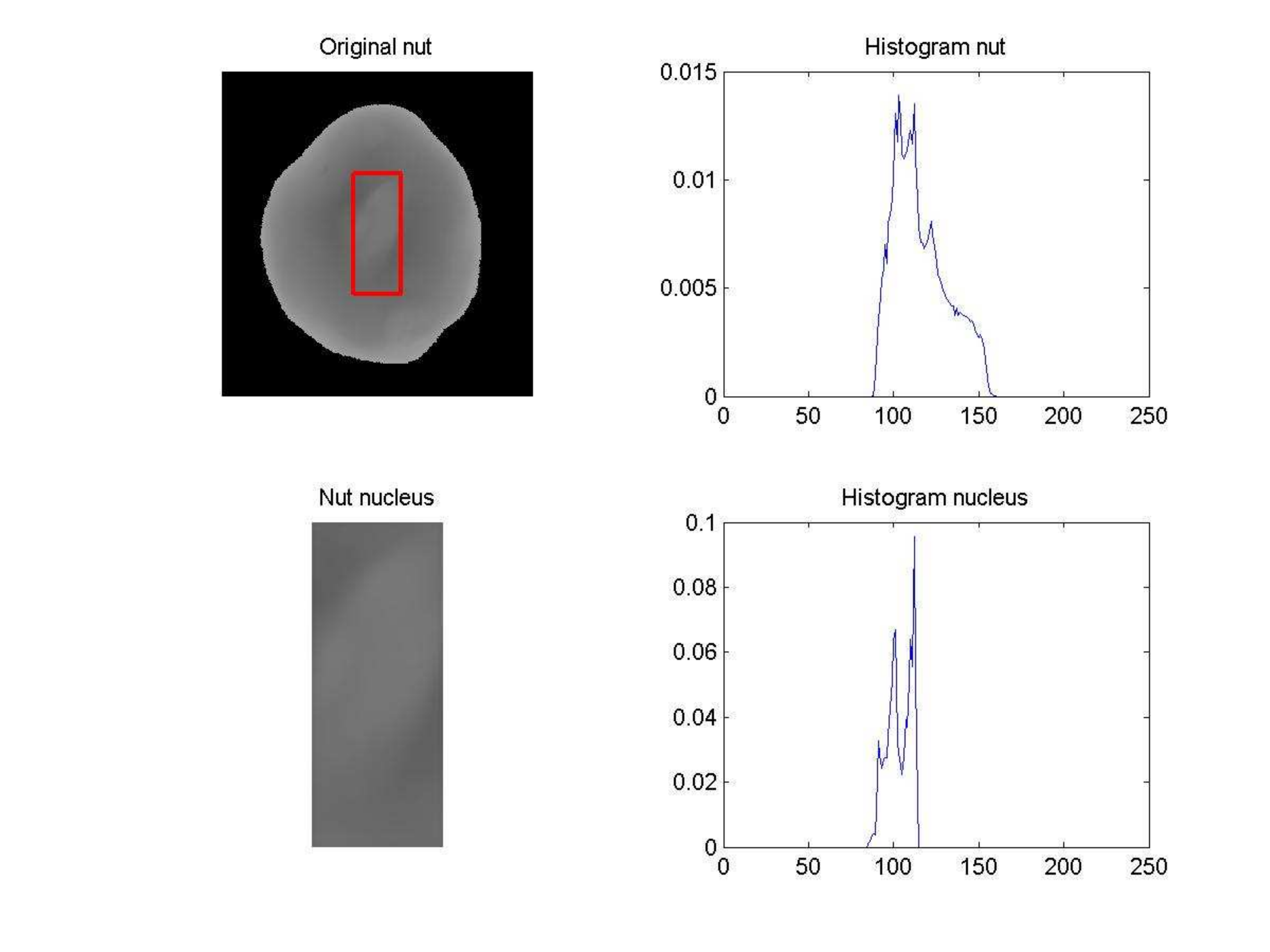}
\caption{Image histograms of good hazelnuts (top row), damaged hazelnuts (middle row) and infected hazelnuts (bottom row). Each image shows the histogram of the entire hazelnut (top histogram) and the histogram referring to the fraction of the image delimited by the thick red rectangles, characterized by $\epsilon=80$ and $\rho=2.5$.}\label{hazelnuts3}
\end{figure}

From the knowledge of $d_{2,i}^{(A)}$ in (\ref{eucl}), the quantity $\langle d \rangle^{(A)}_2$, shown in Tab. \ref{normtot}, is then computed by averaging over $A$:
\be
\langle d \rangle_2^{(A)} =\frac{1}{N_{A}}\sum_{i=1}^{N_{A}}d_{2,i}^{(A)} \label{aver}
\ee 
It is worth noticing, from Tab. \ref{normtot}, that, no matter of what notion of distance is adopted, the magnitude of the fluctuations is not significantly affected by $N_{A}$.
The scale $\langle d \rangle^{(A)}$, which, for any $A \in \mathcal{S}$, is of the order $\langle d \rangle^{(A)}\simeq 10^{-2}$, can be thus regarded as an intrinsic statistical scale pertaining to the set $A$. 
It is worth comparing such scale with another statistical scale, denoted by $\Delta^{(A,B)}$, whose values are listed in Tab. \ref{Deltatot}. The quantity $\Delta^{(A,B)}$ is defined as the distance, computed by using the various notions of distance introduced above, between the pair of mean histograms relative to the sets $A$ and $B$, with $(A,B)\in\mathcal{S}$ and $A \neq B$. The symmetric form of the distances introduced above entails, in particular, that $\Delta^{(A,B)}=\Delta^{(B,A)}$. 
A better interpretation of the meaning of the scales $\langle d \rangle^{(A)}$ and $\Delta^{(A,B)}$ can be achieved by noticing that a large value of $\langle d \rangle^{(A)}$ mirrors the presence of a considerable amount of noise on top of the mean histogram $\overline{p}_i^{(A)}(j)$, which thus blurs the distinctive features of the set $A$. On the contrary, a larger value of $\Delta^{(A,B)}$ reflects a more significant separation between the mean histograms of the two sets $A$ and $B$, which instead favours the pattern recognition. In the sequel of this Section we will focus, therefore, on the ratio of two such scales.
From an inspection of Tabs. \ref{normtot} and \ref{Deltatot}, we first observe that $\Delta^{(A,B)}\sim\langle d \rangle^{(A)}$. That is, the two scales are comparable: the fluctuations, within each set, are comparable with the typical distances between different sets. This entails, hence, that the histograms shown in Fig. \ref{hazelnuts2} can not be regarded as a useful source of information to perform a pattern recognition.
A different route can be pursued by just focusing on a selected portion of the original images. This approach is motivated by the assumption that the distinctive features of each of the three sets are mostly contained in the ``nuclei'' of the hazelnuts. We calculated, therefore, the histograms corresponding to the cropped portions of the original images, delimited by the tick red rectangles shown in Fig. \ref{hazelnuts3}. 

\begin{figure}[H]
\centering
\includegraphics[width=0.45\textwidth]{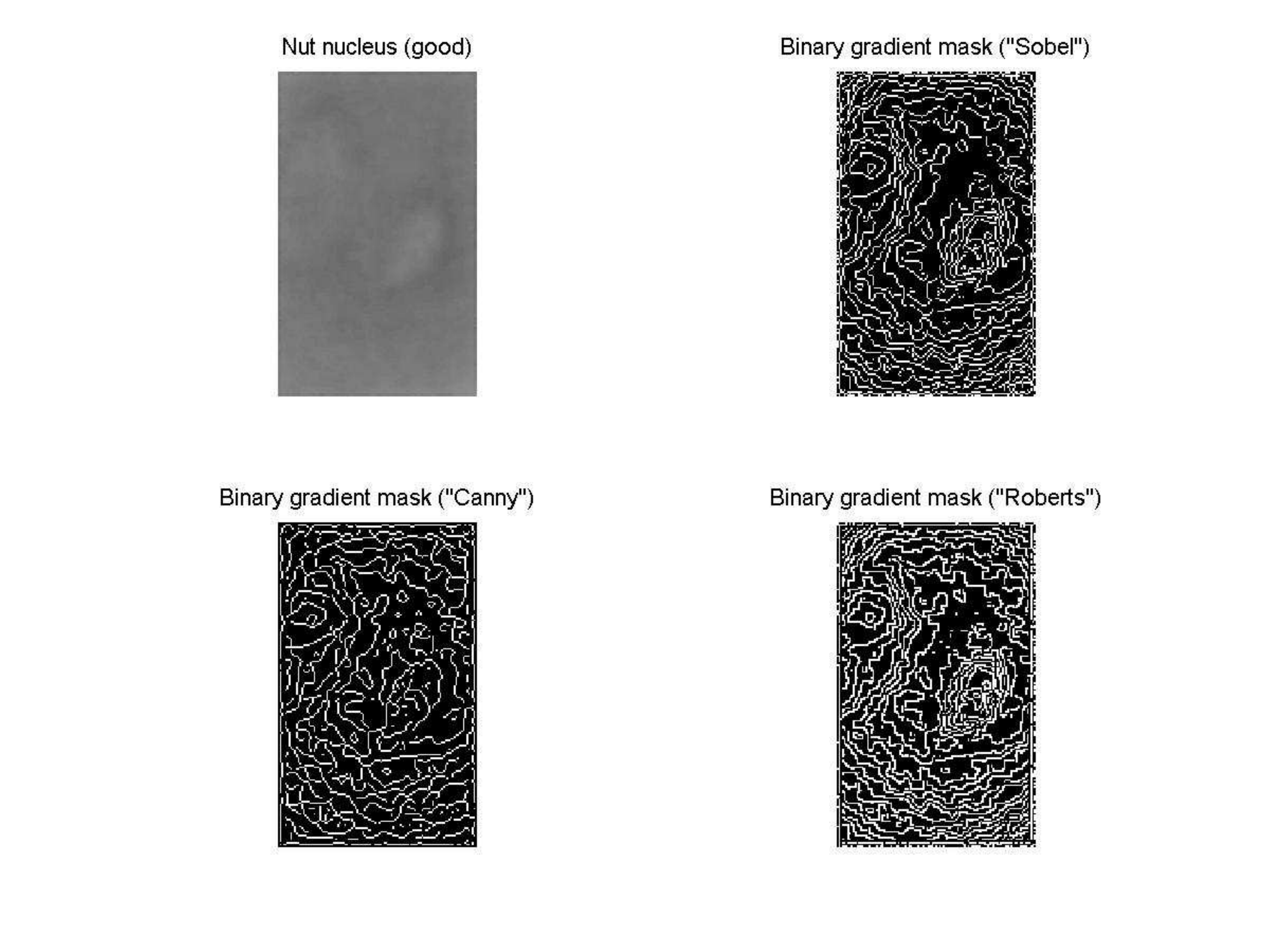}
\includegraphics[width=0.45\textwidth]{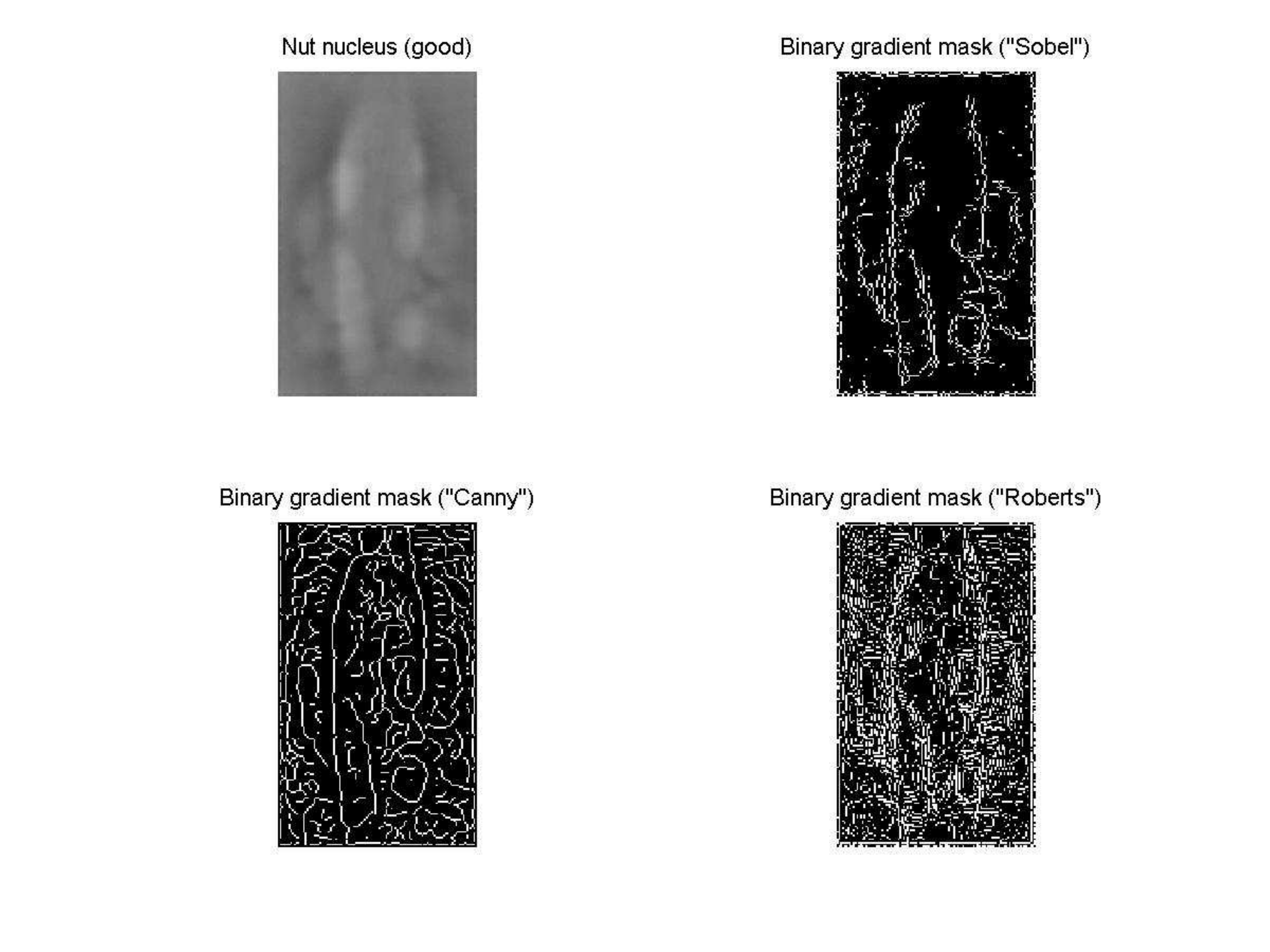}\\
\includegraphics[width=0.45\textwidth]{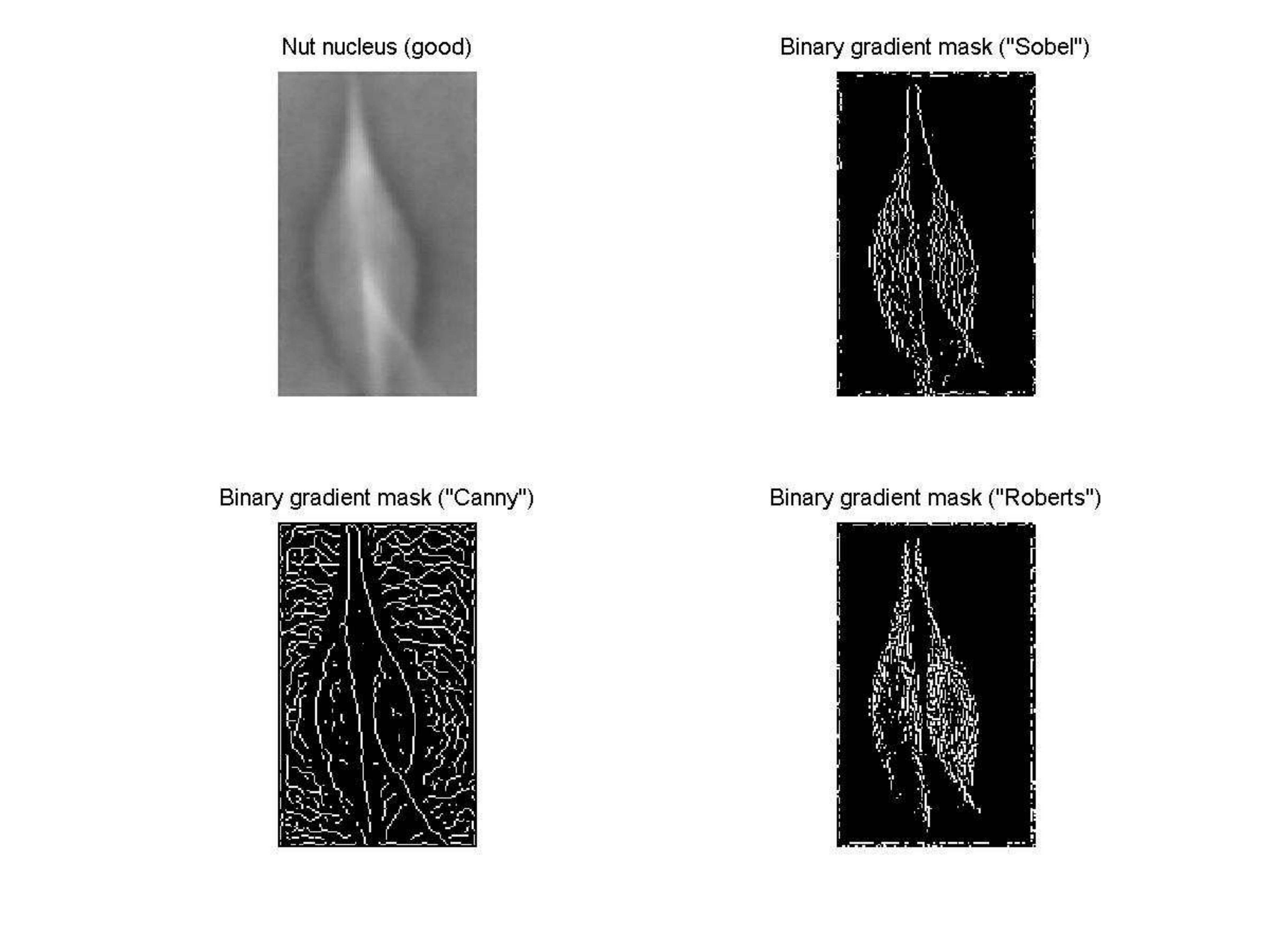}
\includegraphics[width=0.45\textwidth]{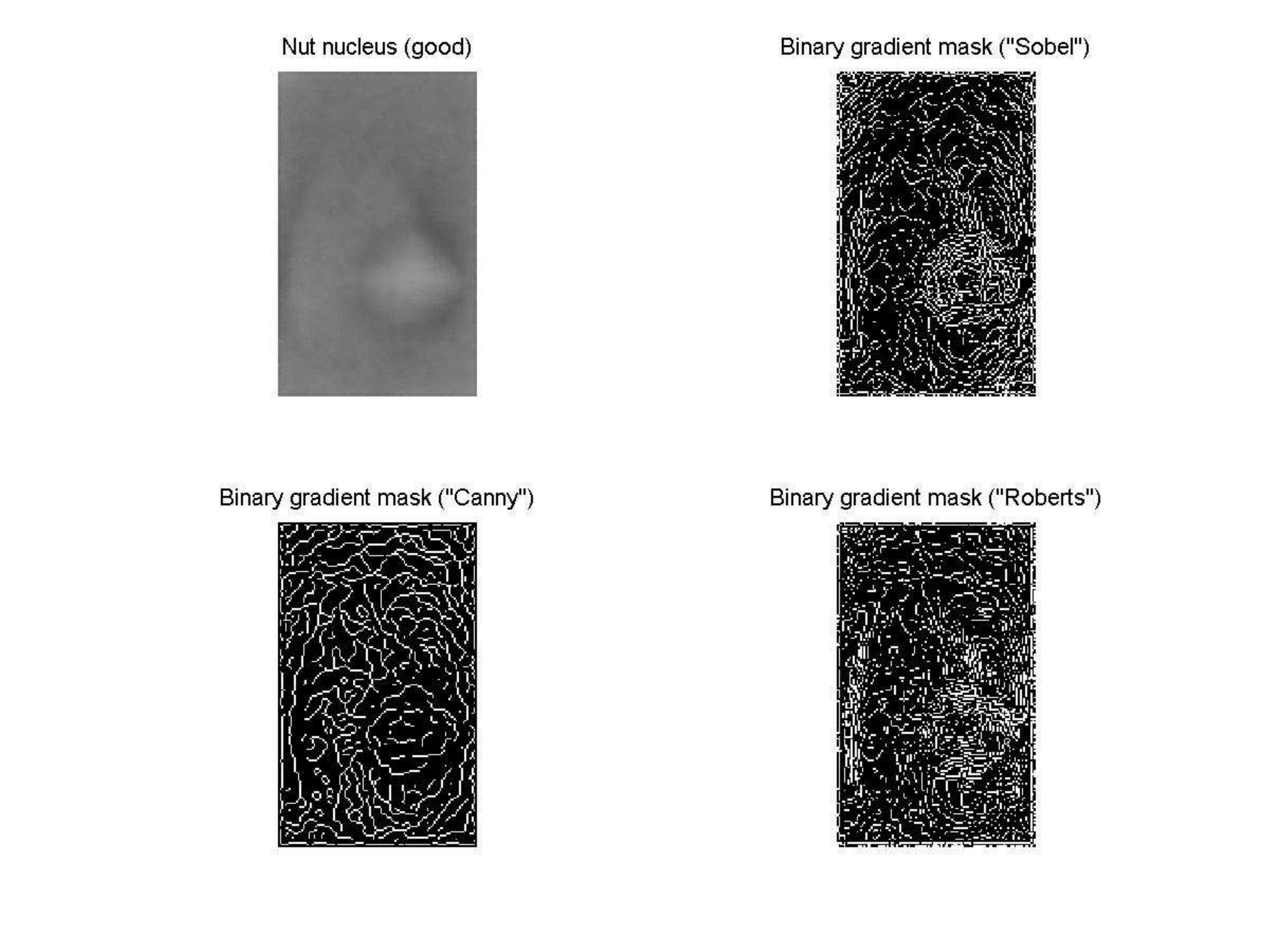}
\caption{Image processing of the hazelnut nuclei belonging to the set $G$, for $\epsilon=100$ and $\rho=1.5$, by means of edge-detection algorithms, respectively: Sobel's algorithm (top right figure) , Canny's algorithm (bottom left figure) and Roberts' algorithm (bottom right figure).}\label{nuclS}
\end{figure}

\begin{figure}[H]
\centering
\includegraphics[width=0.45\textwidth]{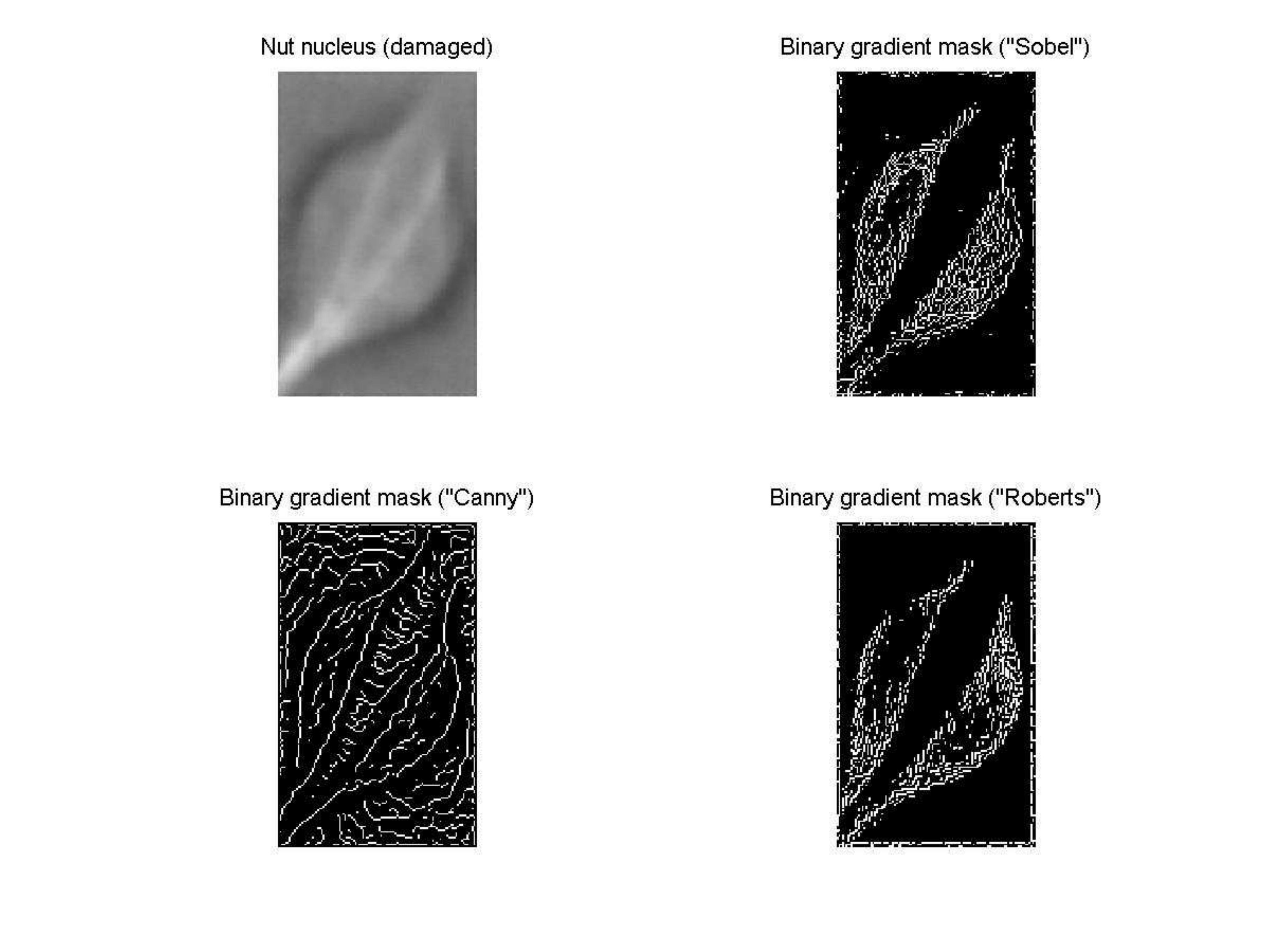}
\includegraphics[width=0.45\textwidth]{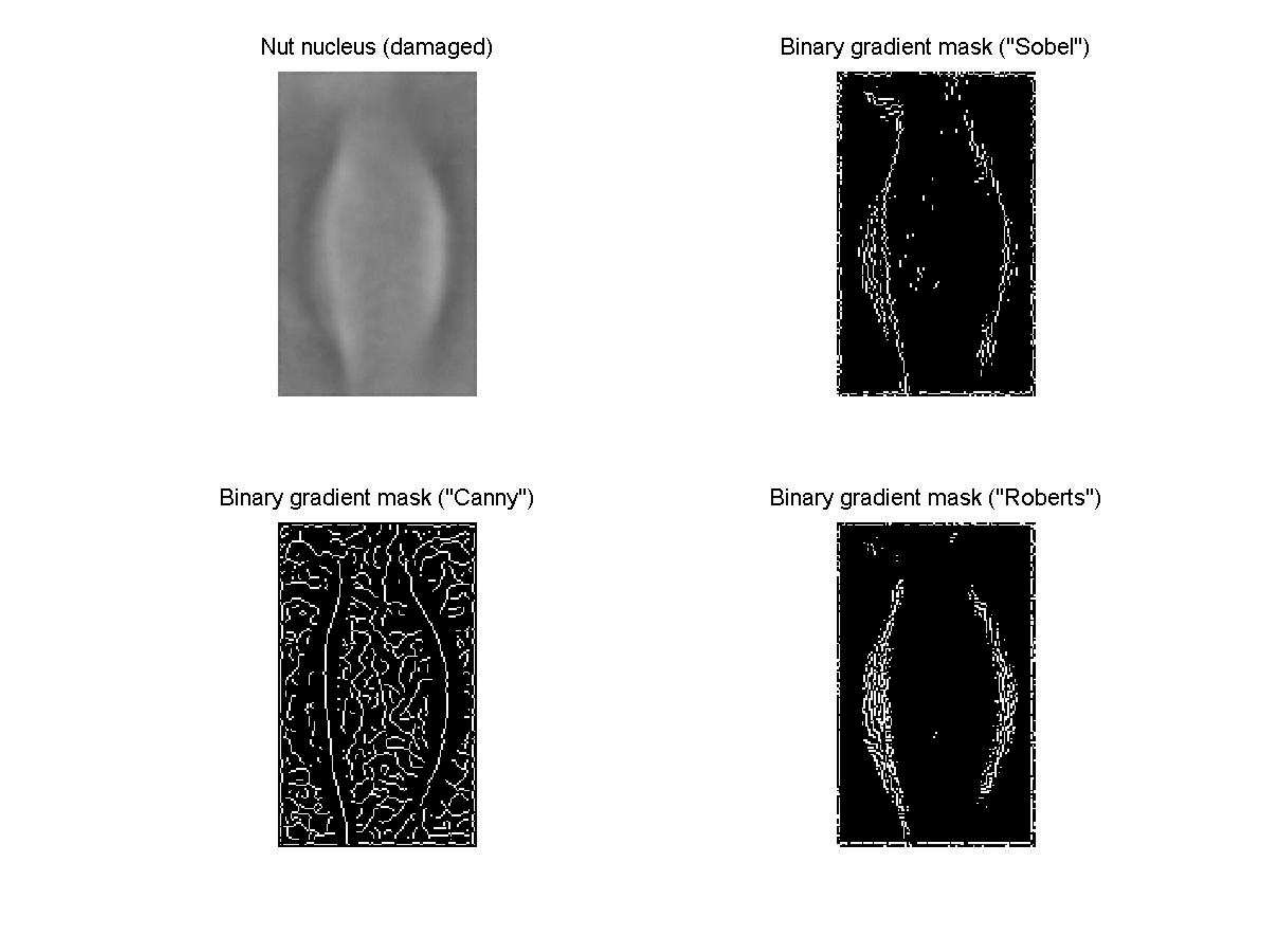}\\
\includegraphics[width=0.45\textwidth]{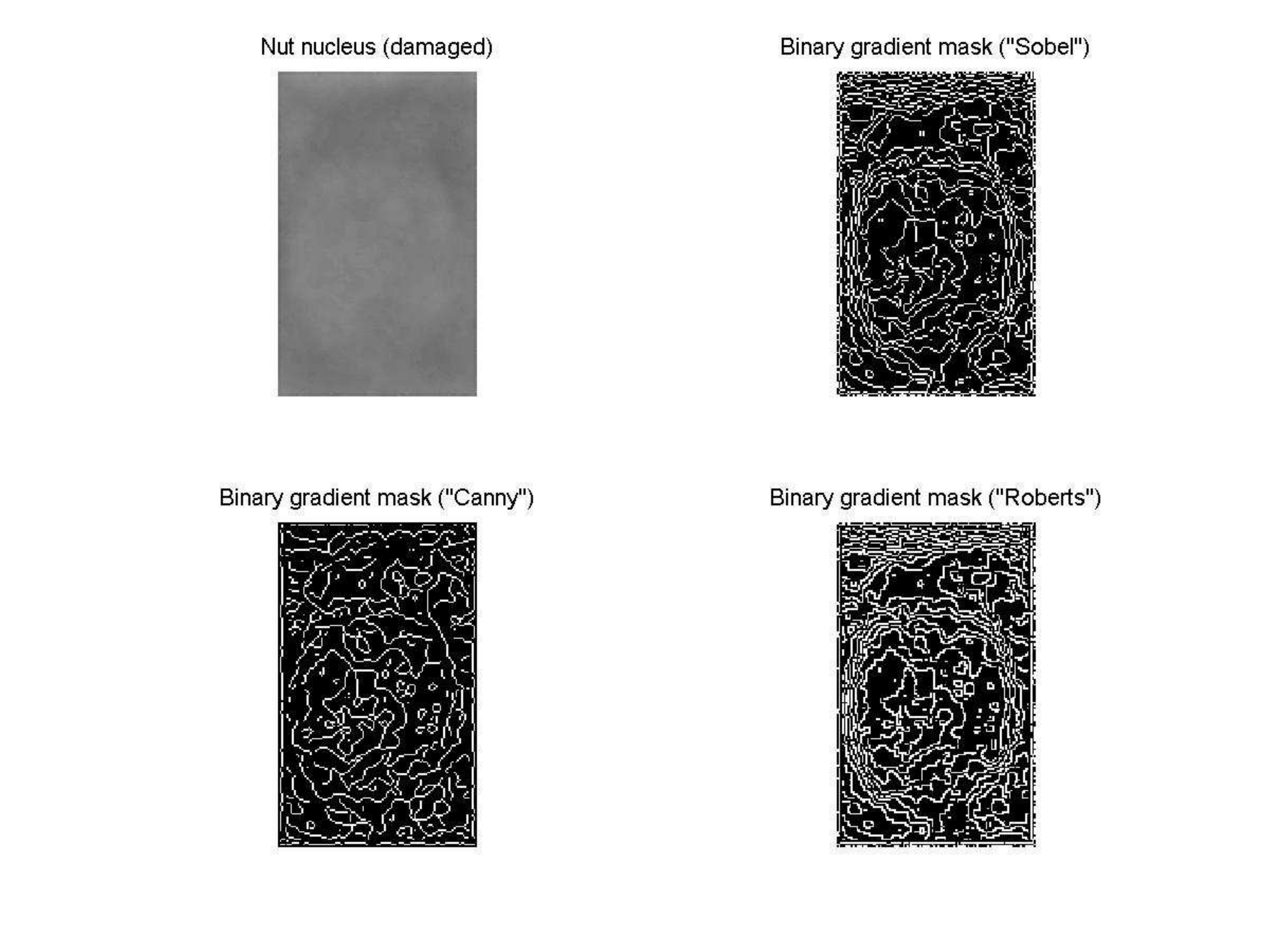}
\includegraphics[width=0.45\textwidth]{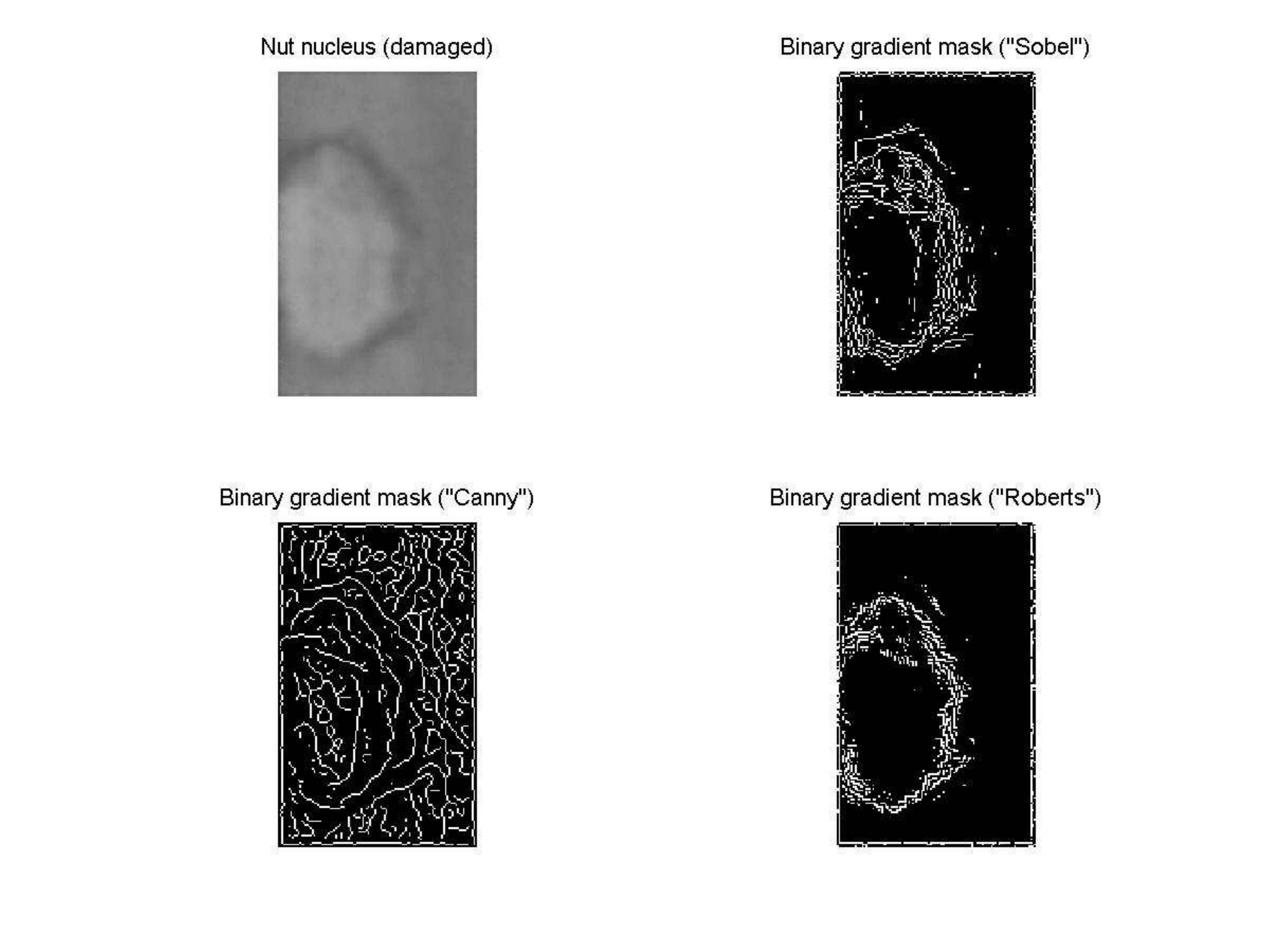}
\caption{Image processing of the hazelnut nuclei belonging to the set $D$, for $\epsilon=100$ and $\rho=1.5$, by means of edge-detection algorithms, respectively: Sobel's algorithm (top right figure) , Canny's algorithm (bottom left figure) and Roberts' algorithm (bottom right figure).}\label{nuclA}
\end{figure}

\begin{figure}[H]
\centering
\includegraphics[width=0.45\textwidth]{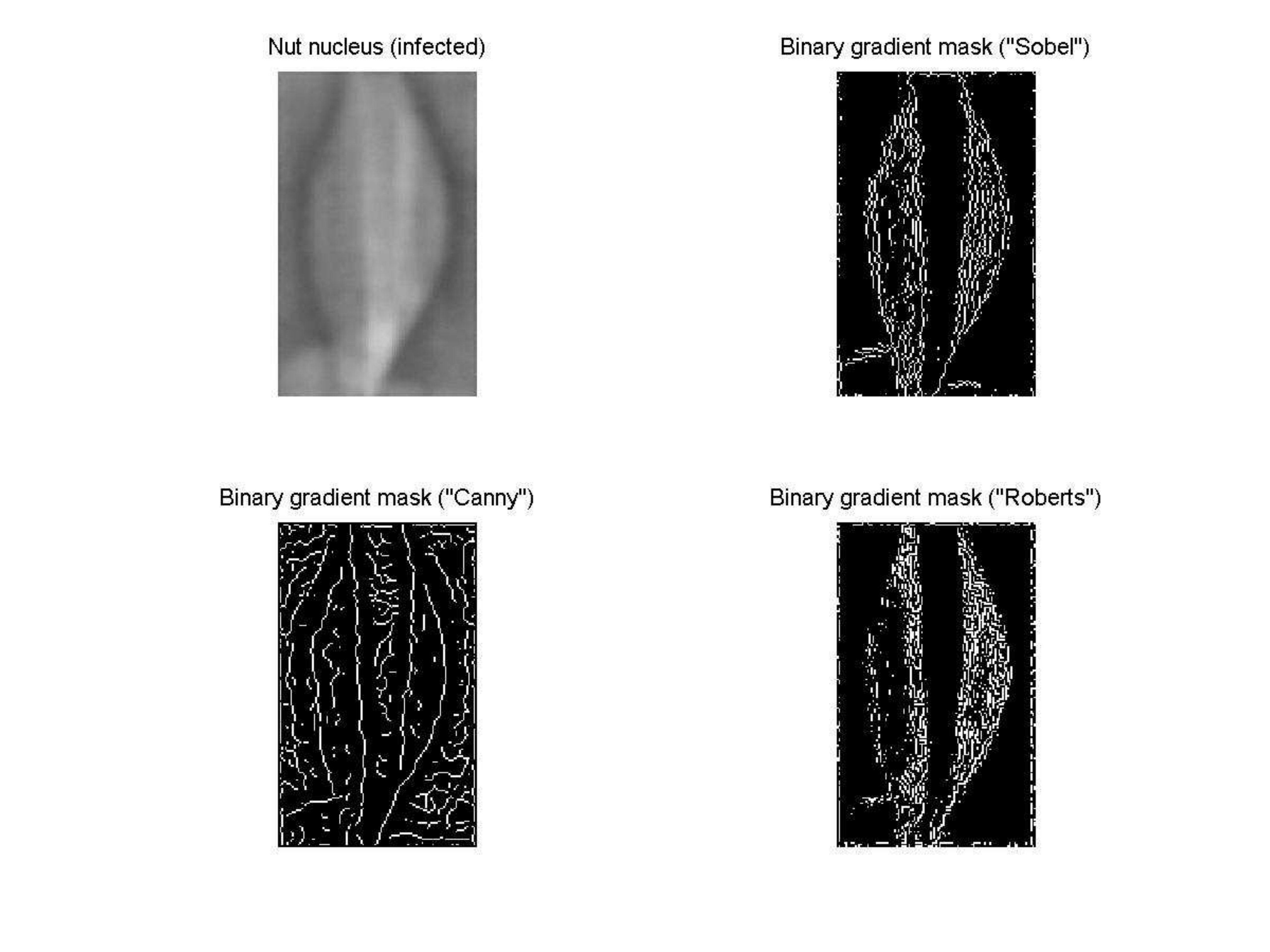}
\includegraphics[width=0.45\textwidth]{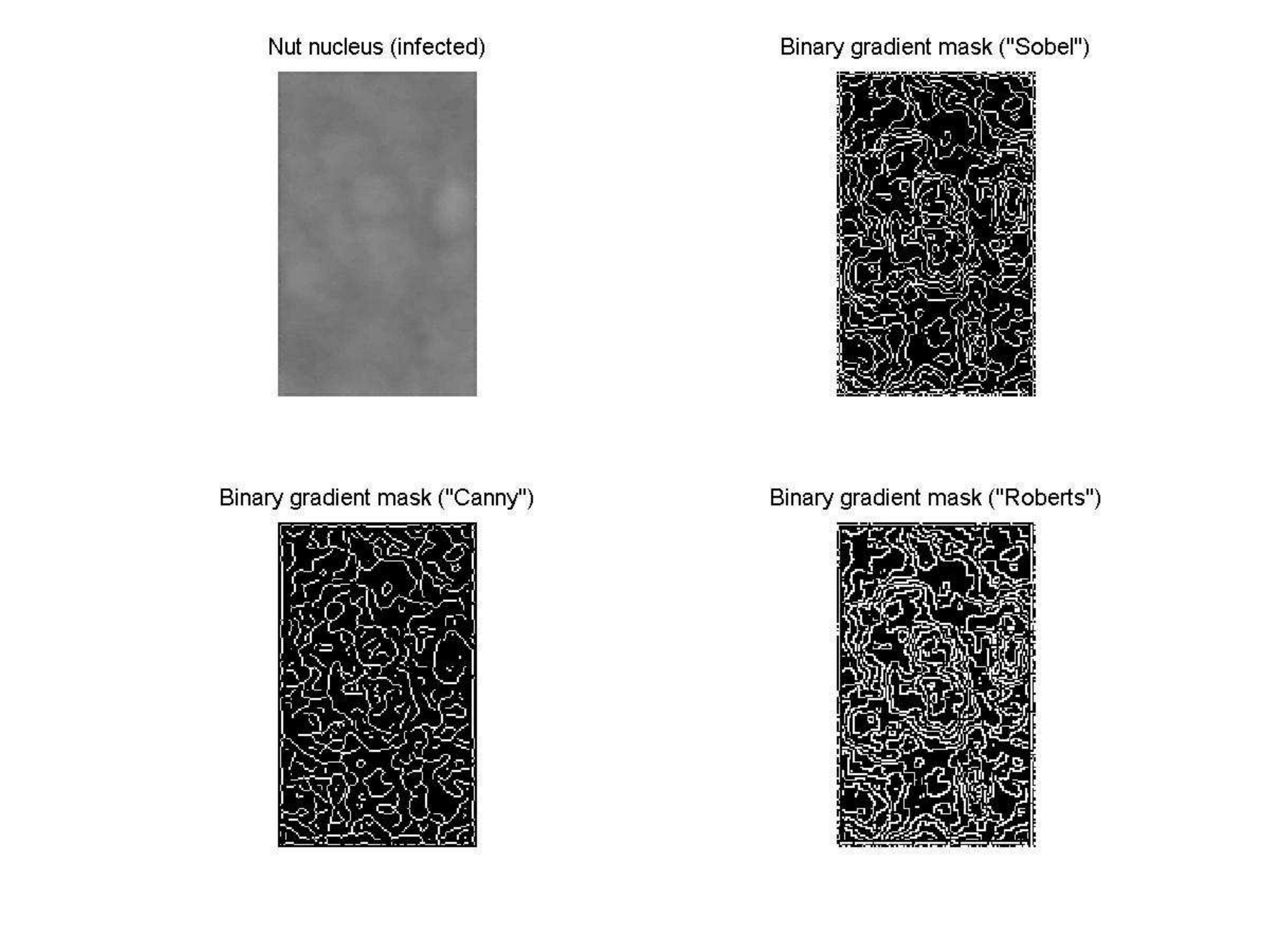}\\
\includegraphics[width=0.45\textwidth]{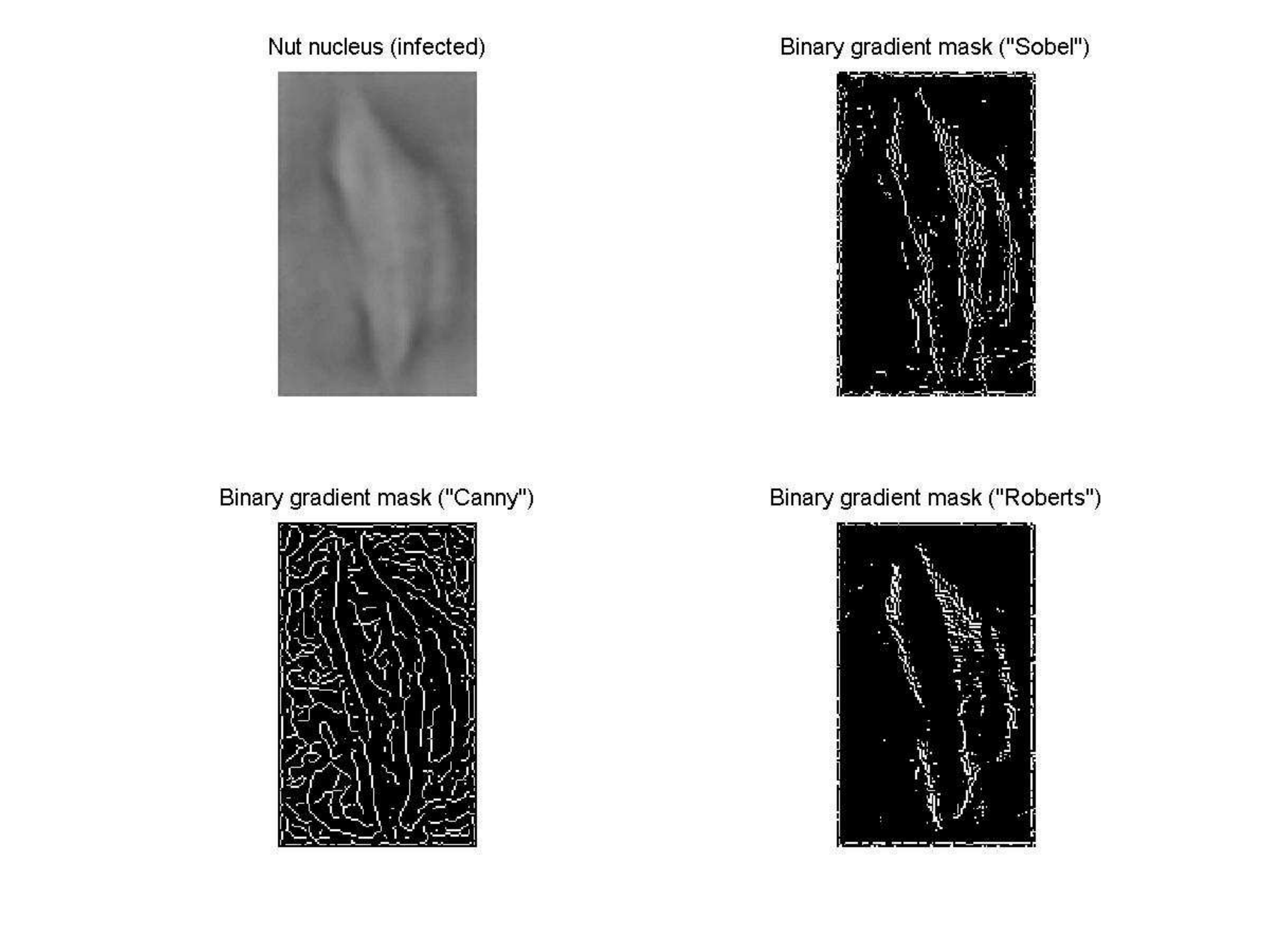}
\includegraphics[width=0.45\textwidth]{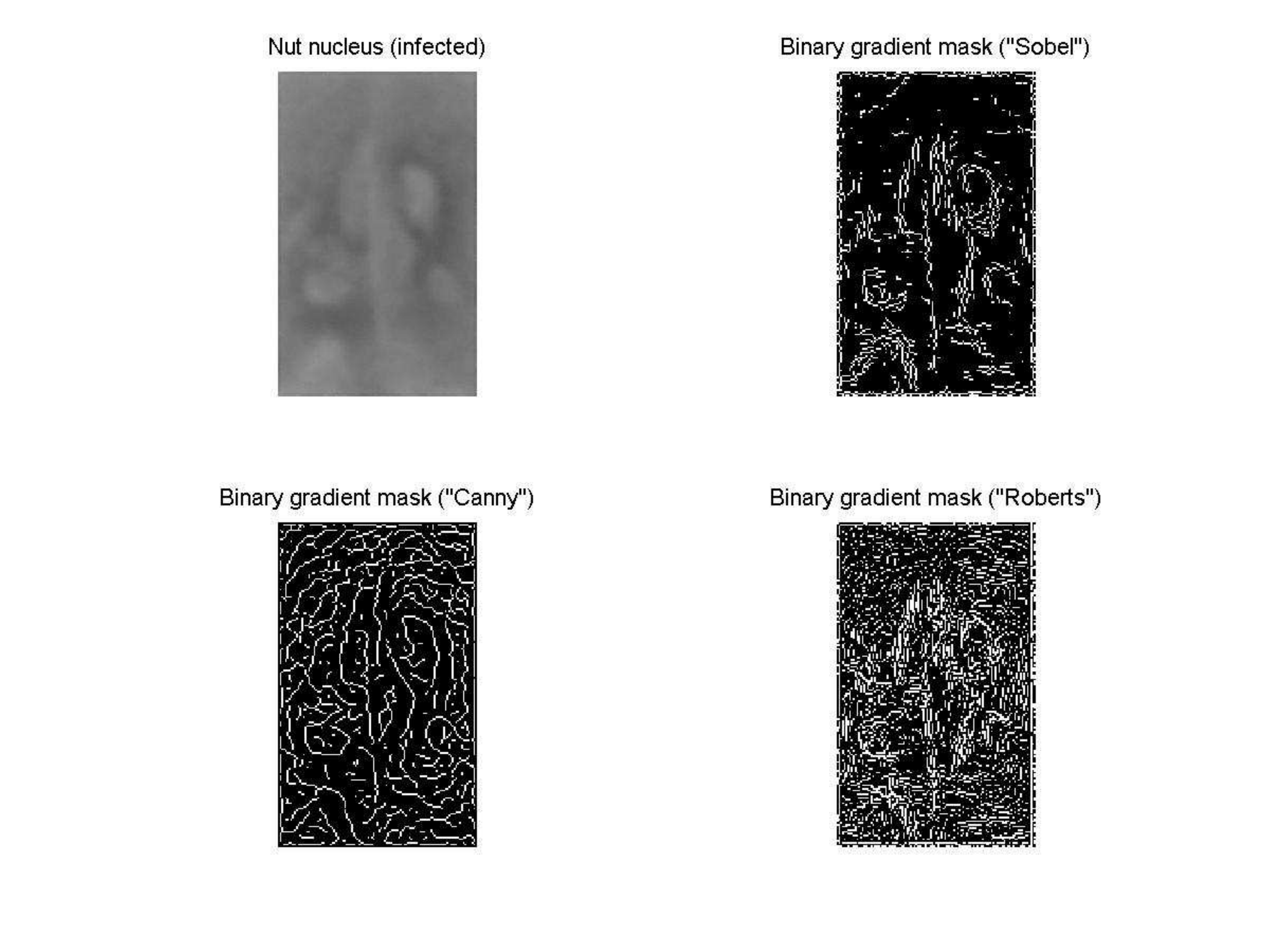}
\caption{Image processing of the hazelnut nuclei belonging to the set $I$, for $\epsilon=100$ and $\rho=1.5$, by means of edge-detection algorithms, respectively: Sobel's algorithm (top right figure) , Canny's algorithm (bottom left figure) and Roberts' algorithm (bottom right figure).}\label{nuclC}
\end{figure}

The red rectangles in Fig. \ref{hazelnuts3} are identified by the pair of parameters $\{\epsilon, \rho\}$, where $\epsilon$, related to the image resolution, is defined as the number of pixels comprised along the horizontal length of the rectangles, while $\rho$ is the ratio of the number of pixels along the vertical length to the corresponding number of pixels along the horizontal one.
In our simulations, the values of the parameters $\{\epsilon, \rho\}$ were kept constant when calculating the histograms relative to different hazelnut nuclei.
Figure \ref{hazelnuts3} refers, for instance, to the case corresponding to $\epsilon=80$ and $\rho = 2.5$. 
In Figs. \ref{nuclS},\ref{nuclA} and \ref{nuclC}, shown is the result of the image processing of the hazelnut nuclei, performed through a noise removal filter (adaptive Wiener filtering) and various edge-detector algorithms. 

\begin{figure}[H]
\centering
\includegraphics[width=0.45\textwidth]{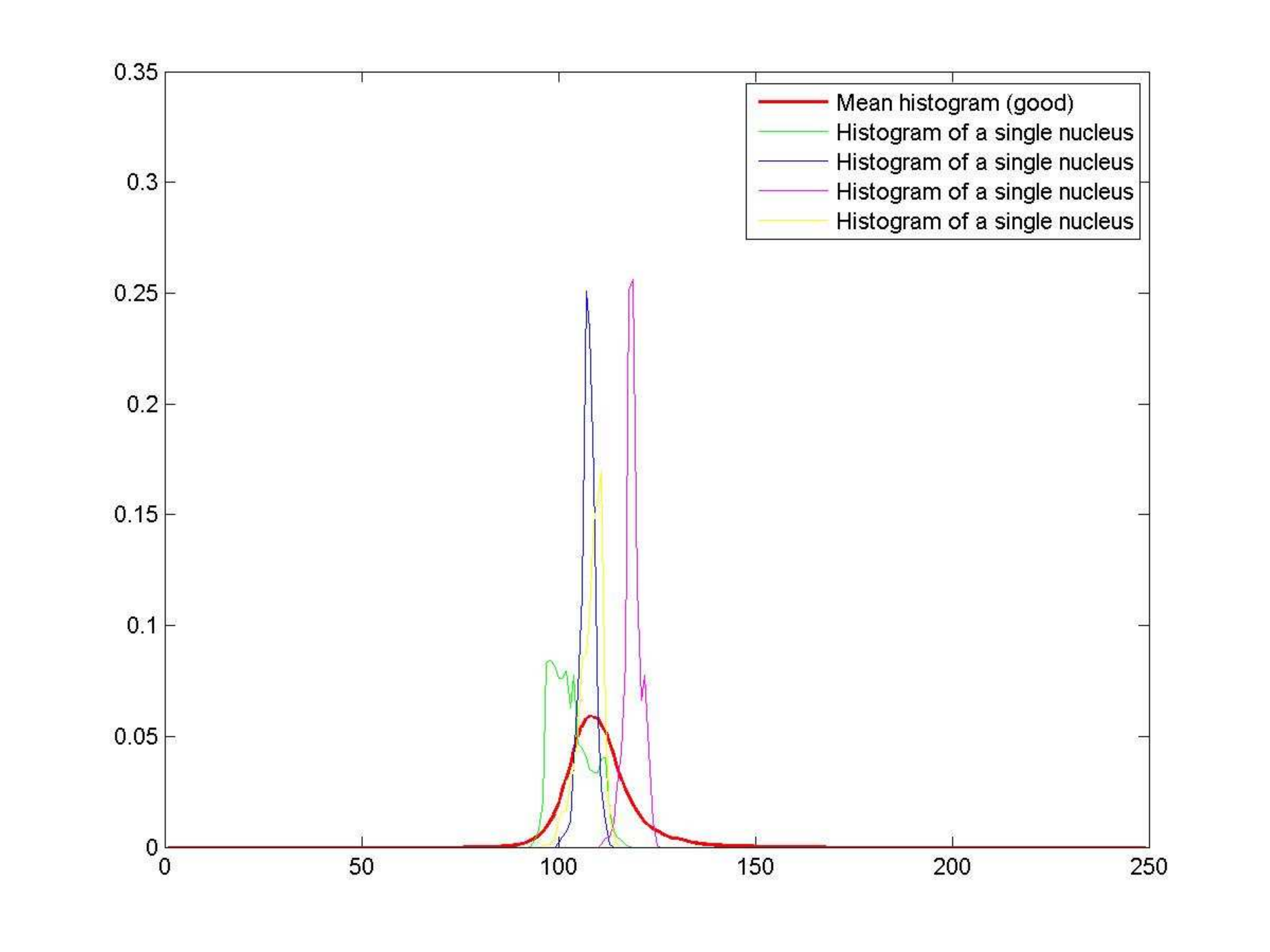}
\hspace{1mm}
\includegraphics[width=0.45\textwidth]{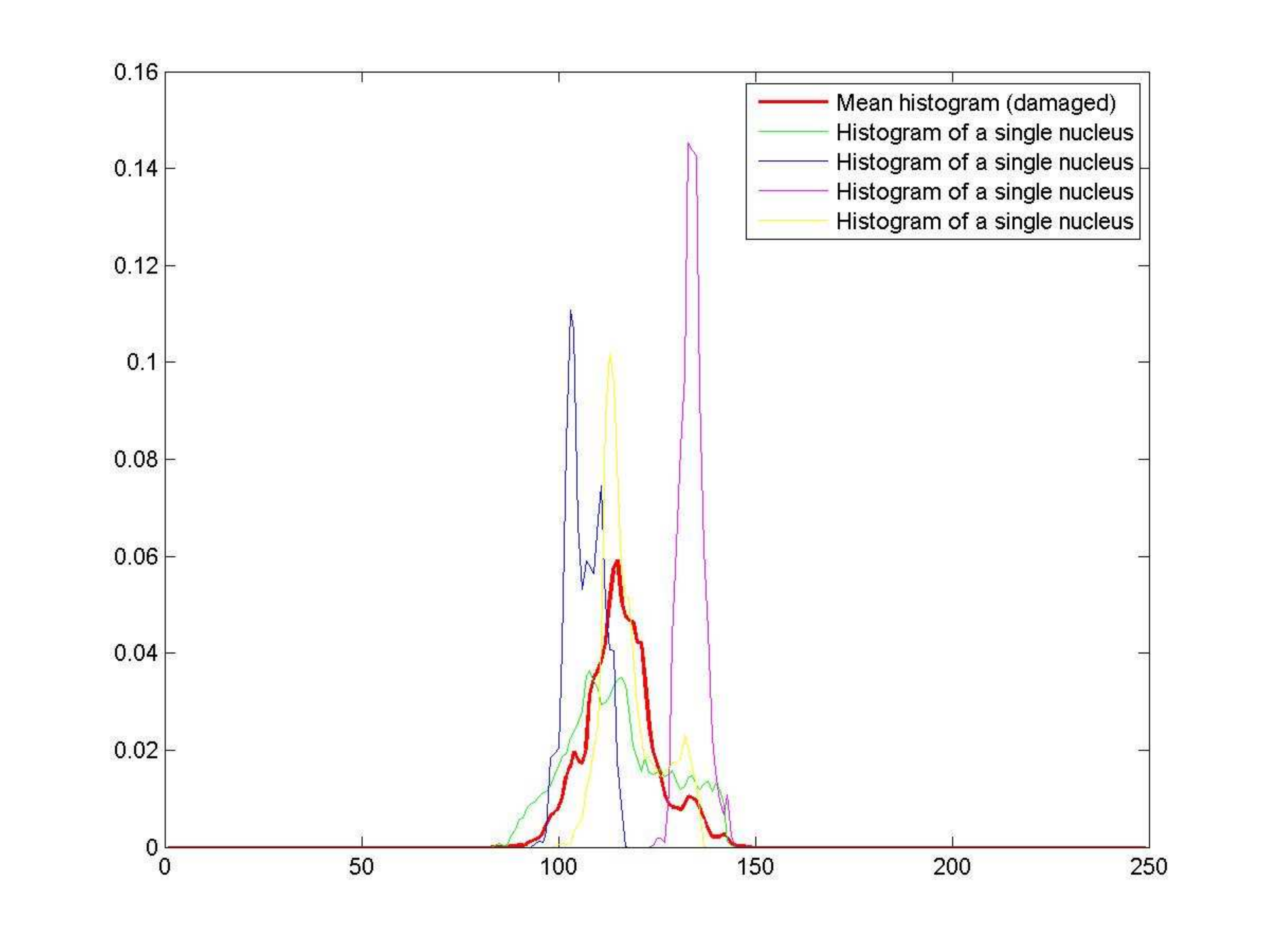}\\
\vspace{1mm}
\includegraphics[width=0.45\textwidth]{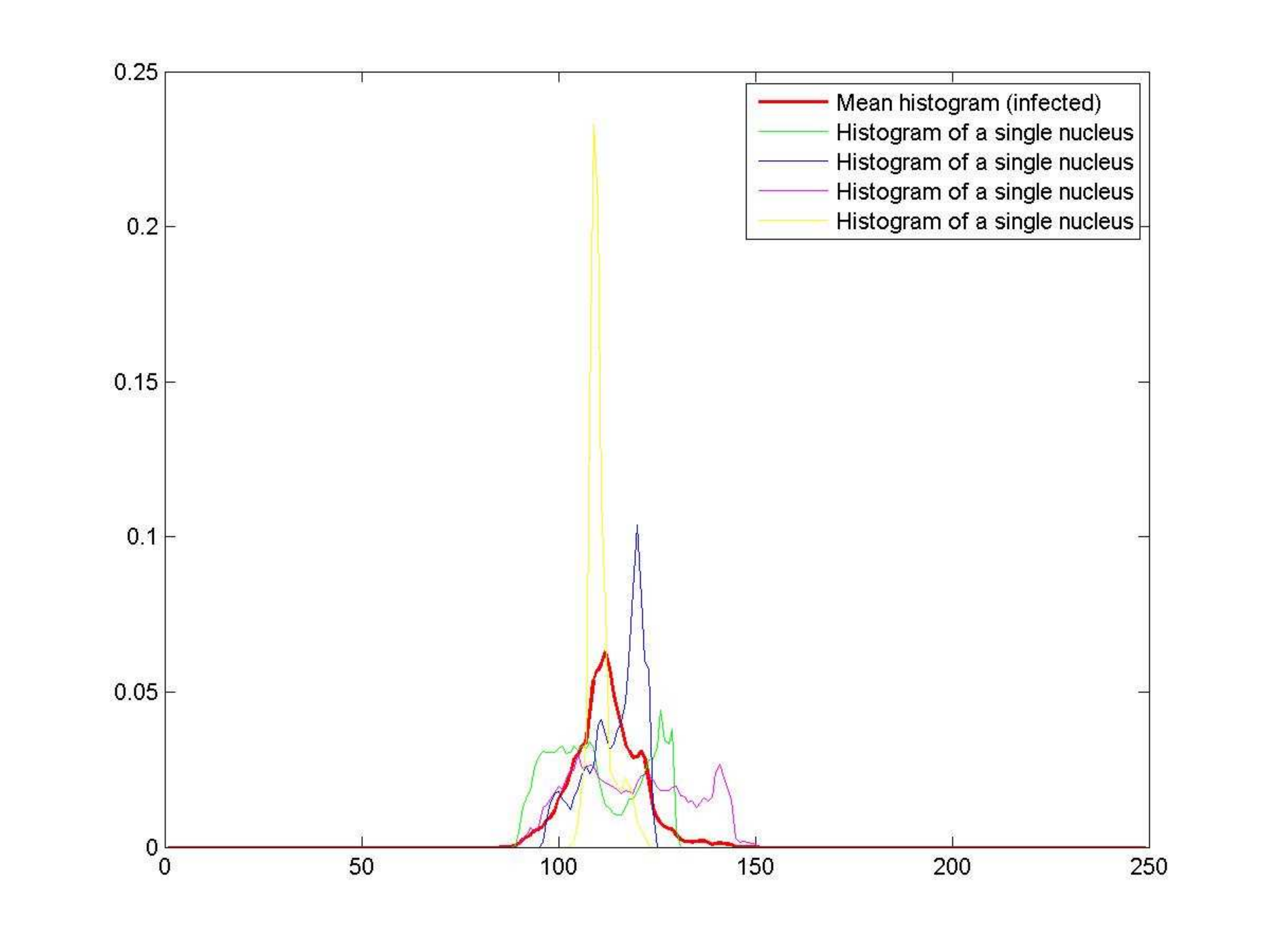}
\hspace{1mm}
\includegraphics[width=0.45\textwidth]{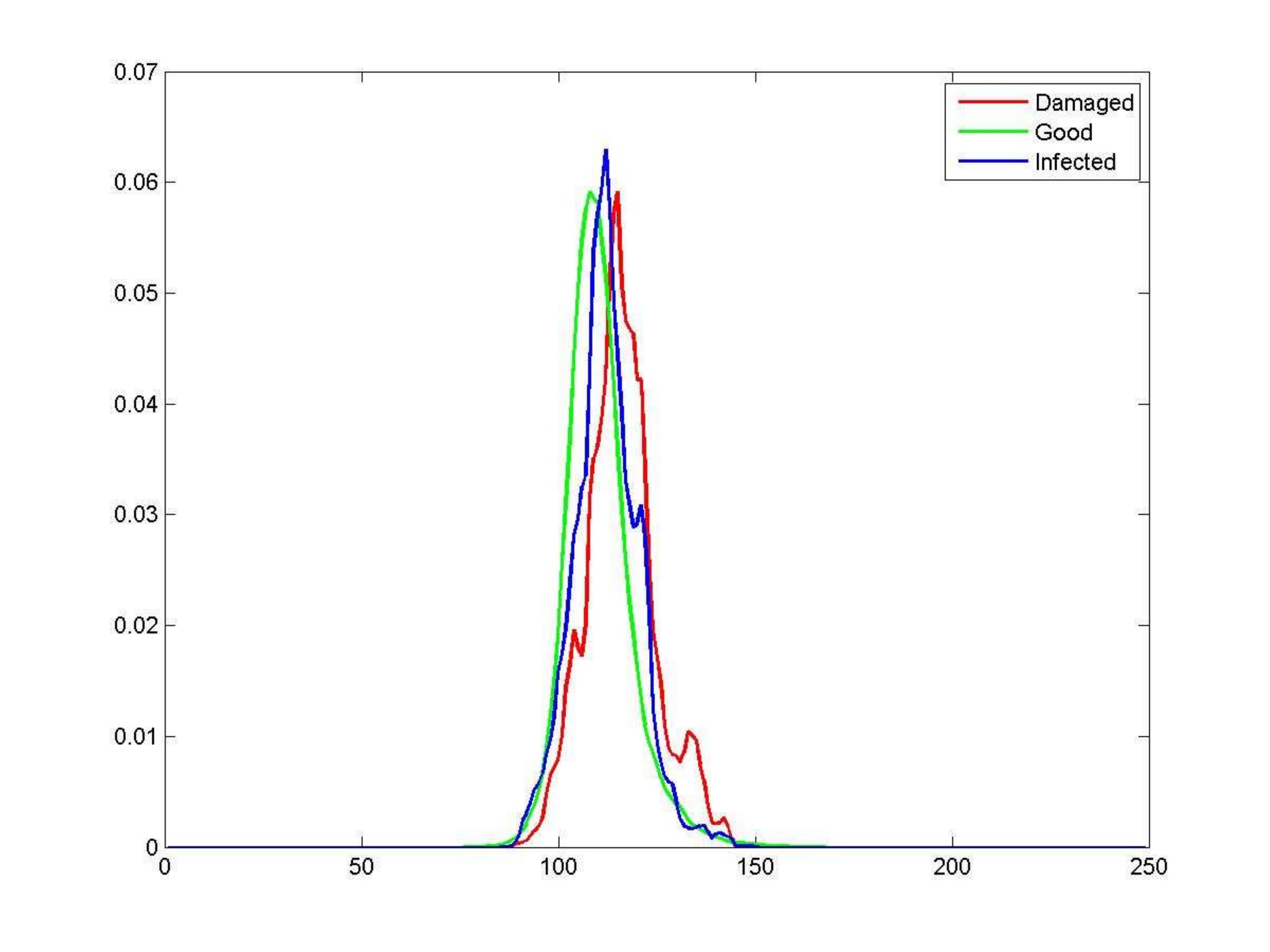}
\caption{Image histograms of the hazelnut nuclei belonging to the sets $G$ (top left), $D$ hazelnuts (top right), $I$ hazelnuts (bottom left). For each of the three sets, the figures display the histograms of single hazelnuts as well as the mean histogram in the corresponding set (mean histogram). On the bottom right corner, the mean histograms of the three sets are compared. All the histograms were obtained by setting $\epsilon = 80$ and $\rho=2.5$.}\label{hazelnutsnucl}
\end{figure}

In Fig. \ref{hazelnutsnucl}, which is worth comparing with Fig. \ref{hazelnuts2}, we plotted the mean histograms relative to the cropped images, with $\epsilon=80$ and $\rho =2.5$. The question arises, then, as to whether the separation between the two scales $\langle d \rangle^{(A)}$ and $\Delta^{(A,B)}$ is amenable to be enhanced by tuning the two parameters $\epsilon$ and $\rho$. 

\begin{figure}[H]
\centering
\includegraphics[width=0.7\textwidth]{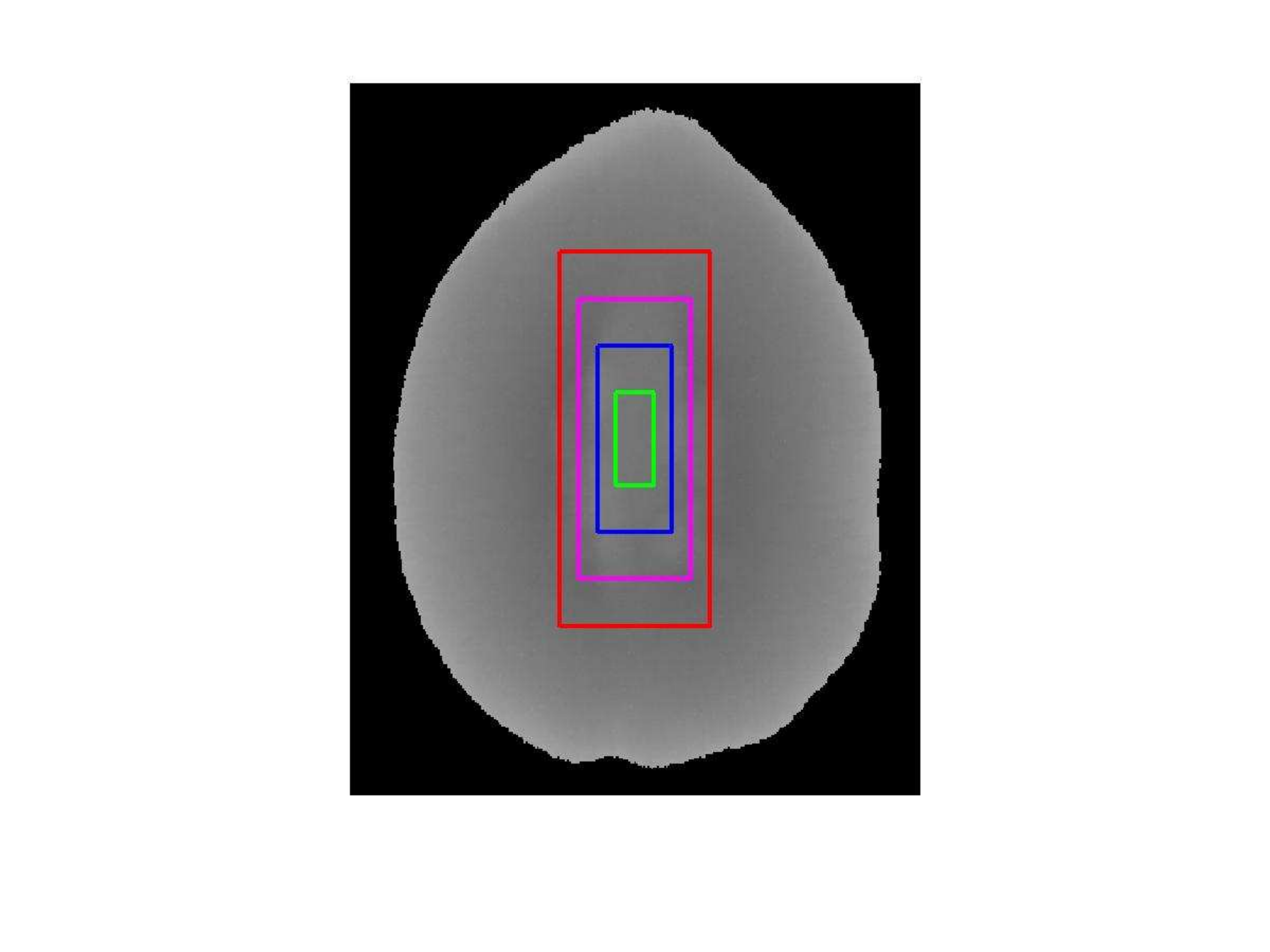}
\caption{Different values of the scale of resolution: $\epsilon=80$ (red rectangle), $\epsilon =60$ (magenta rectangle), $\epsilon=40$ (blue rectangle), $\epsilon=20$ (green rectangle). All the colored rectangles shown in the picture are obtained by setting $\rho =2.5$.}\label{rect}
\end{figure}

We thus studied the behaviour of the mean histograms, shown in Fig. \ref{hazelnutsnucl}, as well as of the typical fluctuations occurring in each set, as a function of $\epsilon$ and $\rho$: in our simulations, $\epsilon$ spans a broad range of values, whereas we let $\rho$ attain the values $1.5$ and $2.5$, cf.  Fig. \ref{rect}. 

\begin{figure}[H]
\centering
\includegraphics[width=0.45\textwidth]{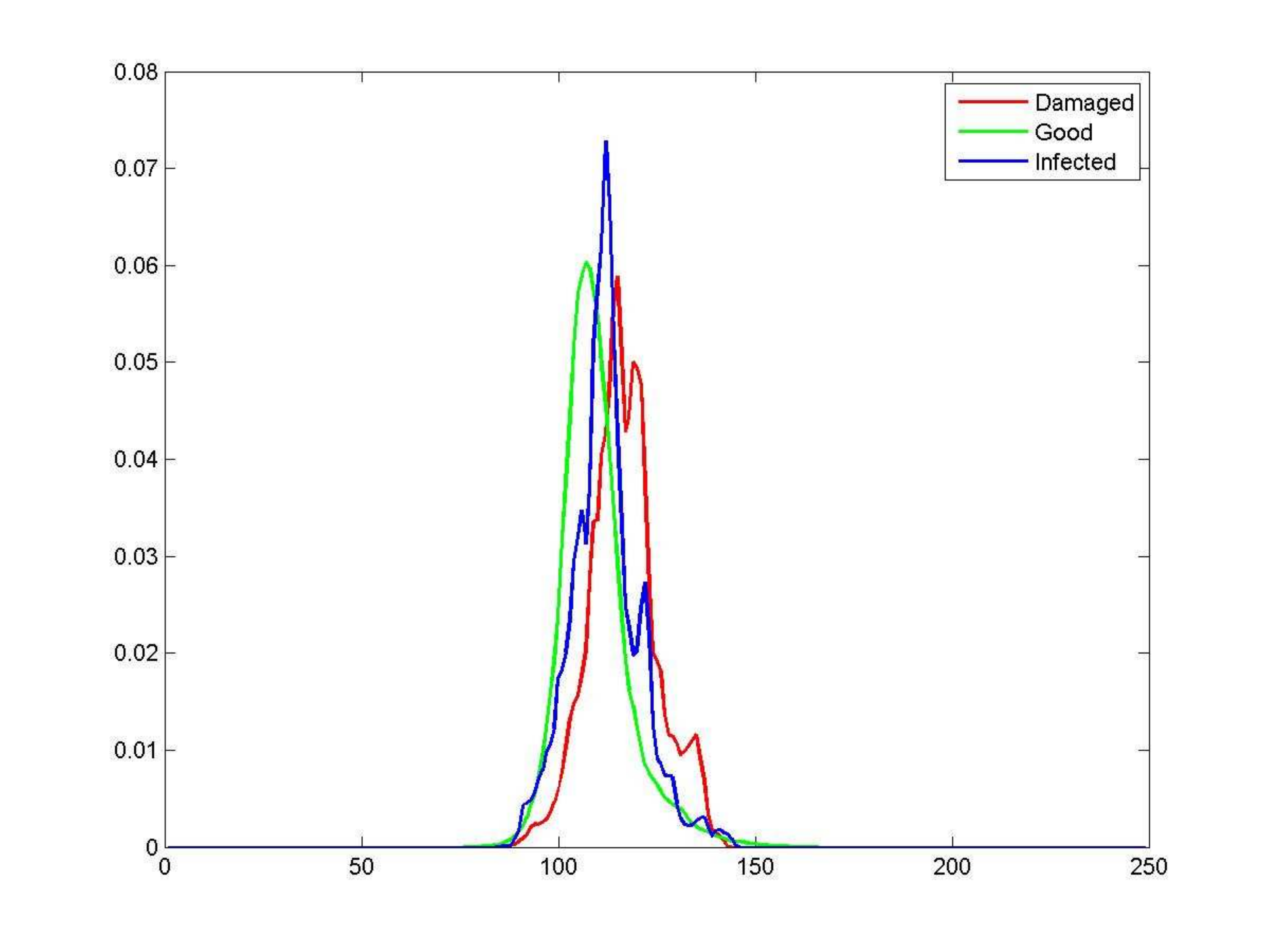}
\hspace{1mm}
\includegraphics[width=0.45\textwidth]{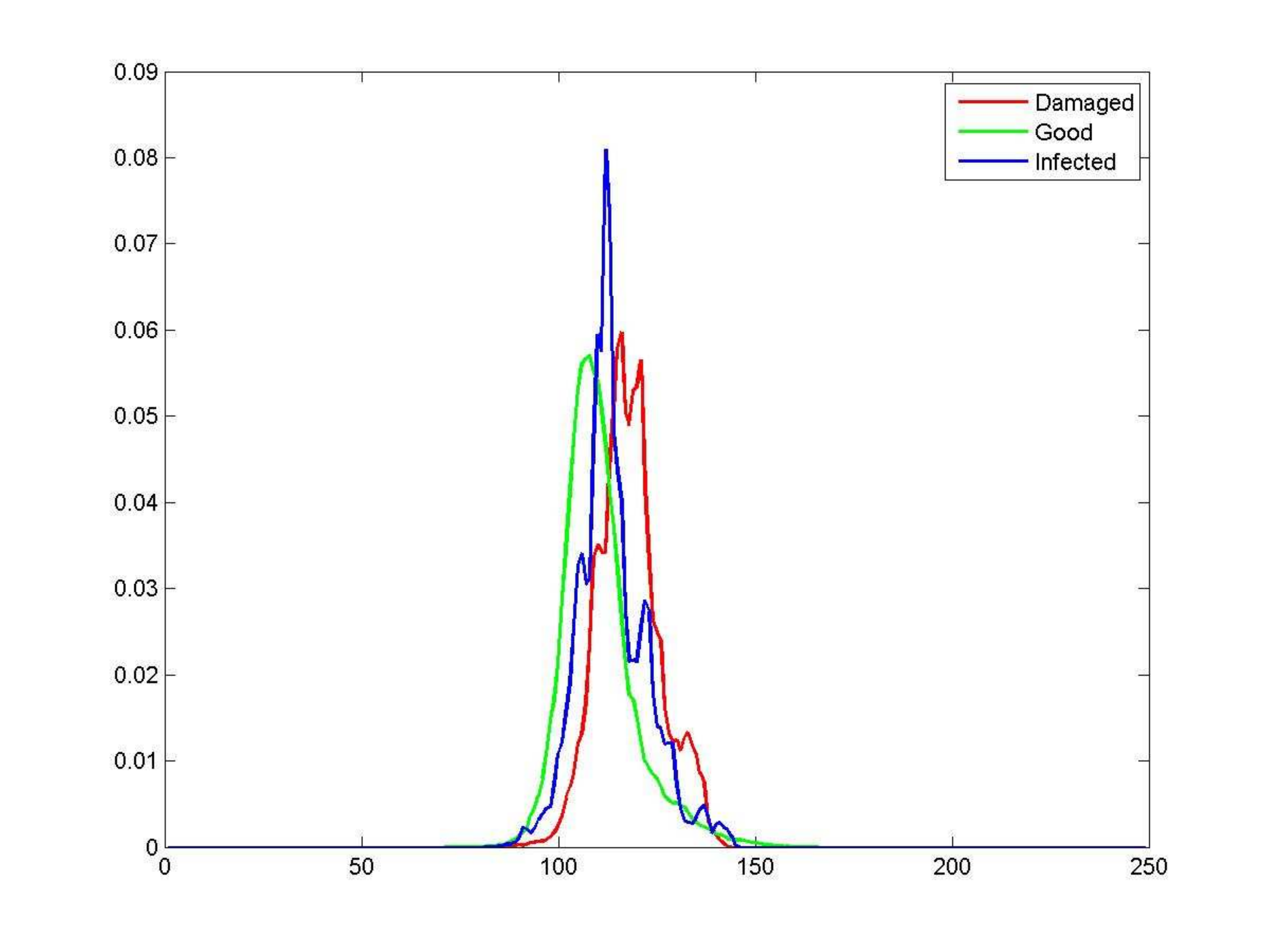}\\
\vspace{1mm}
\includegraphics[width=0.45\textwidth]{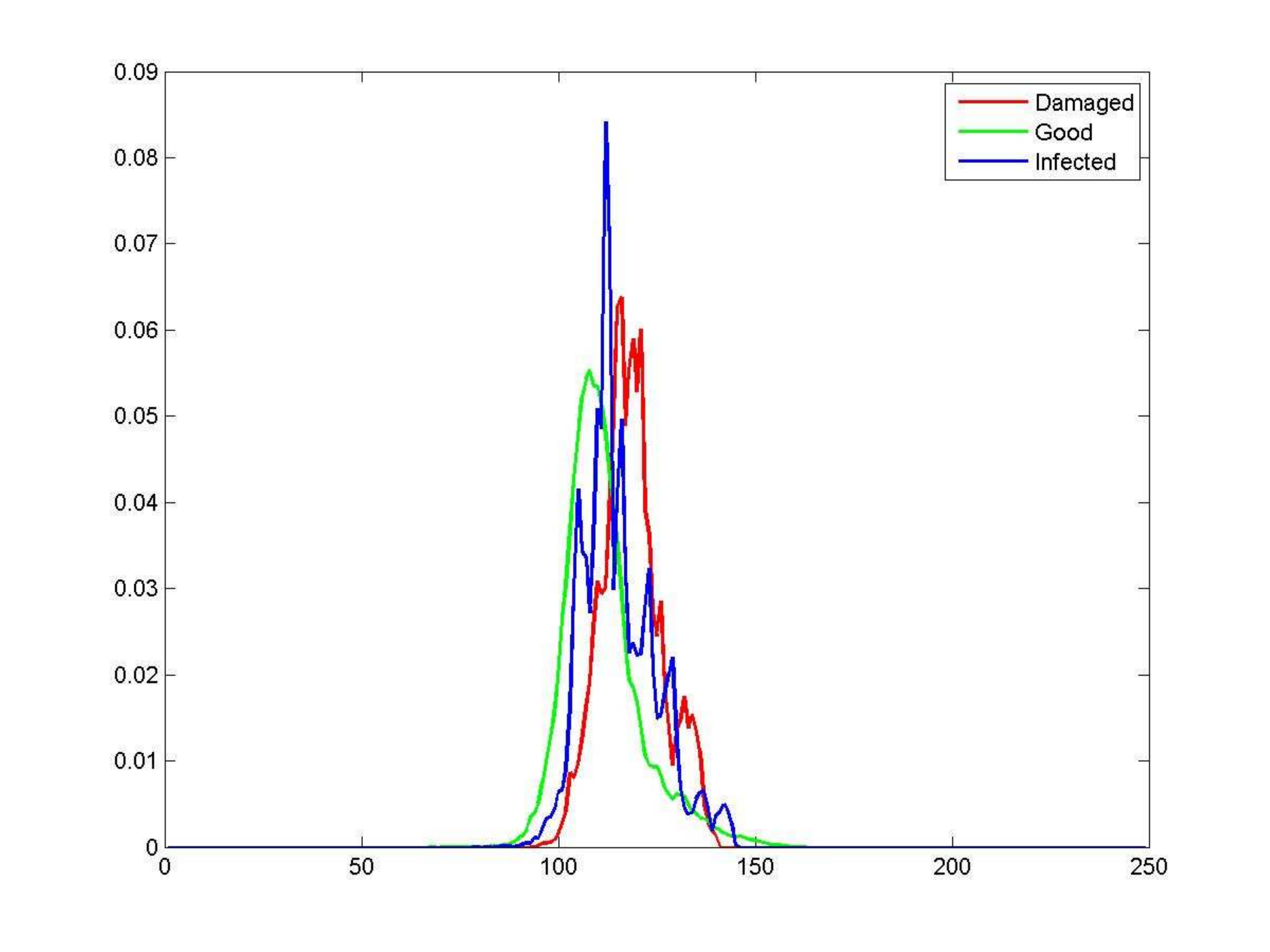}
\hspace{1mm}
\includegraphics[width=0.45\textwidth]{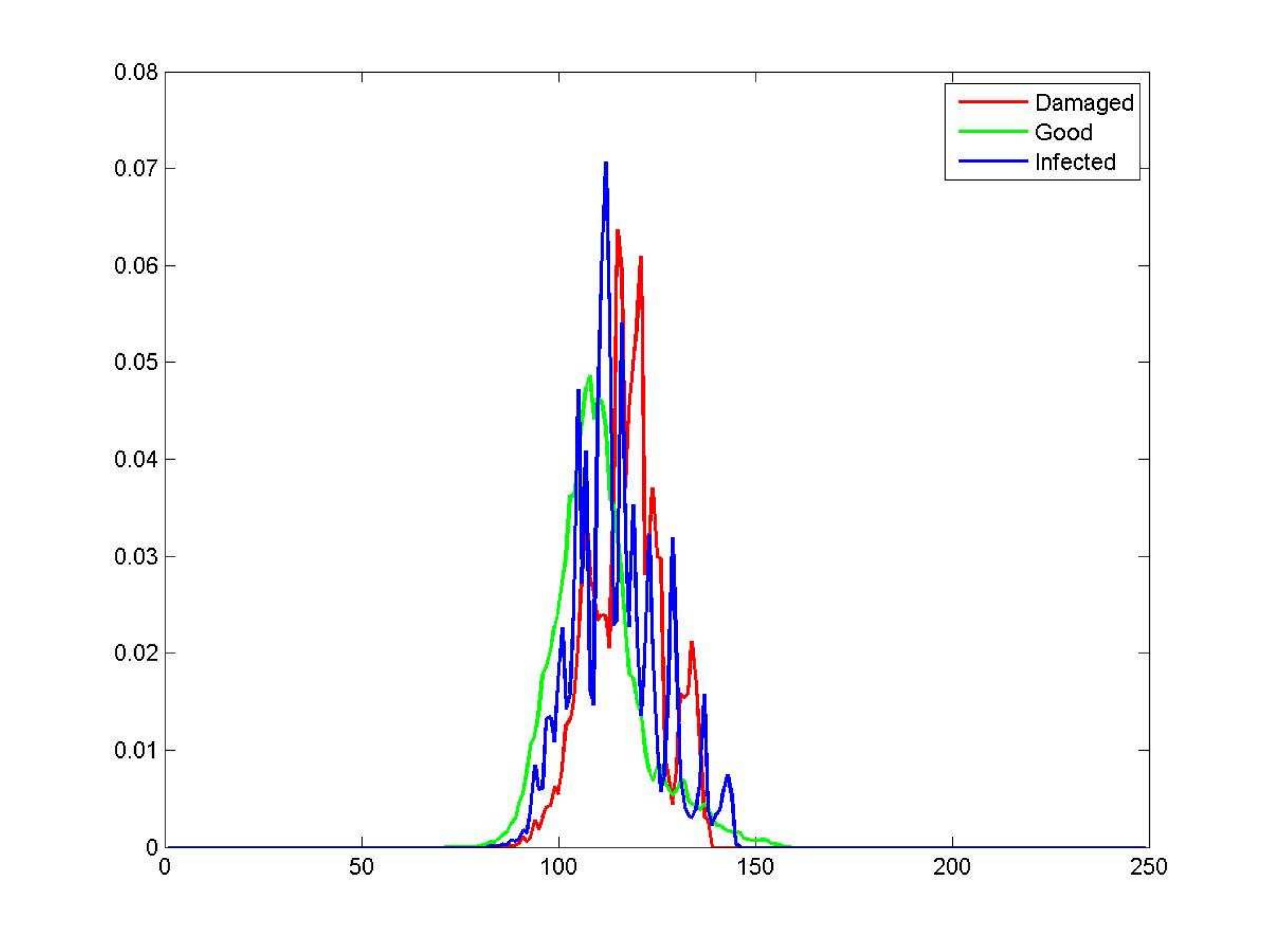}
\caption{Mean histograms of the hazelnut nuclei at different scales of resolution: $\epsilon =80$ (top left), $\epsilon =60$ (top right), $\epsilon =40$ (bottom left) and $\epsilon =20$ (bottom right), with $\rho=1.5$.}\label{nuclhist1}
\end{figure}

\begin{figure}[H]
\centering
\includegraphics[width=0.45\textwidth]{meanhistTotnucl80r25.pdf}
\hspace{1mm}
\includegraphics[width=0.45\textwidth]{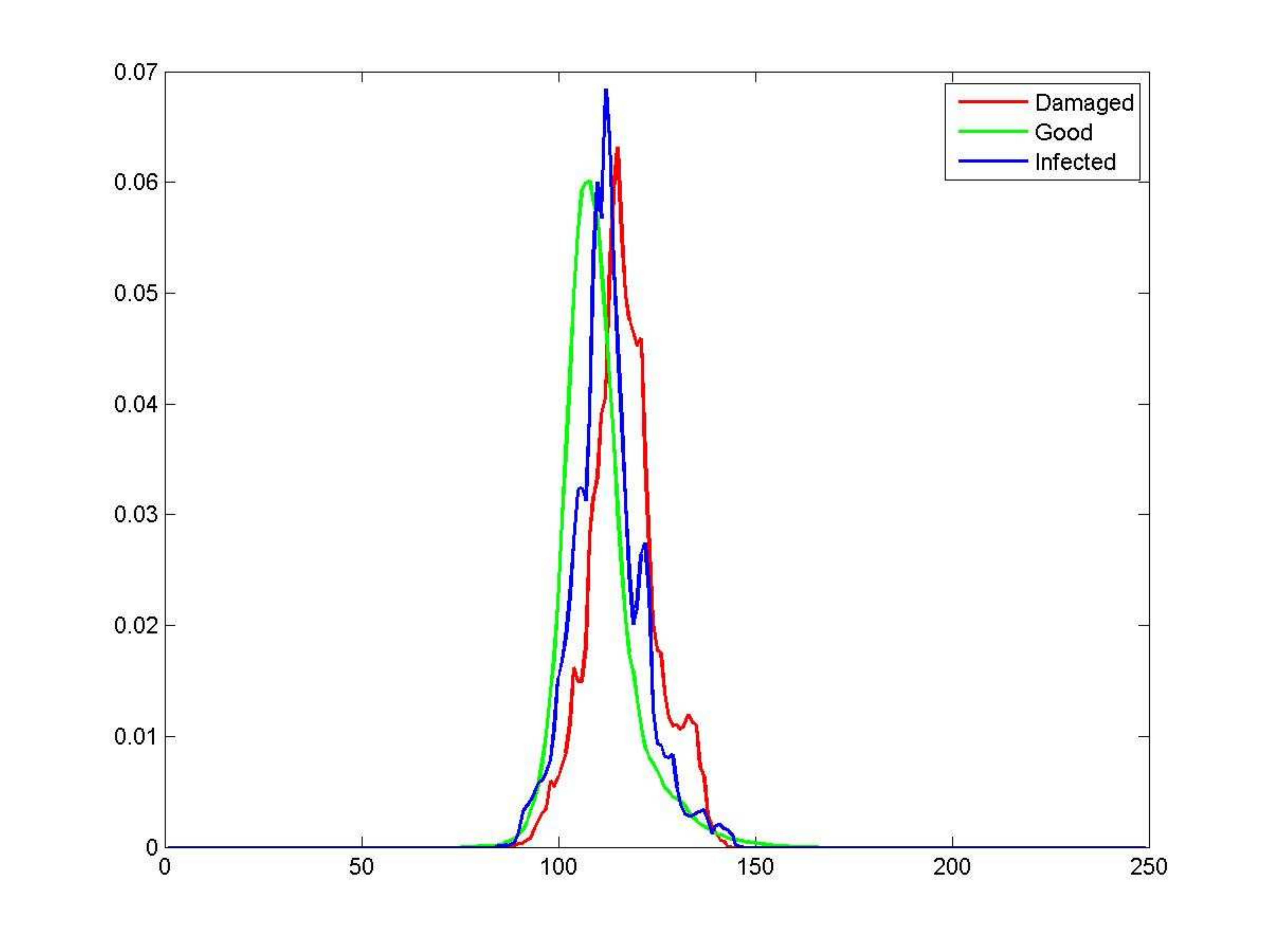}\\
\vspace{1mm}
\includegraphics[width=0.45\textwidth]{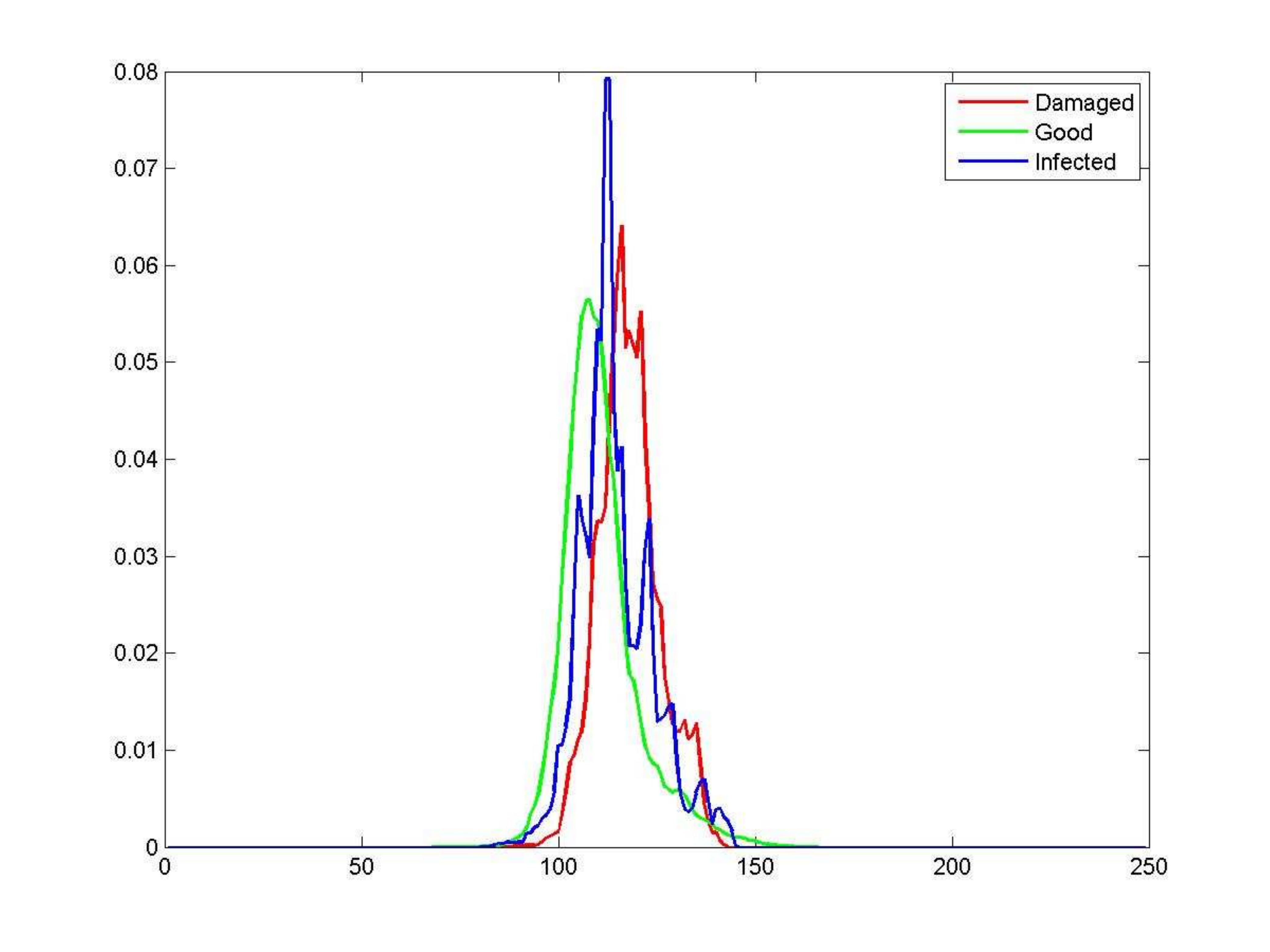}
\hspace{1mm}
\includegraphics[width=0.45\textwidth]{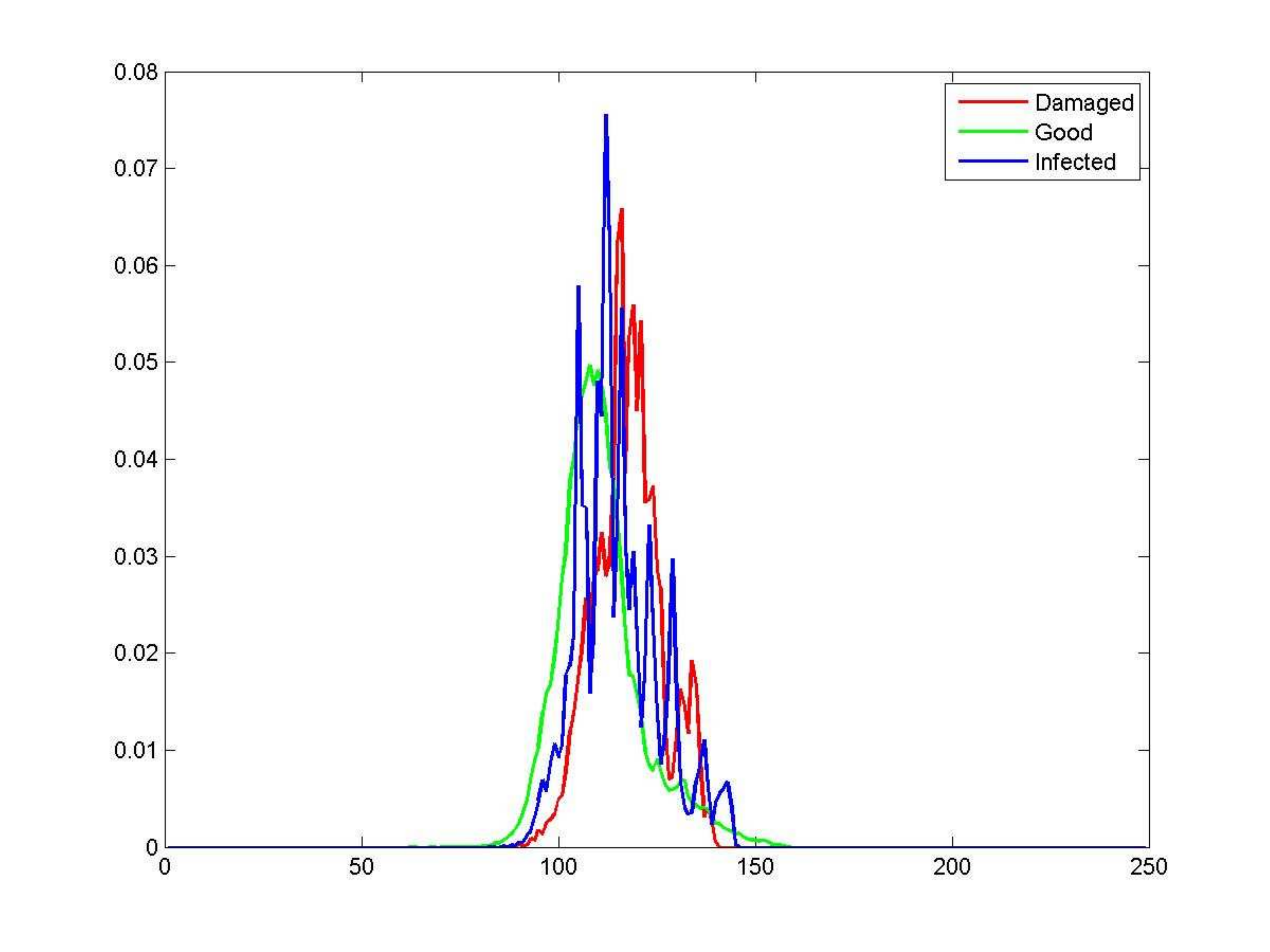}
\caption{Mean histograms of the nuclei of the hazelnuts at different scales of description: $\epsilon =80$ (top left), $\epsilon =60$ (top right), $\epsilon =40$ (bottom left) and $\epsilon =20$ (bottom right), with $\rho=2.5$.}\label{nuclhist2}
\end{figure}

In Fig. \ref{nuclhist1} and \ref{nuclhist2}, the mean histograms of the sets $G$, $D$ and $I$ are shown for different values of $\epsilon$, and for two different values of $\rho$.

We focused, in particular, on the investigation of the dependence of the scales $\langle d \rangle^{(A)}(\epsilon; \rho)$ and $\Delta^{(A,B)}(\epsilon; \rho)$ on the resolution $\epsilon$. 

\begin{figure}[H]
\centering
\includegraphics[width=0.45\textwidth]{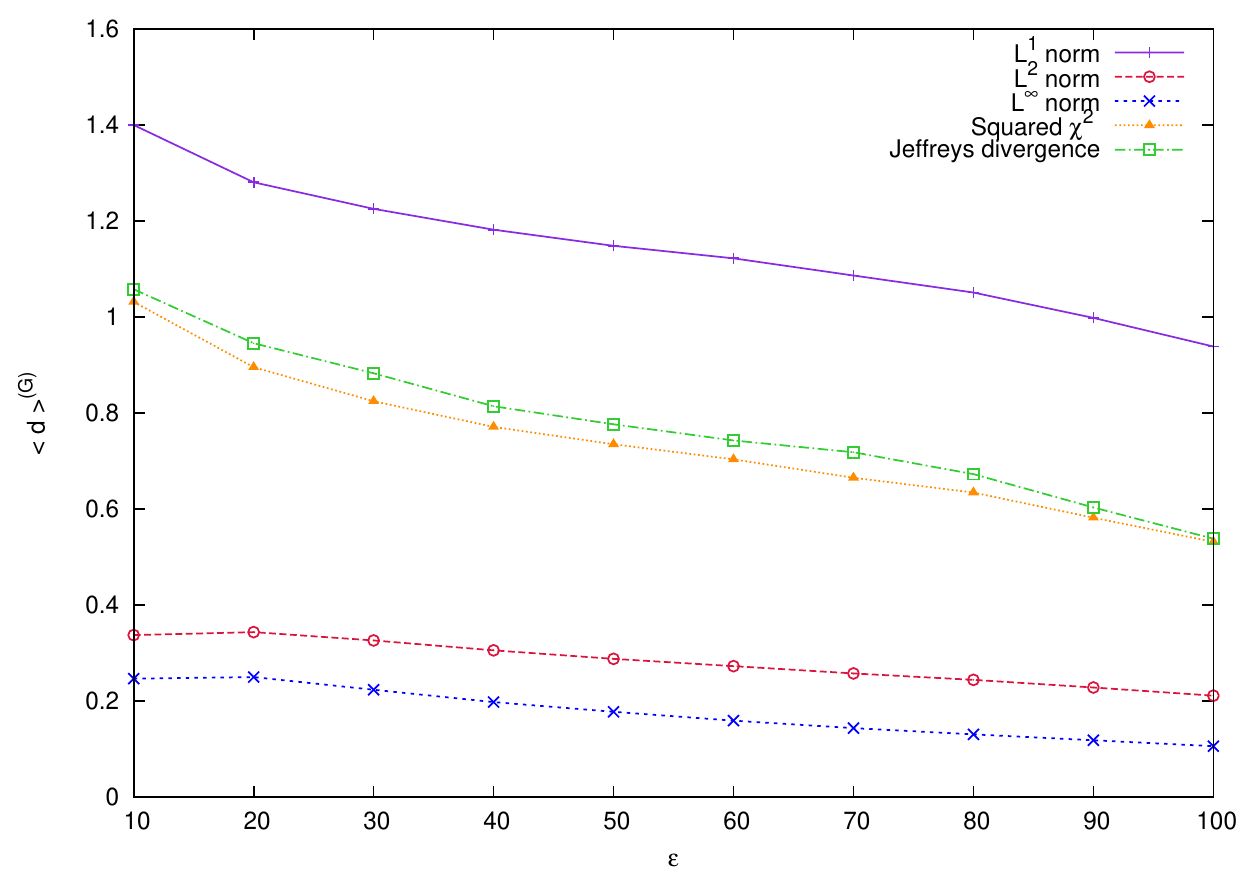}
\includegraphics[width=0.45\textwidth]{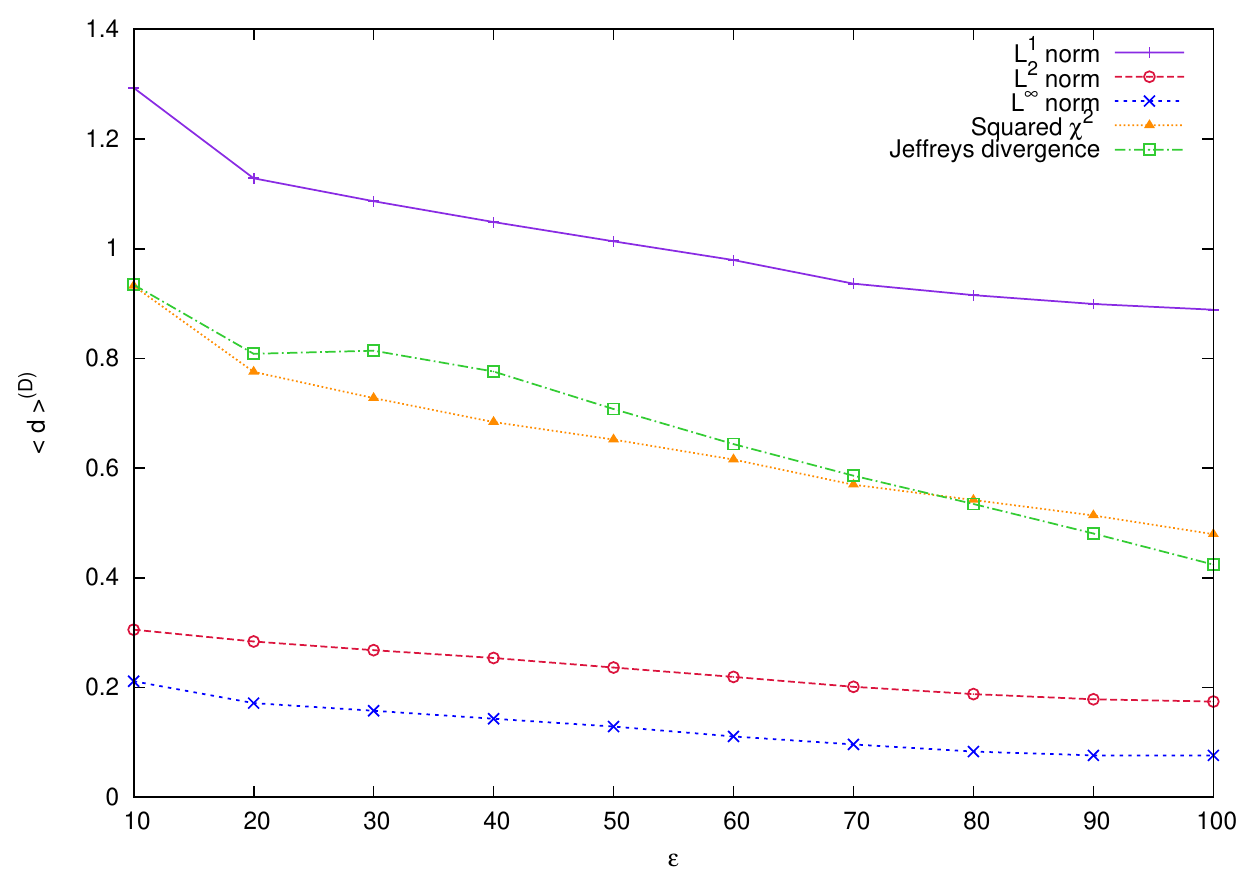}
\includegraphics[width=0.45\textwidth]{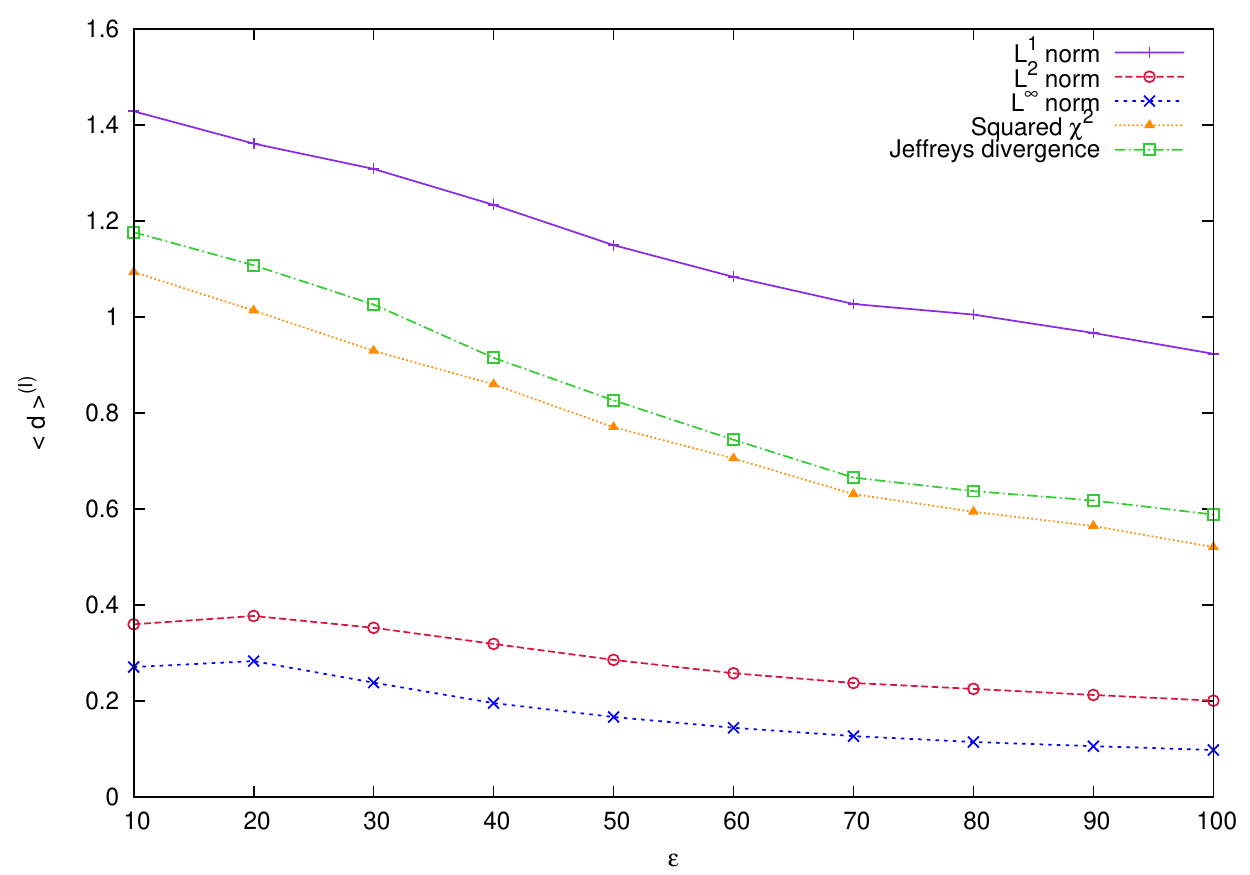}
\caption{Behaviour of the distances $\langle d \rangle^{(A)}$ vs. $\epsilon$, with $\rho=1.5$.}\label{norm1}
\end{figure}

\begin{figure}[H]
\centering
\includegraphics[width=0.45\textwidth]{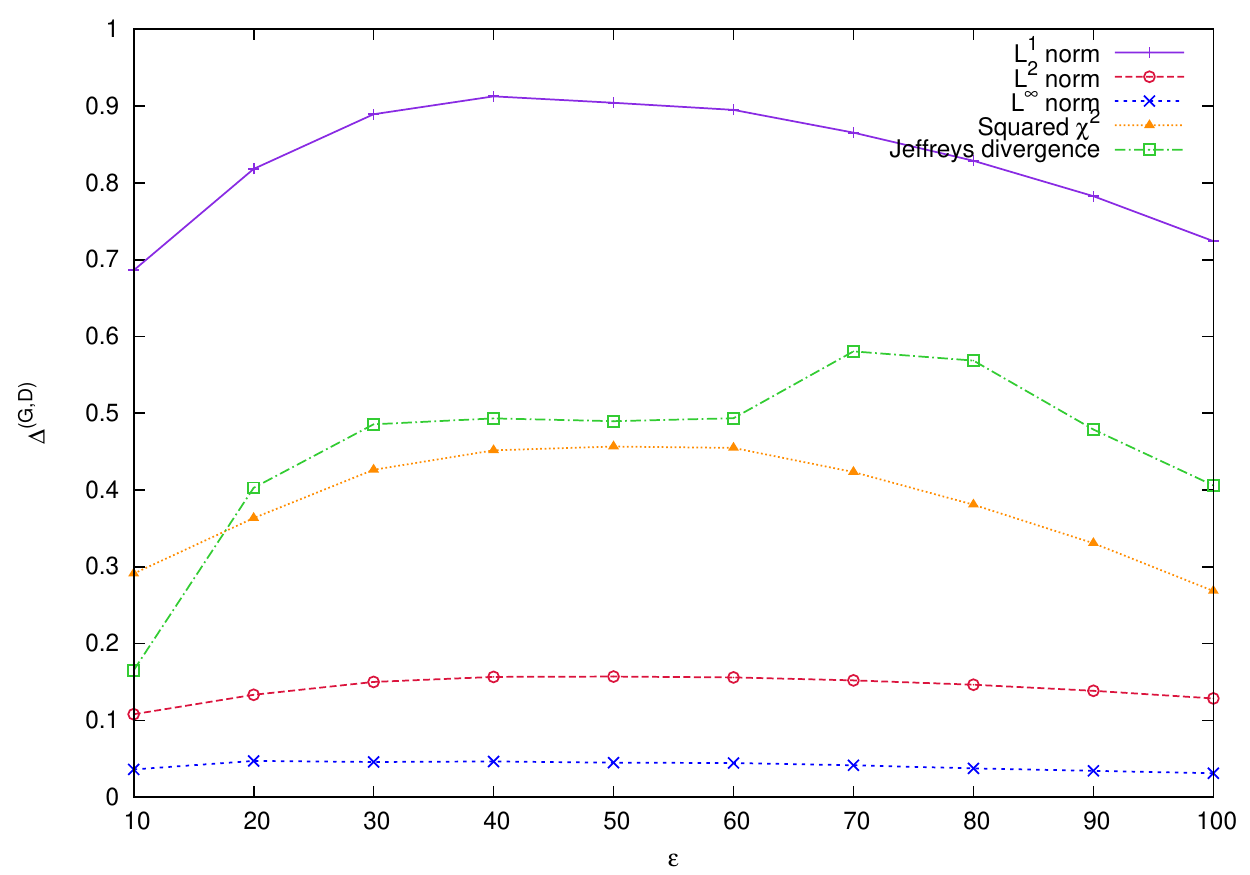}
\hspace{1mm}
\includegraphics[width=0.45\textwidth]{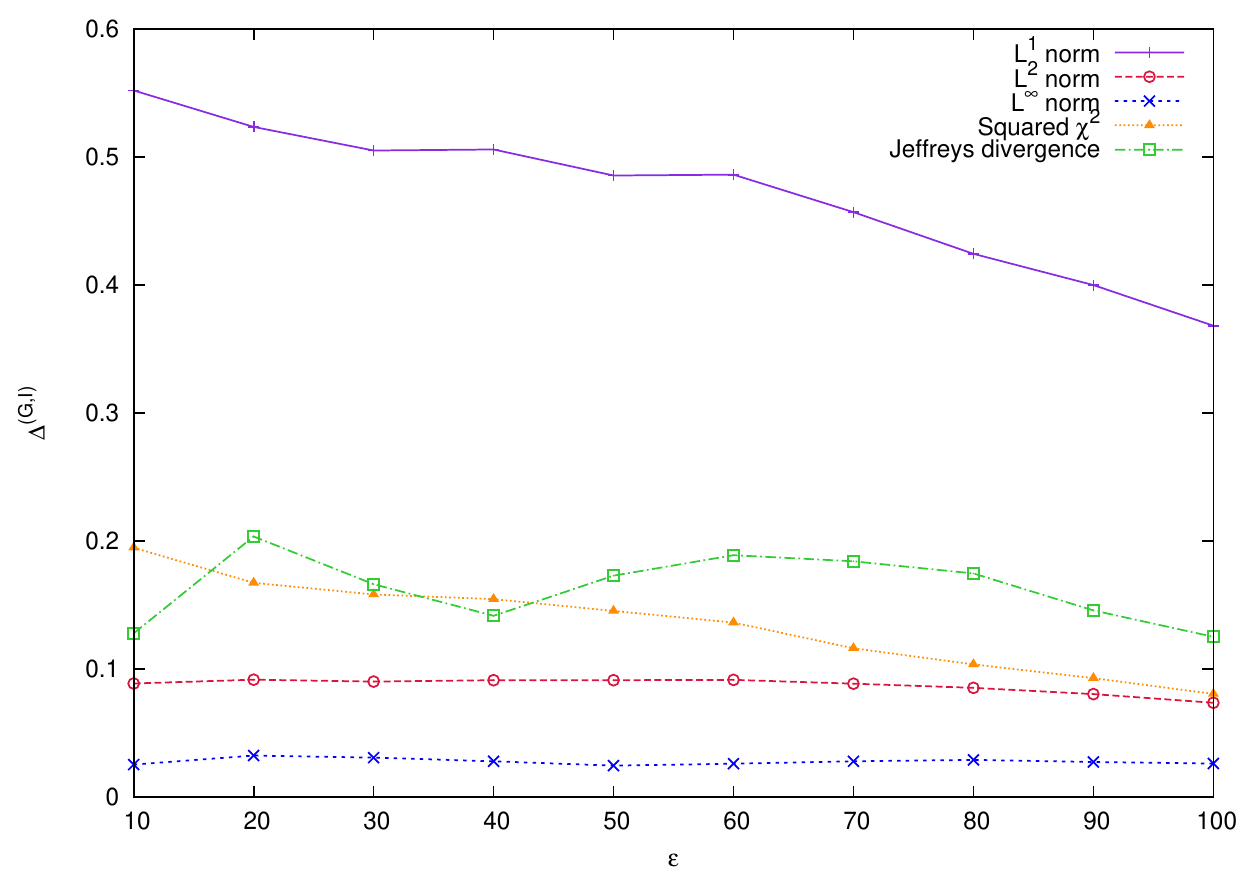}\\
\vspace{2mm}
\includegraphics[width=0.45\textwidth]{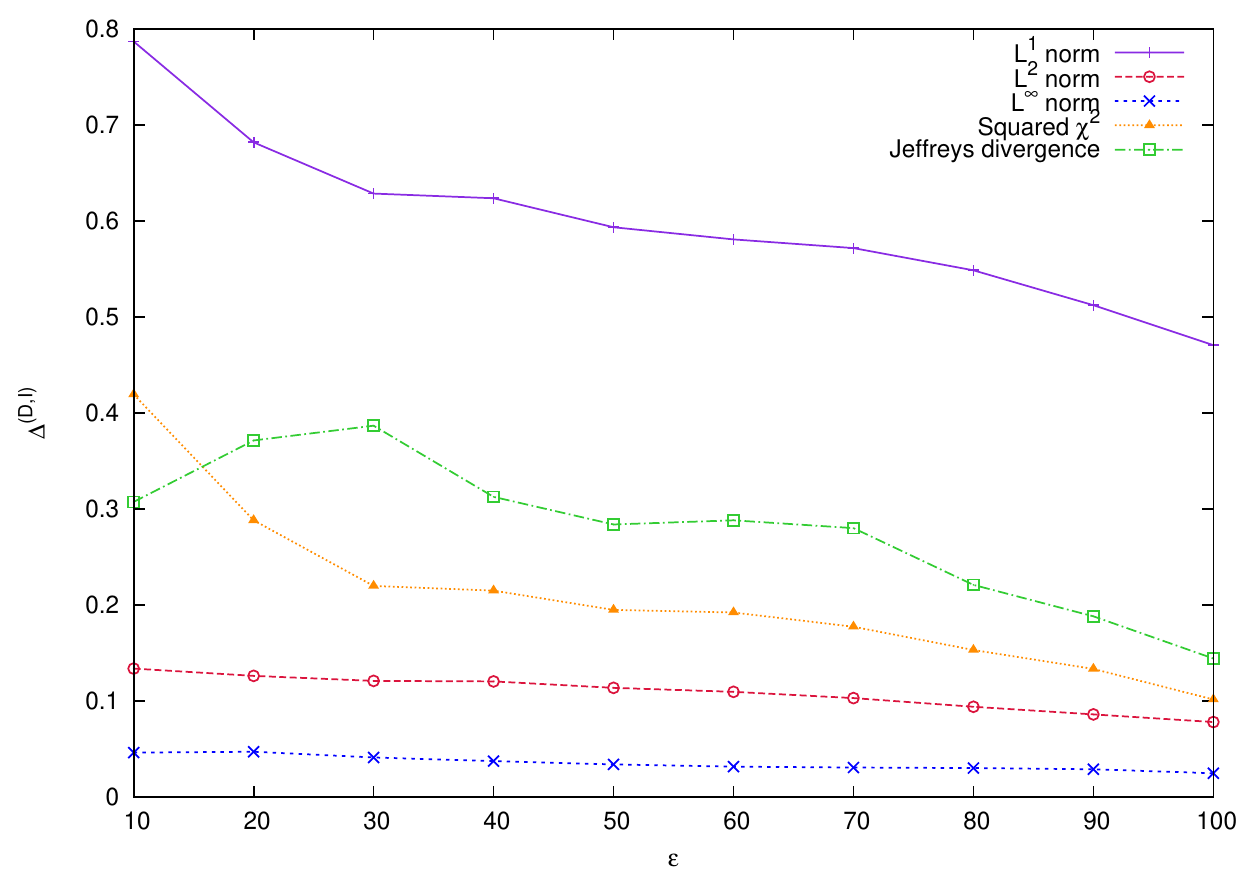}
\caption{Behaviour of the distances $\Delta^{(A,B)}$ vs. $\epsilon$, with $\rho=1.5$.}\label{mean1}
\end{figure}

\begin{figure}[H]
\centering
\includegraphics[width=0.45\textwidth]{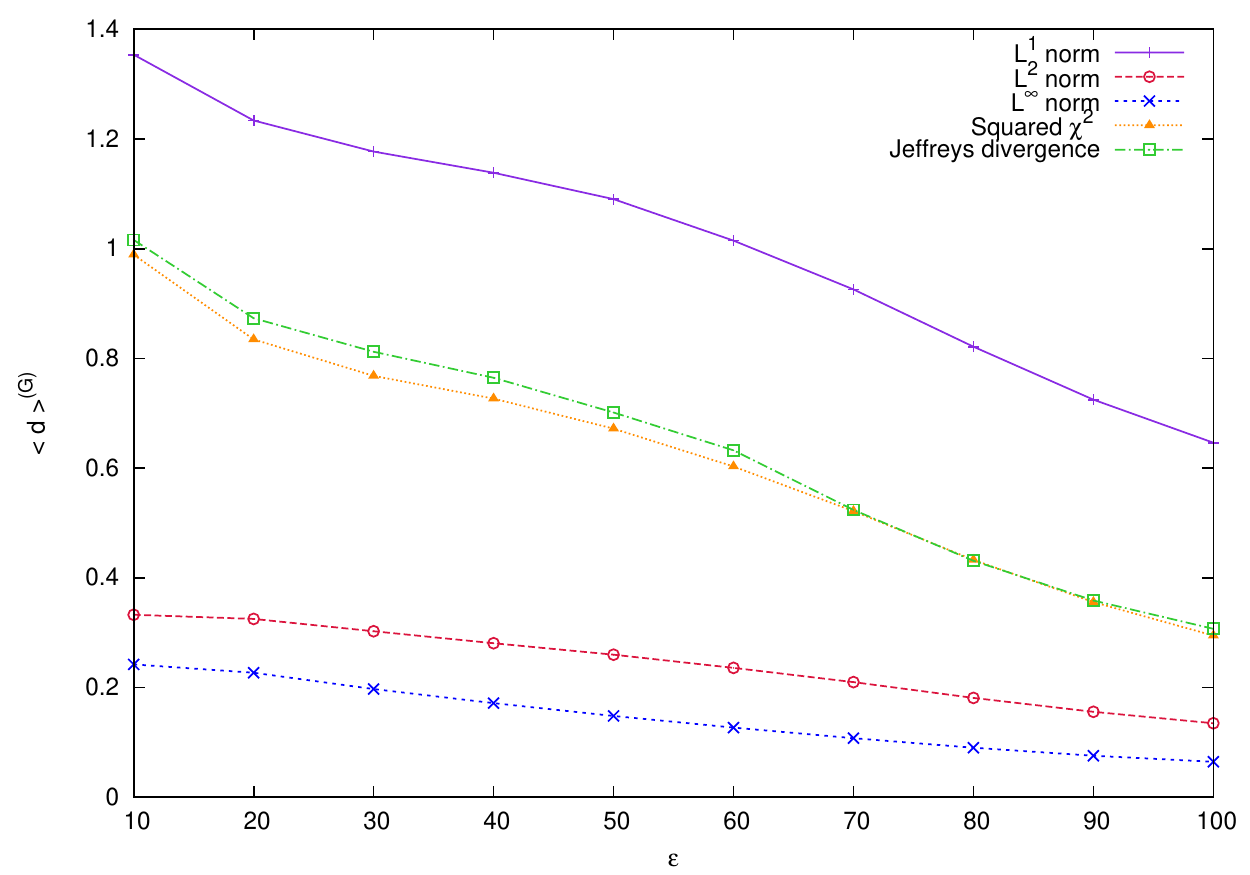}
\hspace{1mm}
\includegraphics[width=0.45\textwidth]{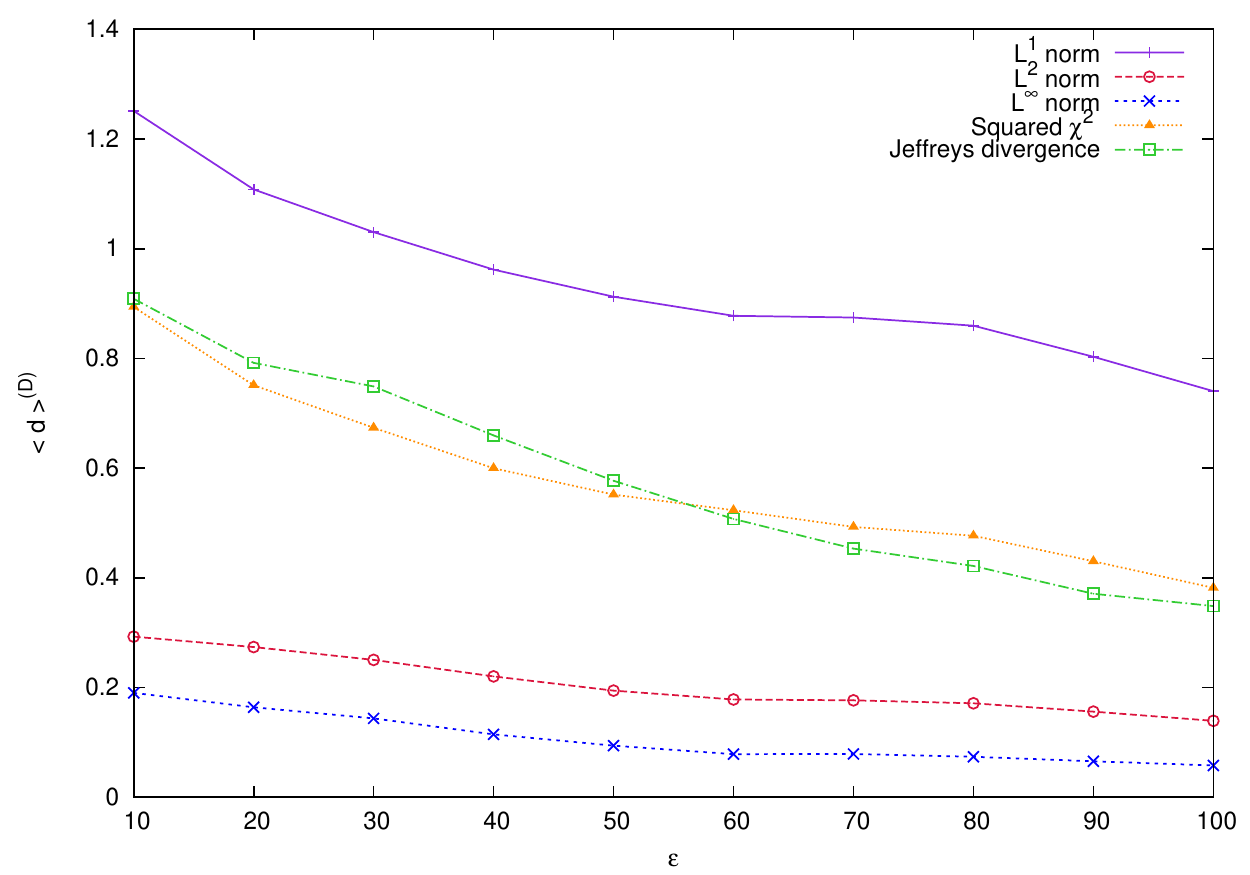}\\
\vspace{2mm}
\includegraphics[width=0.45\textwidth]{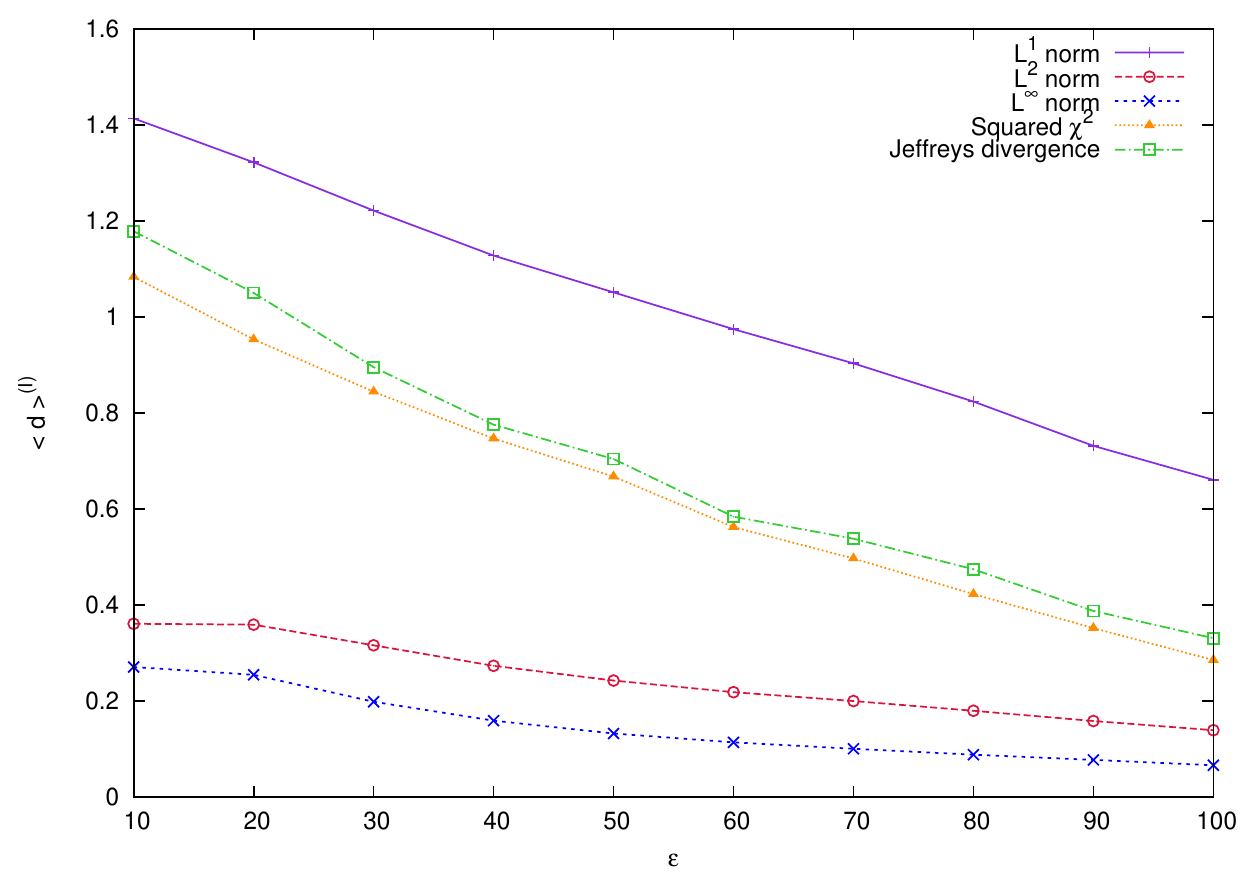}
\caption{Behaviour of the distances $\langle d \rangle^{(A)}$ vs. $\epsilon$, with $\rho=2.5$.}\label{norm2}
\end{figure}

\begin{figure}[H]
\centering
\includegraphics[width=0.45\textwidth]{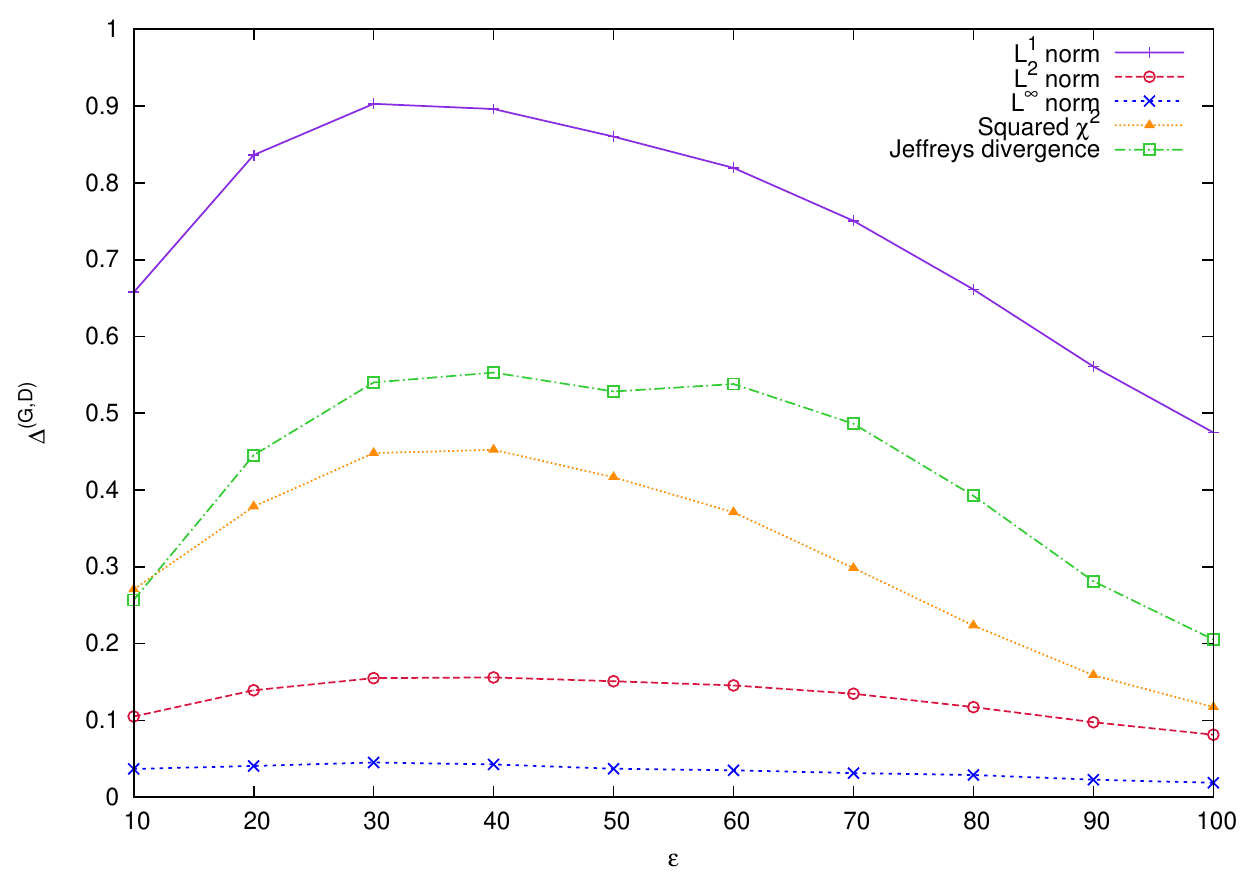}
\hspace{1mm}
\includegraphics[width=0.45\textwidth]{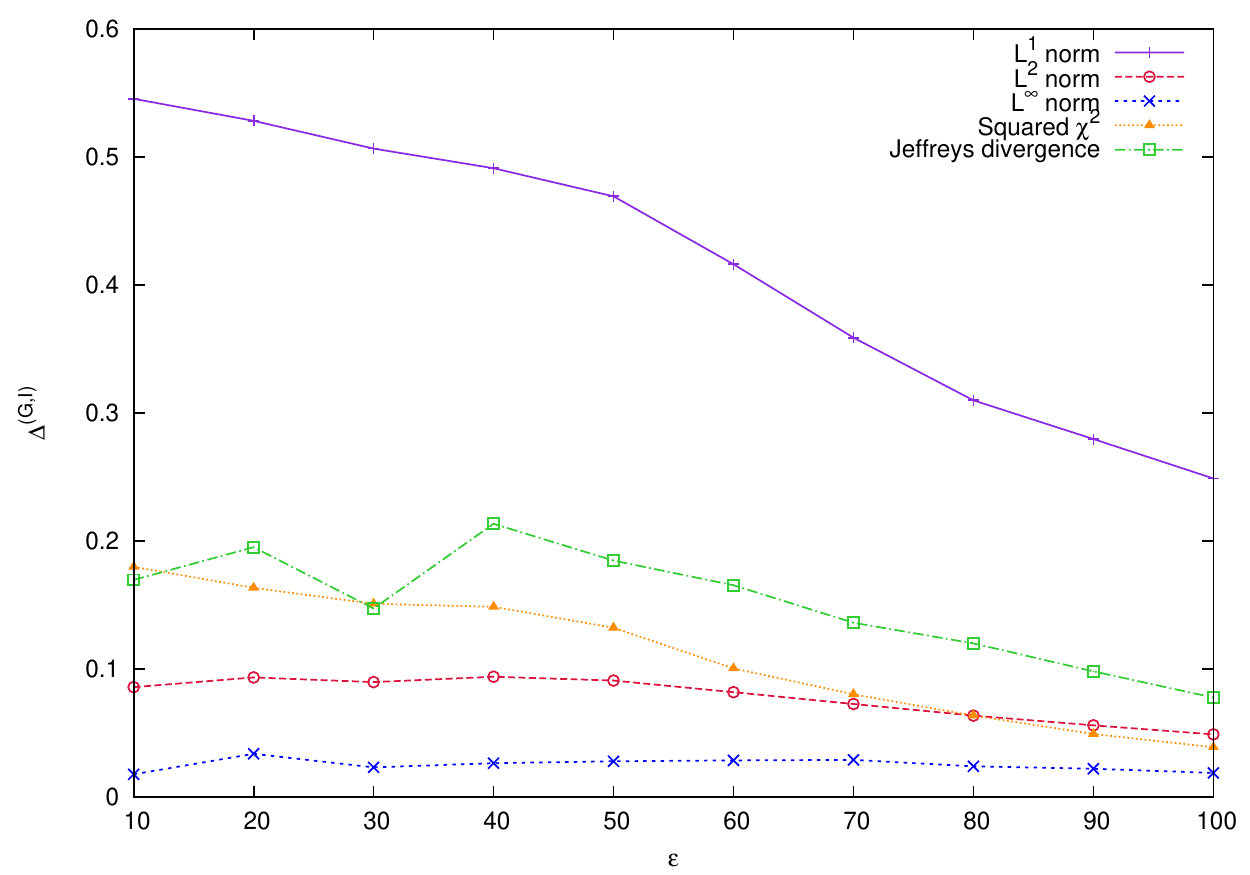}\\
\vspace{2mm}
\includegraphics[width=0.45\textwidth]{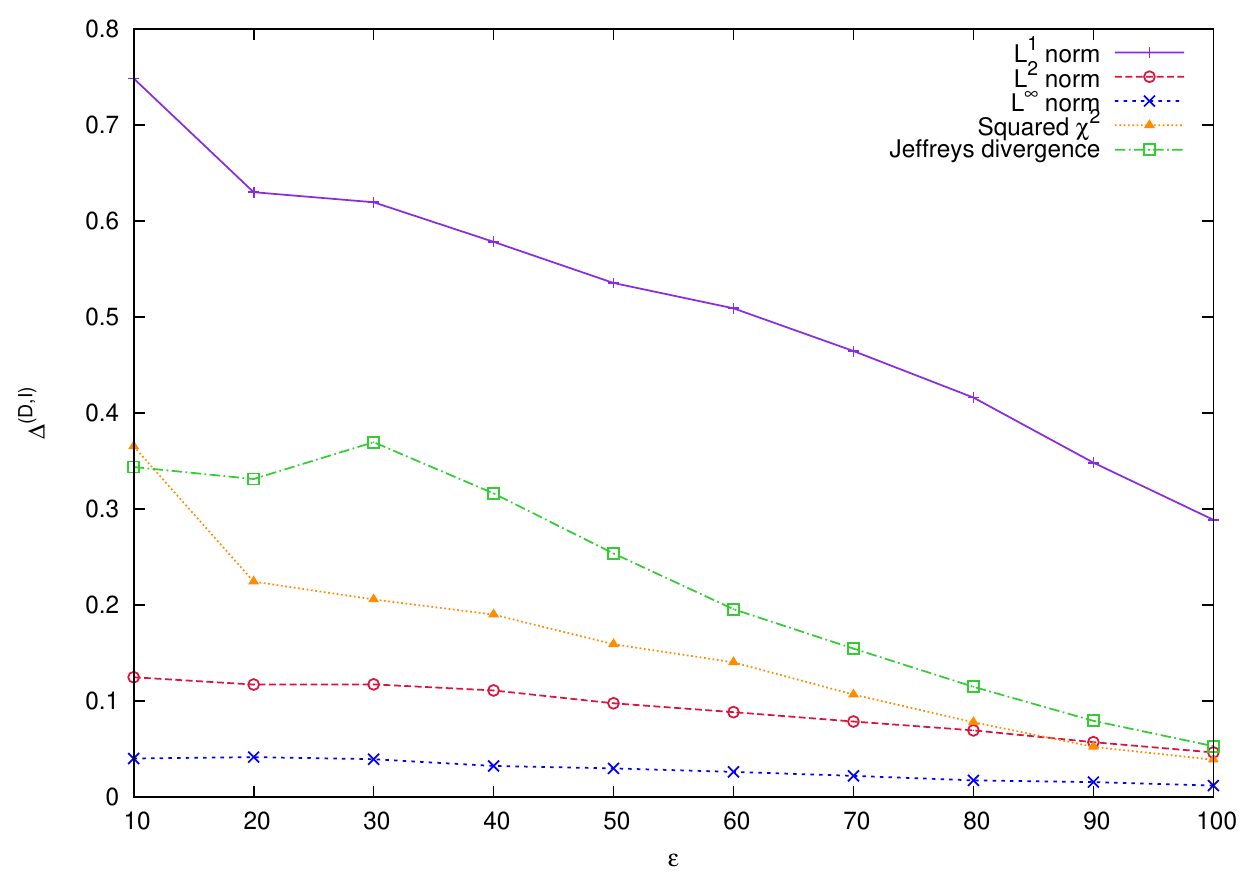}
\caption{Behaviour of the distances $\Delta^{(A,B)}$ vs. $\epsilon$, with $\rho=2.5$.}\label{mean2}
\end{figure}

Figures \ref{norm1} and \ref{mean1} illustrate the behaviour of $\langle d \rangle^{(A)}$ and $\Delta^{(A,B)}$ vs. $\epsilon$ for $\rho=1.5$, whereas 
Figs. \ref{norm2} and \ref{mean2} show the analogous behaviour of $\langle d \rangle^{(A)}$ and $\Delta^{(A,B)}$ vs. $\epsilon$ for $\rho=2.5$

\begin{figure}[H]
\centering
\includegraphics[width=0.45\textwidth]{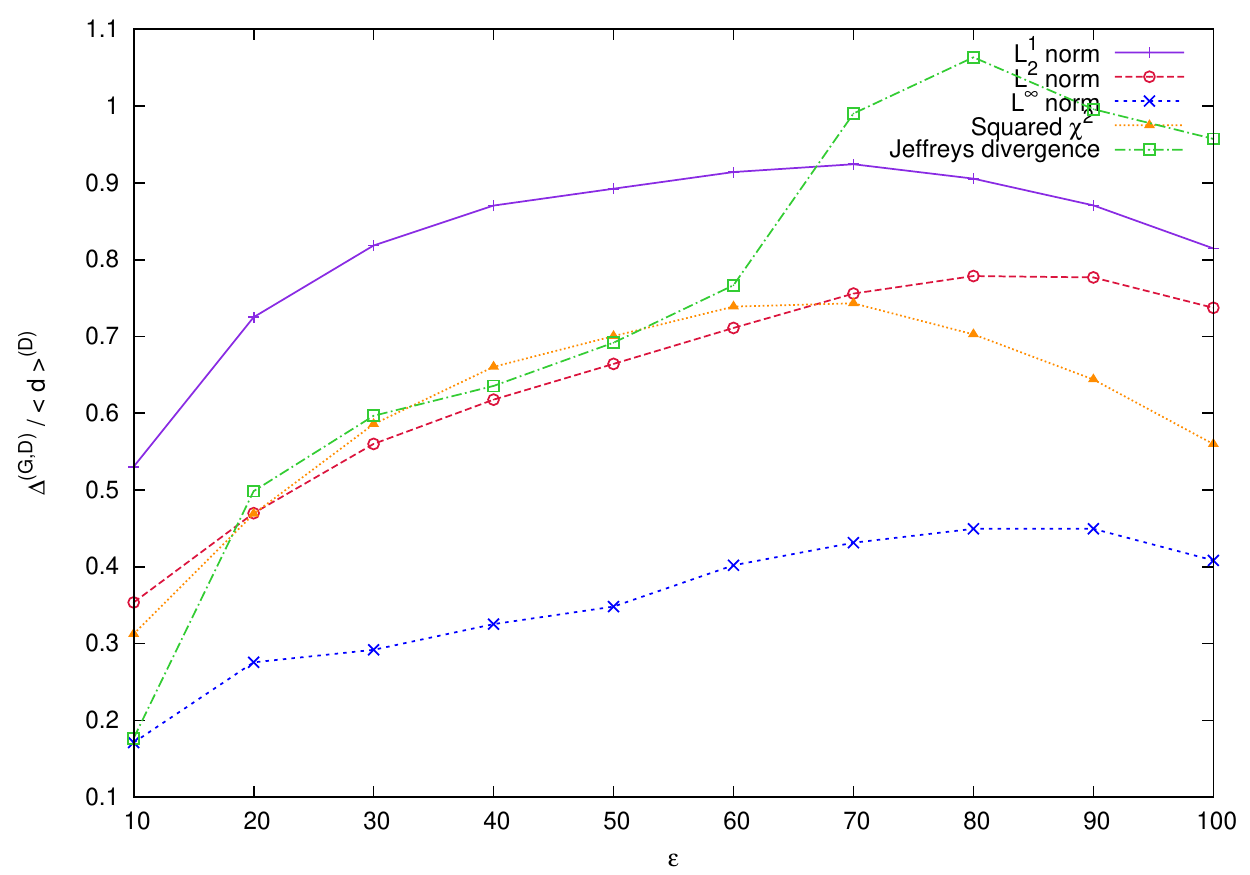}
\hspace{1mm}
\includegraphics[width=0.45\textwidth]{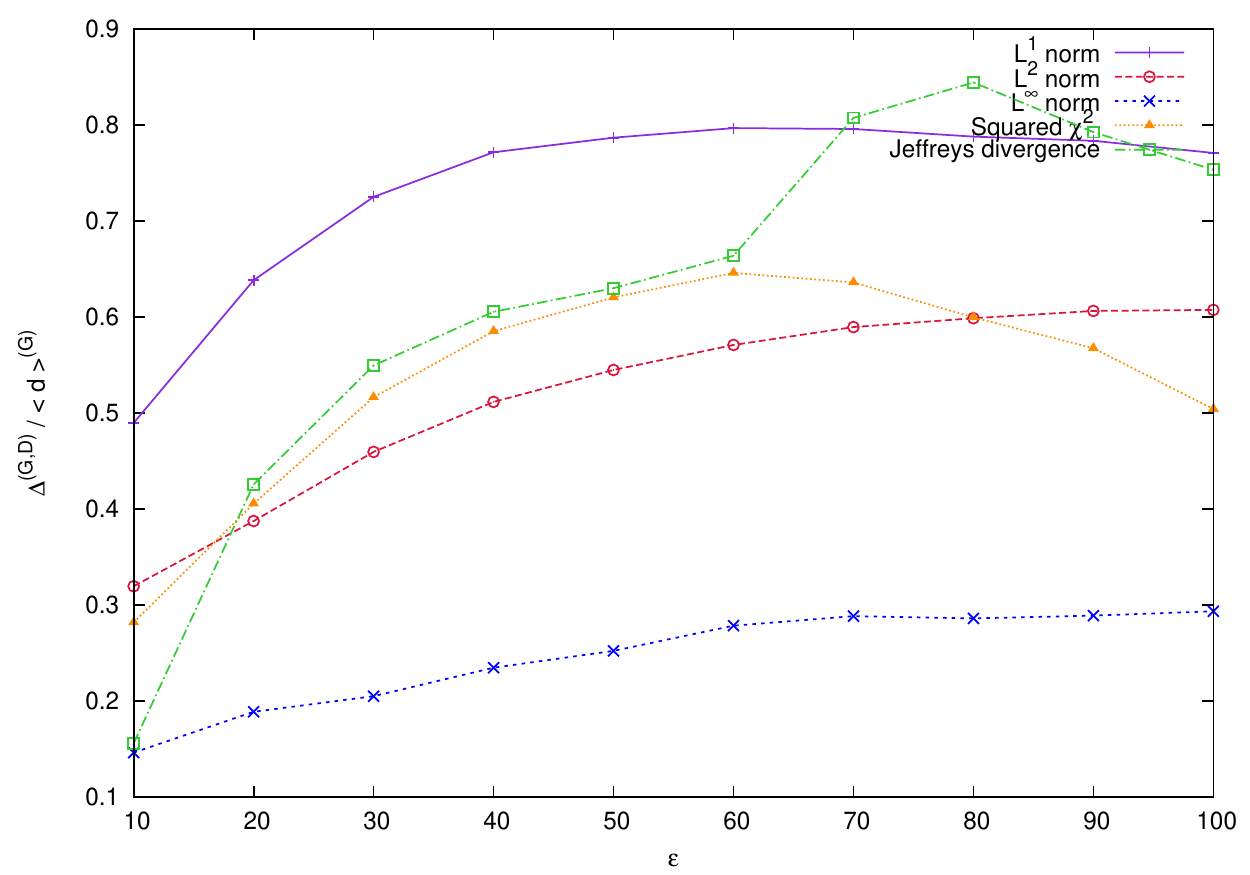}\\
\vspace{2mm}
\includegraphics[width=0.45\textwidth]{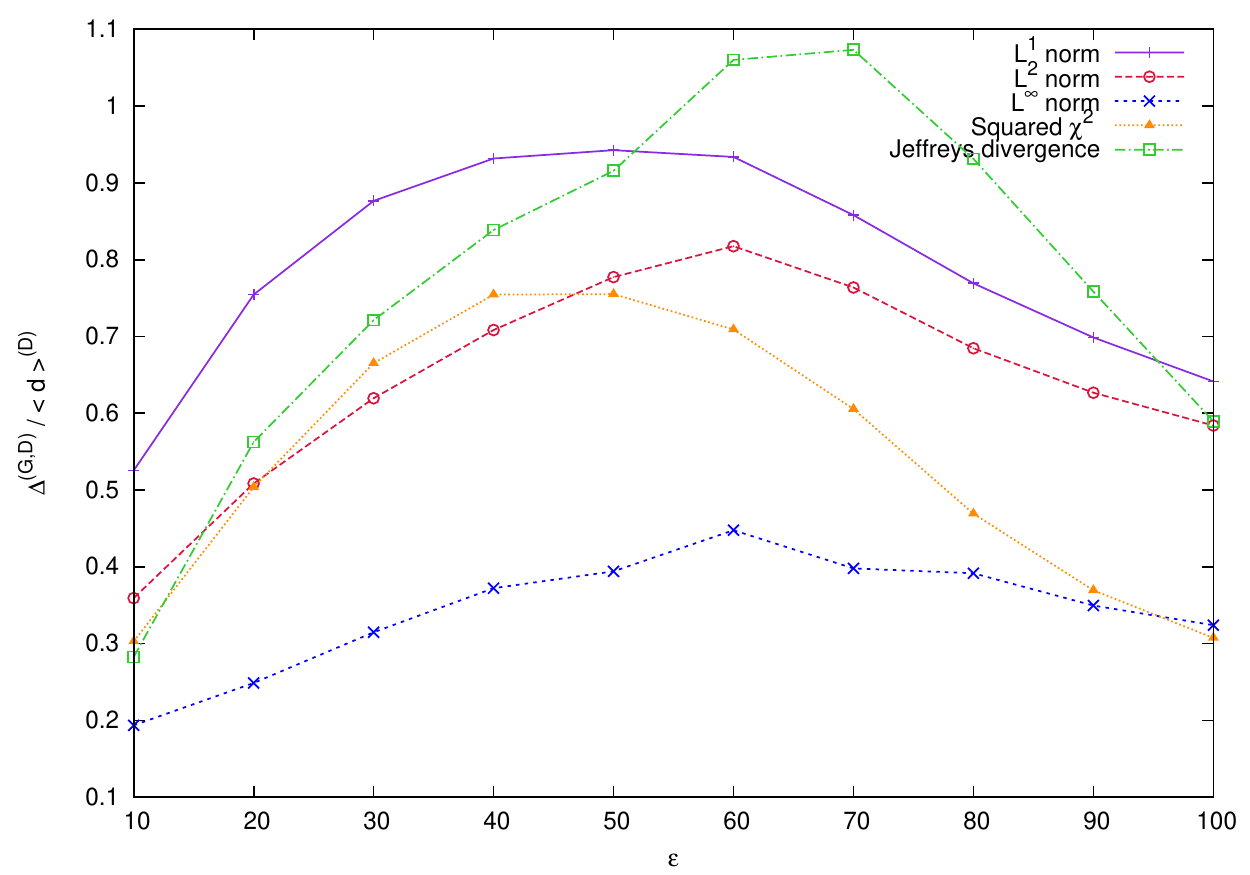}
\hspace{1mm}
\includegraphics[width=0.45\textwidth]{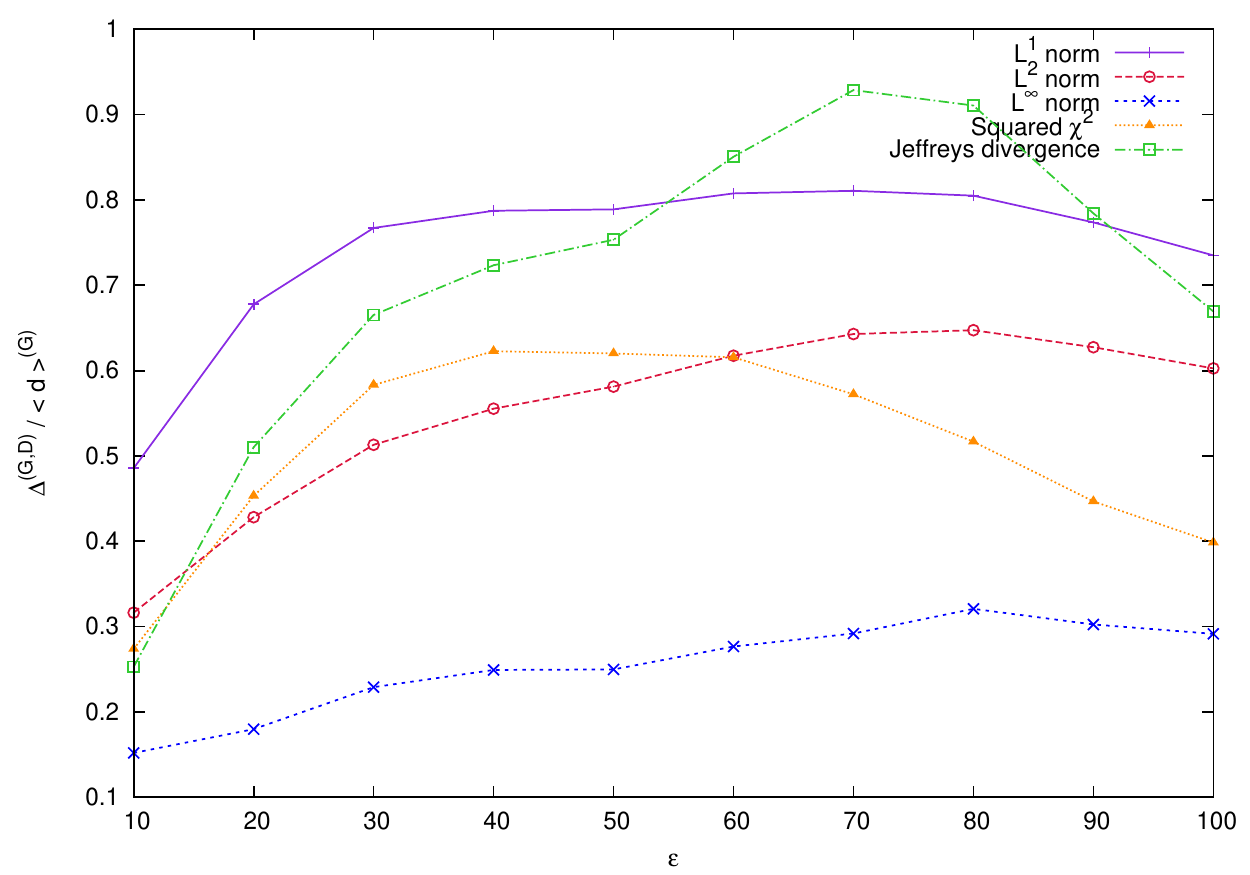}
\caption{Behaviour of the ratio $\Delta^{(G,D)}/\langle d \rangle^{(G)}$ (left column) and $\Delta^{(G,D)}/\langle d \rangle^{(D)}$ (right column) vs. $\epsilon$,  for $\rho=1.5$ (upper row) and $\rho=2.5$ (lower row).}\label{ratio}
\end{figure}

The two plots \ref{mean1} and \ref{mean2} reveal that reducing $\epsilon$ leads, on the one hand, to a remarkable increase of $\Delta^{(A,B)}$, which attains an order of magnitude of about $\Delta^{(A,B)} \simeq 10^{-1}$. On the other hand, this effect is counterbalanced by the simultaneous increase of the scale $\langle d \rangle^{(A)}$, evidenced in Figs. \ref{norm1} and \ref{norm2}, which turns out to be, for both the considered values of $\rho$, of the same order of magnitude of $\Delta^{(A,B)}$. This is more clearly visible in Fig. \ref{ratio}, which illustrates the behaviour of the ratio of $\Delta^{(A,B)}$ to $\langle d \rangle^{(A)}$ and to $\langle d \rangle^{(B)}$, for different values of $\epsilon$, obtained by setting $A = G$ and $B = D$.

The plots in Fig. \ref{ratio} confirm that the two scales $\Delta^{(A,B)}$ and $\langle d \rangle^{(A)}$ remain of the same order, also when reducing $\epsilon$. 
On the contrary, an efficient pattern recognition, based on the analysis of the image histograms, can be obtained if the ratio  $\Delta^{(A,B)} /\langle d \rangle^{(A)} \gg 1$, i.e. when the mean statistical distance between different sets overwhelms the typical size of fluctuations characteristic of each set.
Thus, the study of the behaviour of the two latter scales allows one to predict a poor performance of a machine learning algorithm aiming at classifying the hazelnuts on the basis of the image histograms.
Nevertheless, an interesting aspect can be evinced from an inspection of Fig. \ref{ratio}: despite the similarity of the magnitudes of the two statistical scales, the plot of their ratio vs. $\epsilon$ yields a non monotonic function. 

\begin{figure}[H]
\centering
\includegraphics[width=0.45\textwidth]{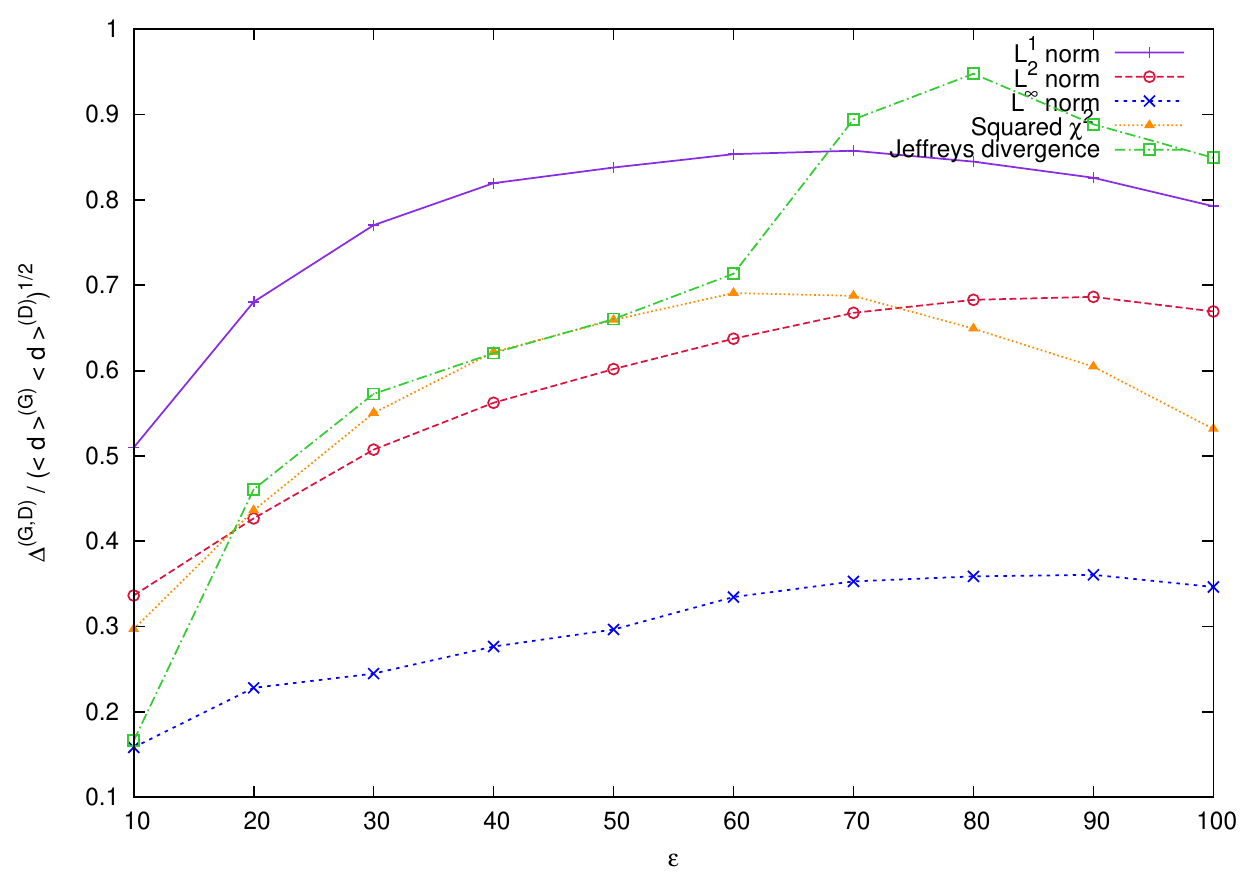}
\hspace{1mm}
\includegraphics[width=0.45\textwidth]{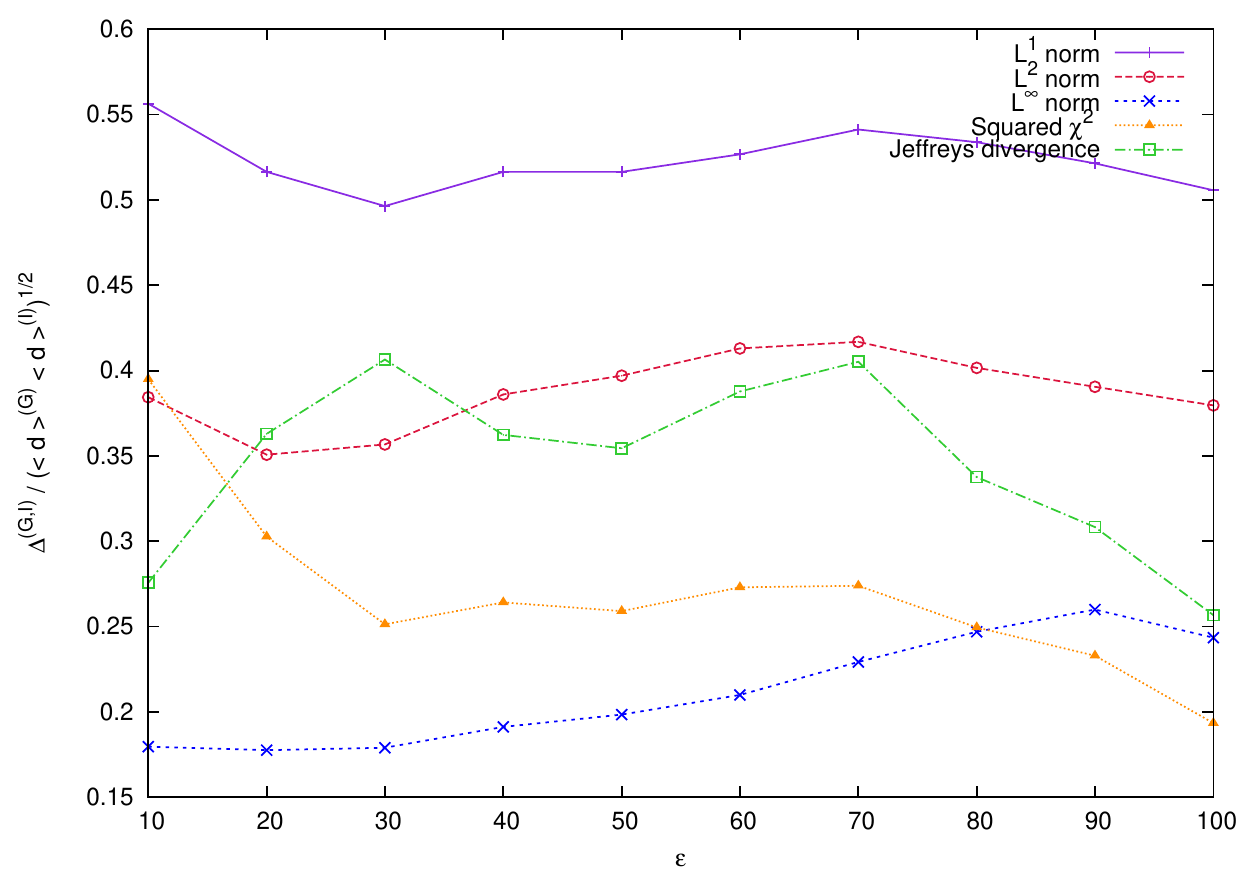}
\caption{\textit{Left panel:} Behaviour of the ratio $\Delta^{(G,D)}/\sqrt{\langle d \rangle^{(G)} \langle d \rangle^{(D)}}$ vs. $\epsilon$, for $\rho=1.5$. \textit{Right panel:} Behaviour of the ratio $\Delta^{(G,I)}/\sqrt{\langle d \rangle^{(G)} \langle d \rangle^{(I)}}$ vs. $\epsilon$, for $\rho=1.5$.}\label{ratio2}
\end{figure}

To better evidence this point, we plotted, in Fig. \ref{ratio2}, the ratio of the scale $\Delta^{(A,B)}$ to the geometric mean $\sqrt{\langle d \rangle^{(A)} \langle d \rangle^{(B)}}$, where we set $A=G, B=D$ (left panel) and $A=G, B=I$ (right panel). 

\begin{figure}[H]
\centering
\includegraphics[width=0.70\textwidth]{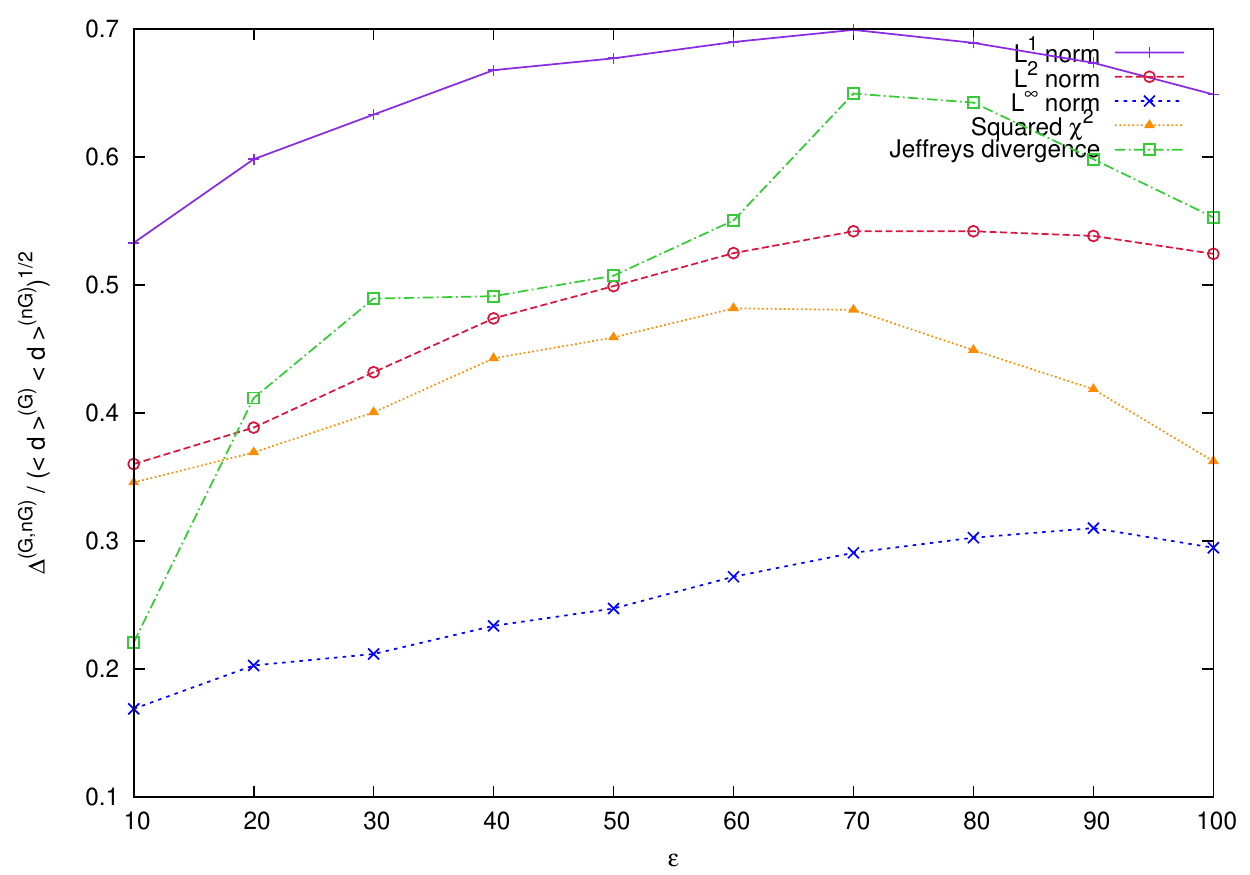}
\caption{\textit{Left panel:} Behaviour of the ratio $\Delta^{(G,nG)}/\sqrt{\langle d \rangle^{(G)} \langle d \rangle^{(nG)}}$ vs. $\epsilon$, for $\rho=1.5$. The plot evidences the onset of an optimal scale $\epsilon^*$ at which the ratio of the statistical scales is maximized.}\label{ratio3}
\end{figure}

In Fig. \ref{ratio3}, instead, for reasons to be further clarified in Sec. \ref{sec:sec2}, we show the results, analogous to those portrayed in Fig. \ref{ratio2}, obtained by merging the two sets $D$ and $I$ into one single set, labeled as $nG$ (``not good'' hazelnuts). The plot in Fig. \ref{ratio3} shows that, for $\rho=1.5$, the value $\epsilon^*=70$ maximizes the ratio of the aforementioned statistical scales wrt almost all the various notions of ``statistical distance'' we considered. In Sec. \ref{sec:sec2} we will show that such optimal value $\epsilon^*$, here obtained by only relying on information theoretic methods, can be also recovered by using Support Vector Machines numerical algorithms, by averaging their performance over a set of training samples.

\section{Support Vector Machines}
\label{sec:sec2}

\begin{figure}[H]
\centering
\includegraphics[width=0.45\textwidth]{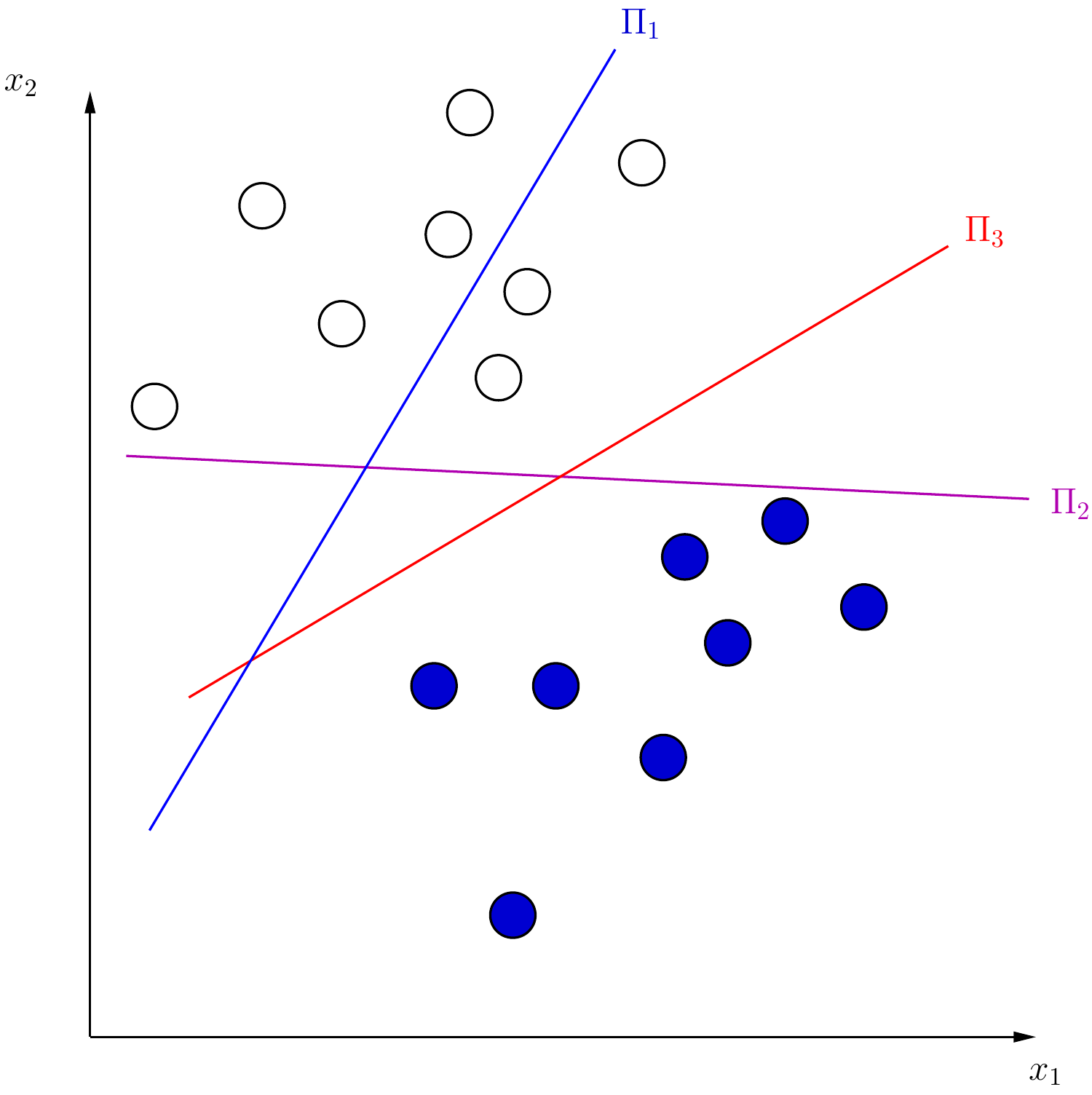}
\hspace{1mm}
\includegraphics[width=0.45\textwidth]{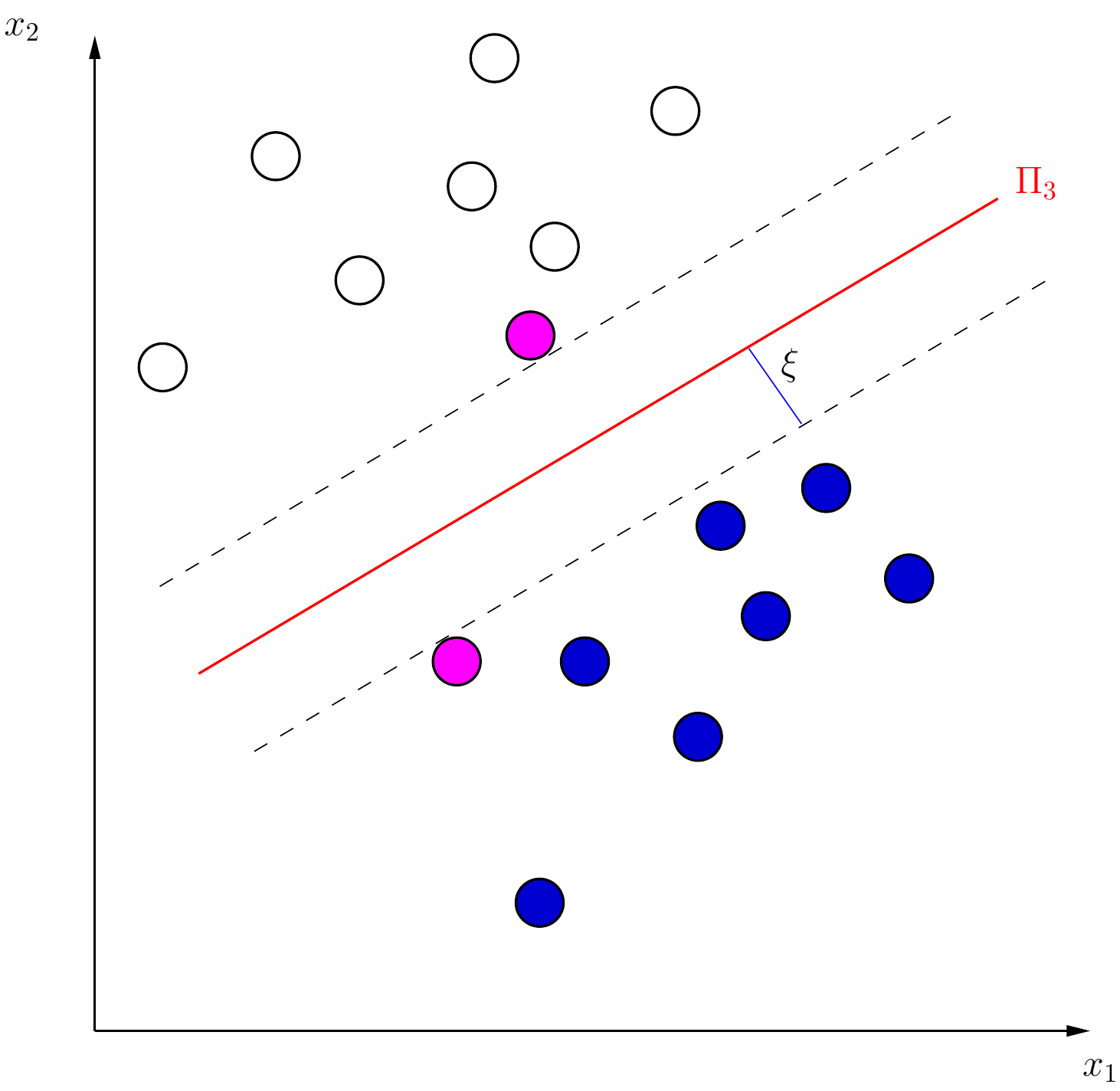}
\caption{\textit{Left panel:} Example of a linear discriminant analysis based on the SVM algorithm. Shown are three different hyperplanes: $\Pi_1$, which does not separate the two classes, $\Pi_2$ which separates the classes but only with a small margin, and $\Pi_3$, which corresponds to the best separating hyperplane. \textit{Right panel:} Illustration of the best separating hyperplane (red straight line), the canonical hyperplanes (black dashed lines), the support vectors (magenta circles) and the margin of separation $\xi$.}\label{SVM}
\end{figure}

In this Section, we discuss the results obtained by elaborating our data through a supervised learning method known as Support Vector Machines (SVM) \cite{Haykin,Boser, Cortes,Vapnik95,Vapnik98}. The SVM constitute a machine learning algorithm which seeks a separation of a set of data into two classes, by determining the \textit{best separating hyperplane} (BSH) (also referred to, in the literature, as the ``maximal margin hyperplane'' \cite{Webb}), cf. Fig. \ref{SVM}. It is worth recapitulating the basic notions underpinning the numerical algorithm we used.

Let $\{\textbf{x}\}$ denote the set of data (input pattern) to be classified, with $\textbf{x} \in E\subseteq\mathbb{R}^N$, and consider a given training set $\mathcal{T}=\{\textbf{x}_k,d_k\}_{k=1}^{N_T}$, where $N_T$ denotes the dimensionality of $\mathcal{T}$. Let, then, $d_k=\{+1,-1\}$ denote the \textit{desired response} parameter corresponding to $\textbf{x}_k$, whose value depends on which of the two classes $\textbf{x}_k$ belongs to.
The equation of a hyperplane $\Pi$ in $\mathbb{R}^N$ reads:
\be
\textbf{w}^T \cdot \textbf{x} + b=0 \quad , \nonumber
\ee
with $\textbf{w}$ and $b$ denoting, respectively, a $N$-dimensional adjustable weight vector and a bias. The BHS is the hyperplane characterized by the pair $(\textbf{w}_o,b_o)$ which, for linearly separable patterns, fulfills the following conditions \cite{Haykin}:

\bea
\textbf{w}_o^T \cdot \textbf{x}_k+b_o\ge 1 \quad \text{for $d_k = +1$} \quad ,\nonumber\\
\textbf{w}_o^T \cdot \textbf{x}_k+b_o\le -1 \quad \text{for $d_k = -1$} \quad .\label{suppvec}
\eea

\begin{figure}[H]
\centering
\includegraphics[width=0.75\textwidth]{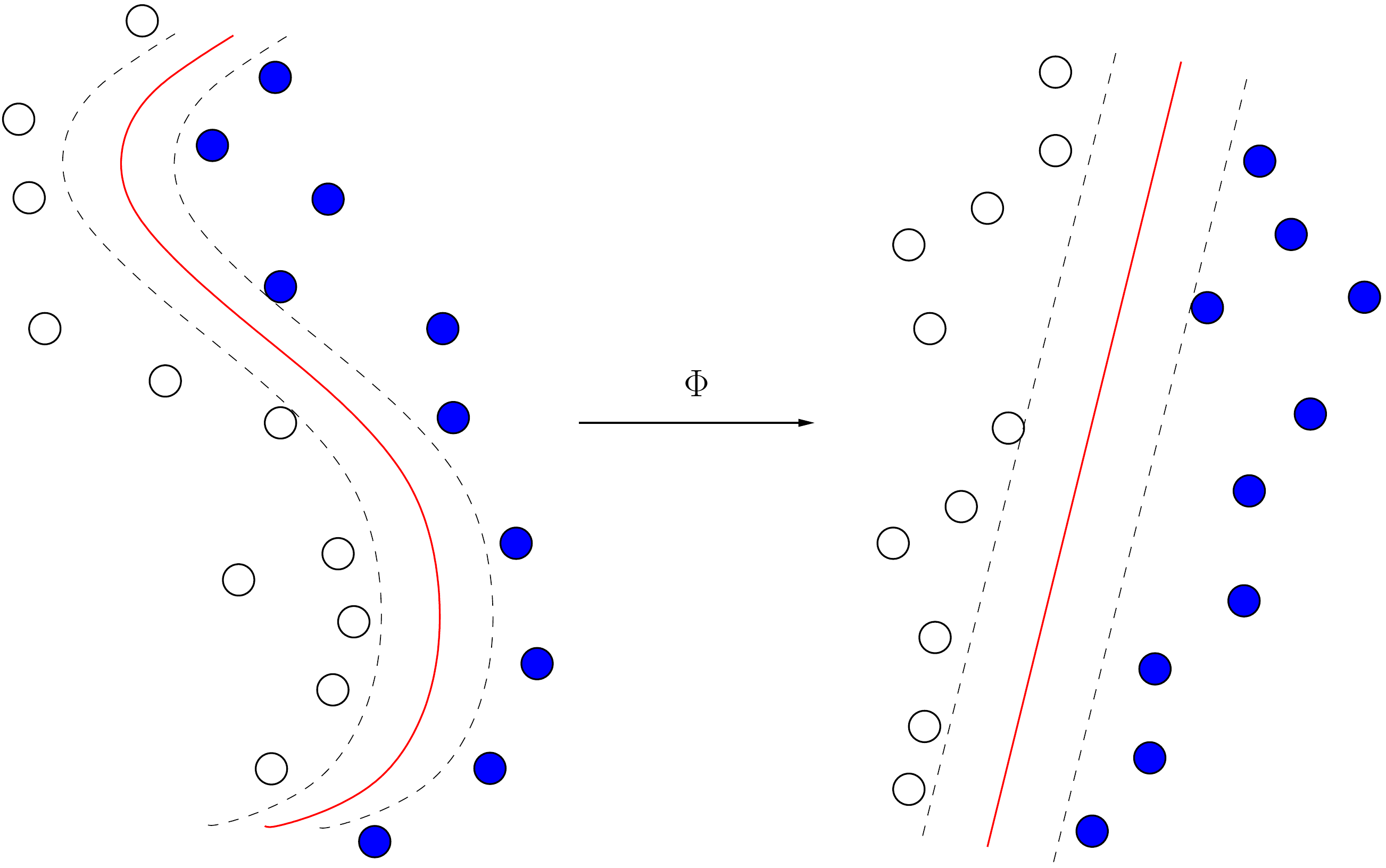}
\caption{Patterns which are not linearly separable can be mapped, via a function $\Phi$, into a \textit{feature space} where a linear separation of the classes can be achieved}\label{nlSVM}
\end{figure}

The data points, portrayed in magenta color in the right panel of Fig. \ref{SVM}, for which Eqs. (\ref{suppvec}) are satisfied with the equality sign, are called \textit{support vectors}, and lie on the so-called \textit{canonical hyperplanes} \cite{Webb}, represented by the black dashed lines in the right panel of Fig. \ref{SVM}. Figure \ref{SVM} also illustrates the so-called \textit{margin of separation}, defined as the distance $\xi=1/\|\textbf{w}_o\|$ between the support vectors and the BSH.
The BSH, which maximizes $\xi$ under the constraints (\ref{suppvec}), can be found by determining the saddlepoint of the Lagrangian function $d\mathcal{L}(\textbf{w},b,\lambda_1,...,\lambda_{N_T})=0$, given by:
\be
\mathcal{L}(\textbf{w},b,\lambda_1,...,\lambda_{N_T})=\frac{1}{2}\textbf{w}^T\cdot \textbf{w}-\sum_{k=1}^{N_T} \lambda_k[d_k(\textbf{w} \cdot \textbf{x}_k+b)-1] \label{lagr} \quad .
\ee 
The solution of such variational problem is easily found in the form \cite{Haykin}:
\be
\mathbf{w}_o=\sum_{k=1}^{N_T} \lambda_k d_k \textbf{x}_k \label{sol1}
\ee
where the Lagrange multipliers $\lambda_k$ satisfy the conditions:
\bea
\sum_{k=1}^{N_T} \lambda_k d_k&=&0 \quad ,\nonumber\\
\lambda_k[d_k(\textbf{w} \cdot \textbf{x}_k+b)-1]&=&0 \quad \text{for   $k=1,...,N_T$}  \quad ,\nonumber
\eea
(the latter being known as the ``Karush-Kuhn-Tucker complementarity condition'' \cite{Webb}) whereas $b_o$ can be determined, once $\textbf{w}_o$ is known, using Eqs. (\ref{suppvec}).
When the two classes are not linearly separable, a possible strategy consists in introducing a suitable (nonlinear) function $\Phi:E\rightarrow F$ , which makes it possible to map the original pattern inputs into a \textit{feature space} $F\subseteq\mathbb{R}^M$, in which a linear separation can be performed, cf. Fig. \ref{nlSVM} \cite{Webb}.

\begin{figure}[H]
\centering
\includegraphics[width=0.45\textwidth]{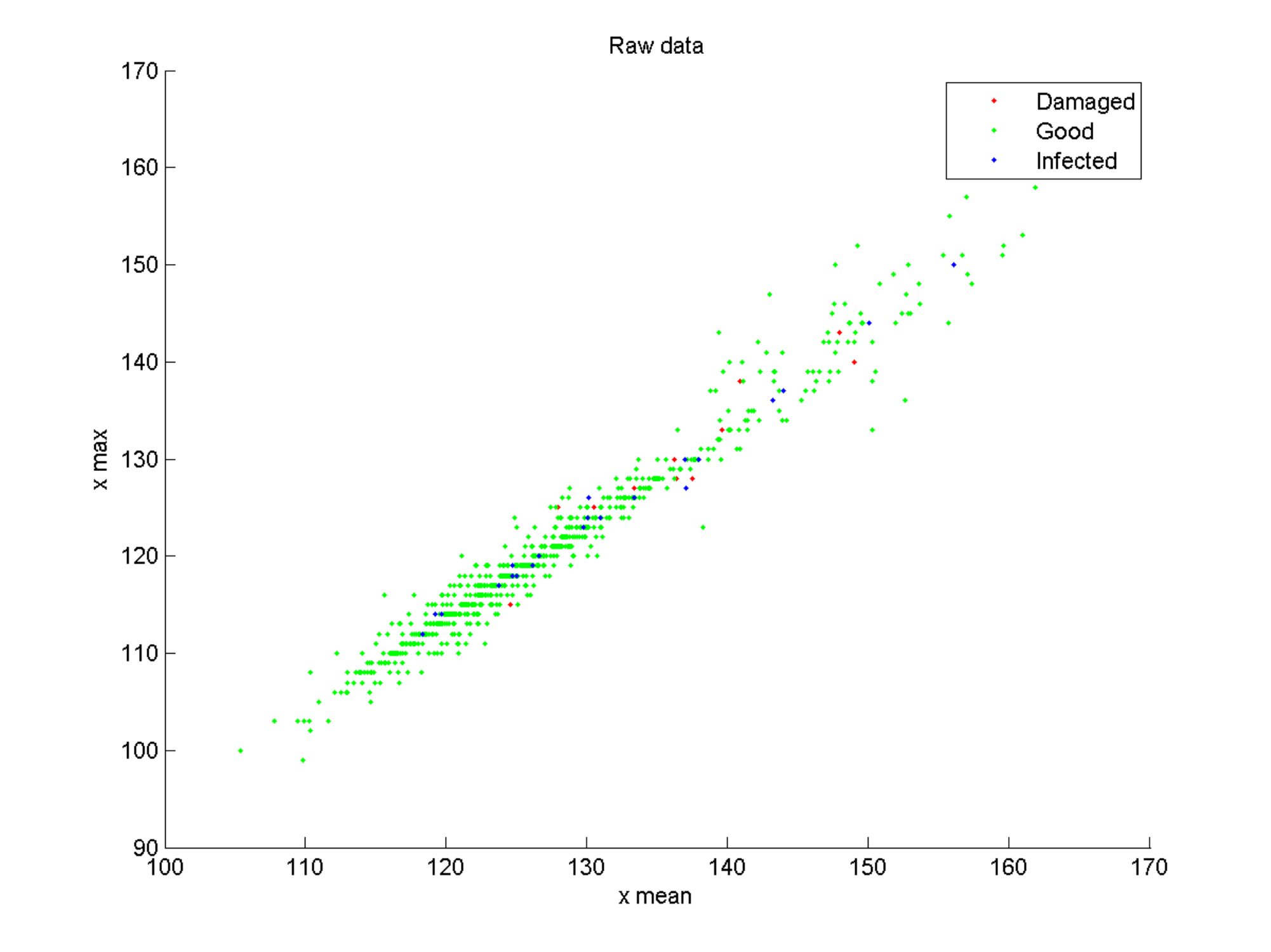}
\hspace{1mm}
\includegraphics[width=0.45\textwidth]{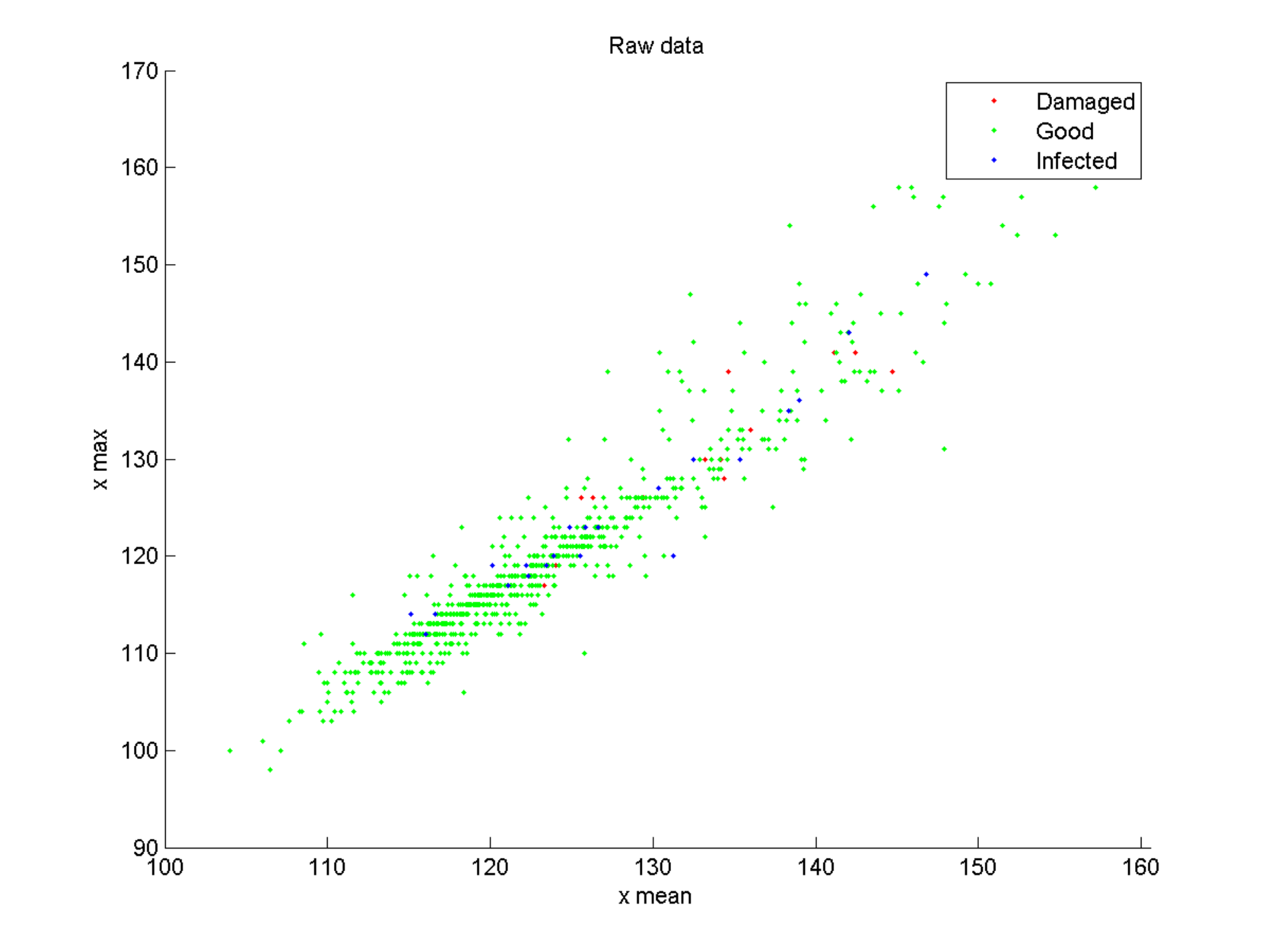}\\
\vspace{2mm}
\includegraphics[width=0.45\textwidth]{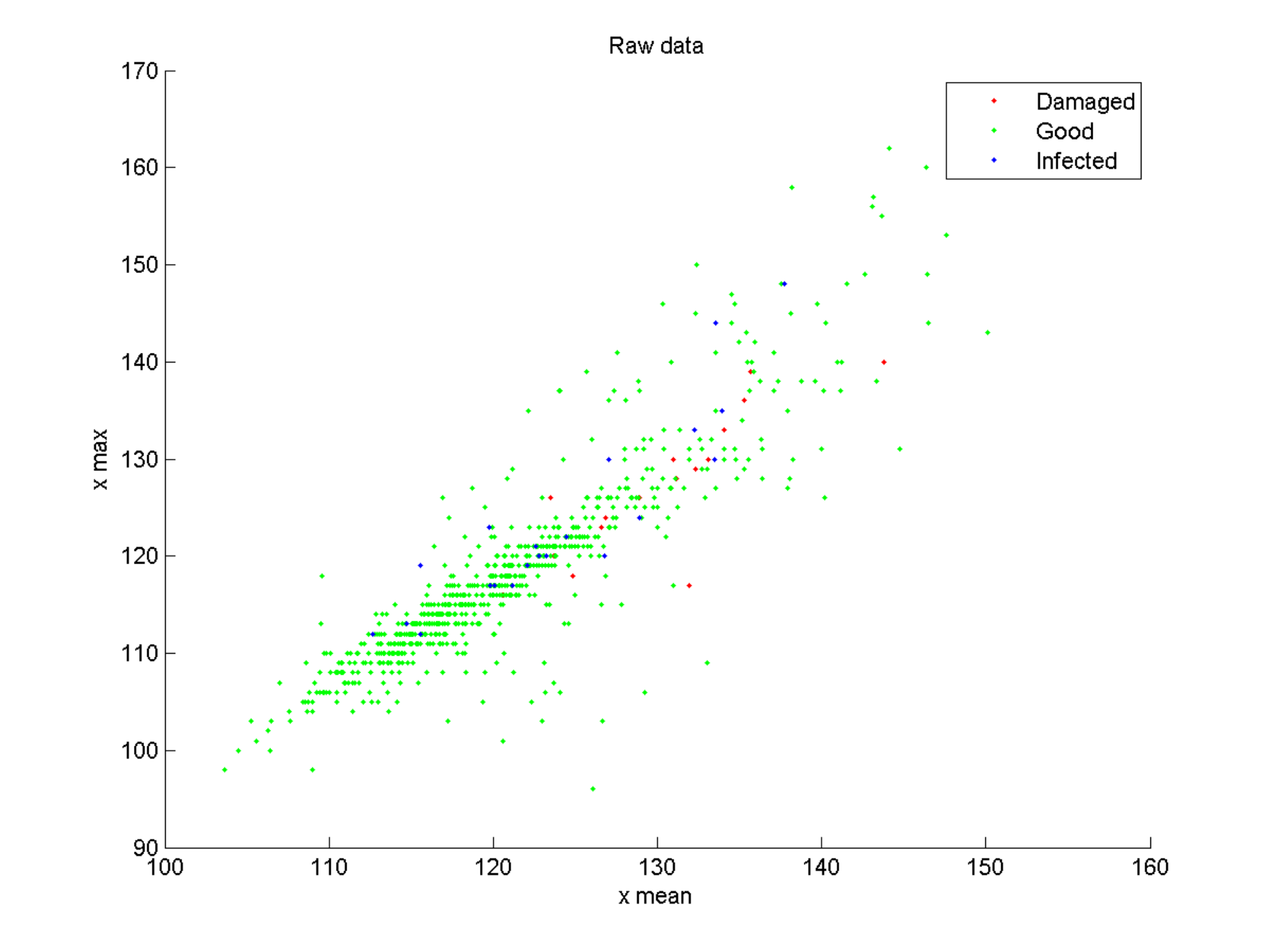}
\hspace{1mm}
\includegraphics[width=0.45\textwidth]{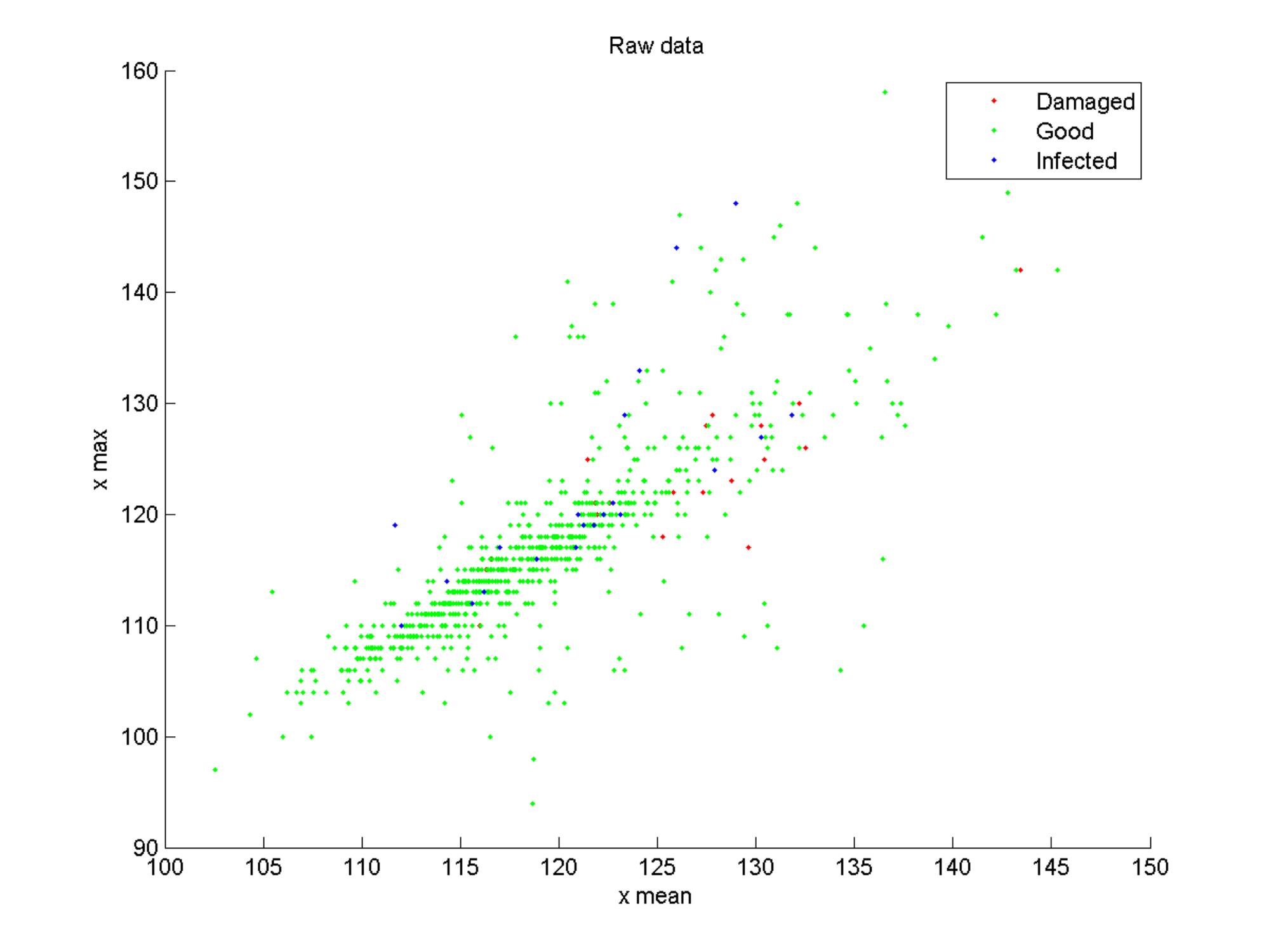}
\caption{Classification of the data in the 2D space spanned by the values of the observables $x_{mean}$ (horizontal axis) and $x_{max}$ (vertical axis), for $\epsilon=20$ (top left), $\epsilon=40$ (top right), $ \epsilon=60$ (bottom left), and $ \epsilon=80$ (bottom right), with $\rho=1.5$.}\label{raw3a}
\end{figure}

\begin{figure}[H]
\centering
\includegraphics[width=0.45\textwidth]{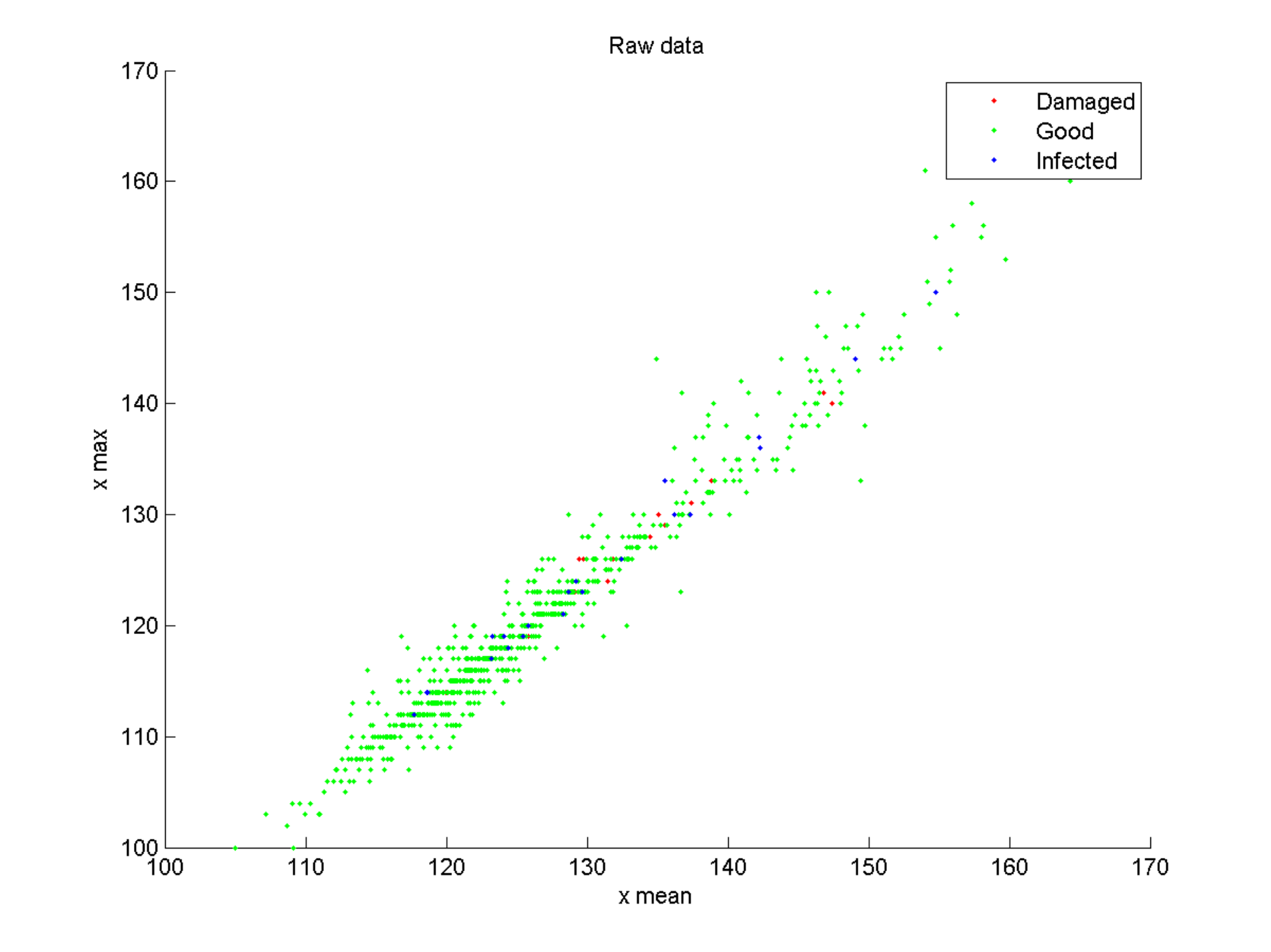}
\hspace{1mm}
\includegraphics[width=0.45\textwidth]{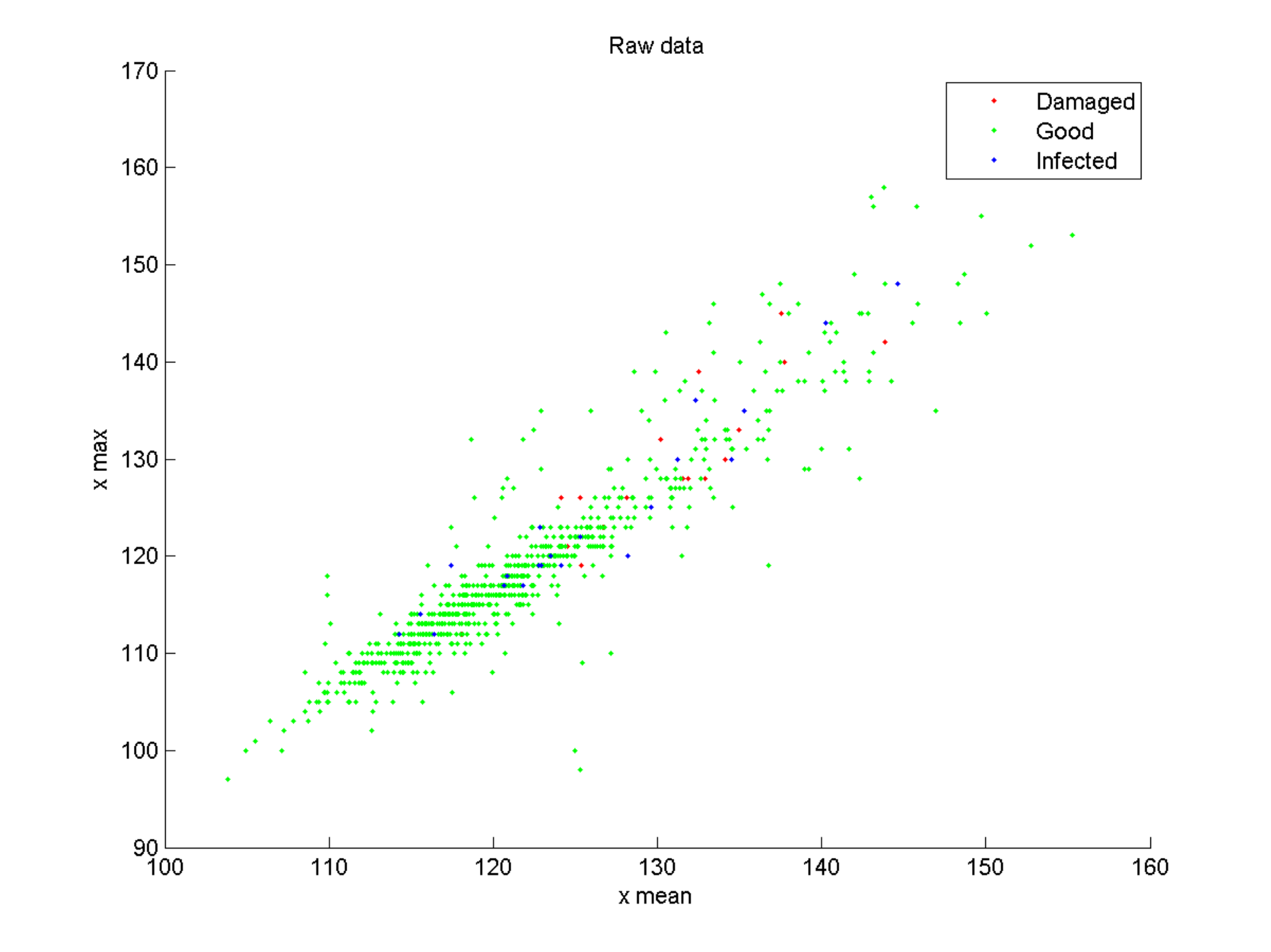}\\
\vspace{2mm}
\includegraphics[width=0.45\textwidth]{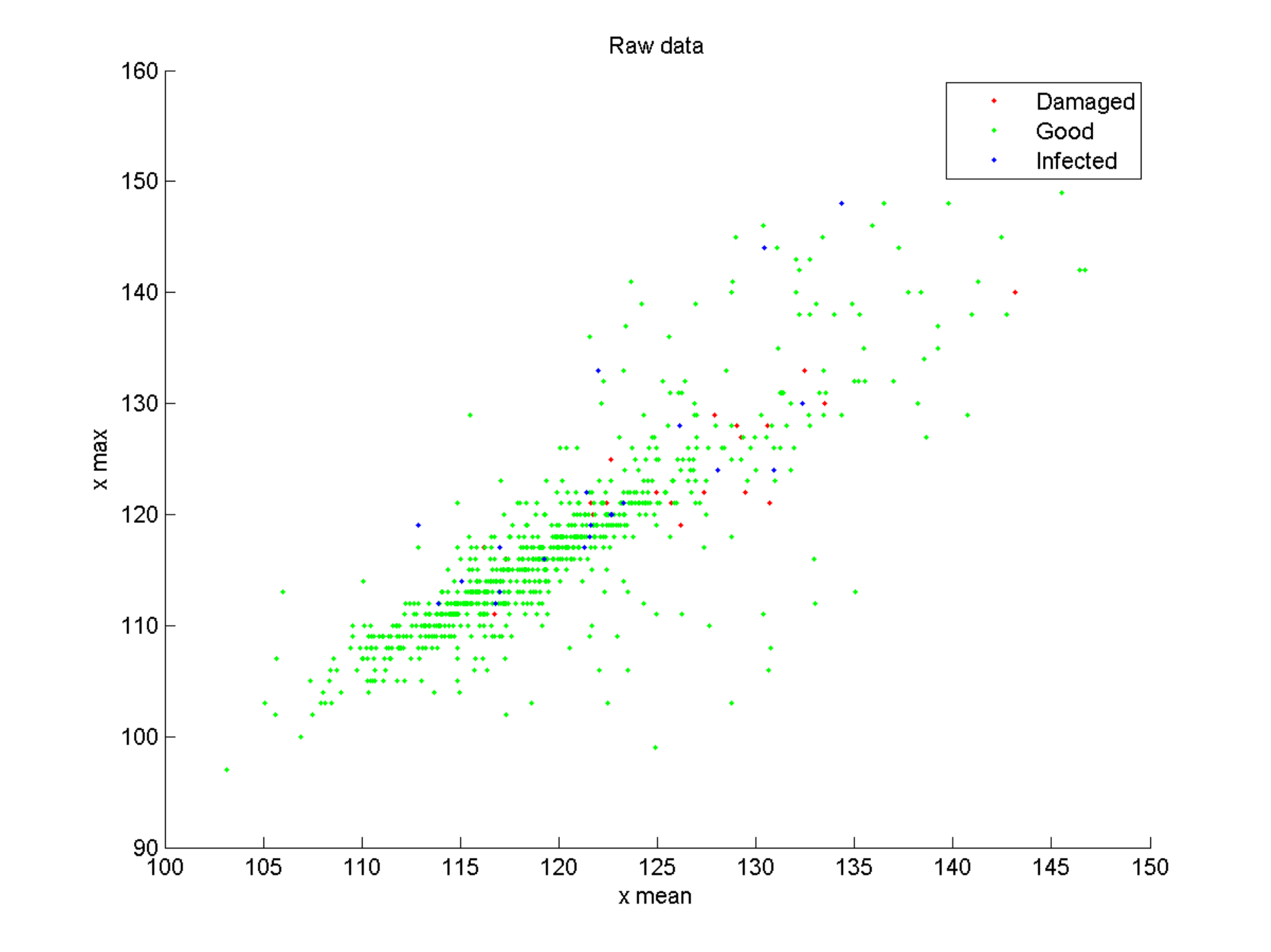}
\hspace{1mm}
\includegraphics[width=0.45\textwidth]{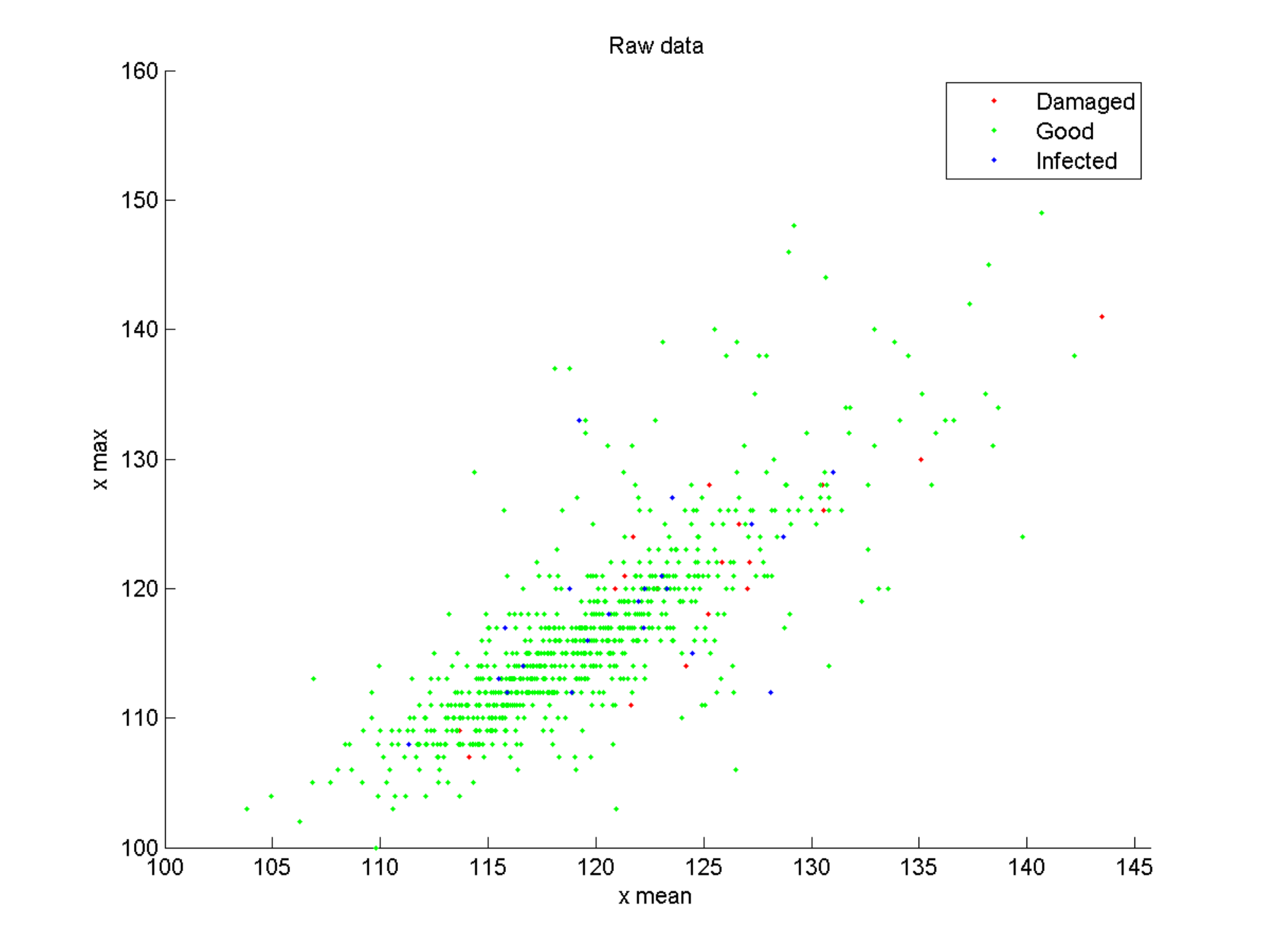}
\caption{Classification of the data in the 2D space spanned by the values of the observables $x_{mean}$ (horizontal axis) and $x_{max}$ (vertical axis), for $\epsilon=20$ (top left), $\epsilon=40$ (top right), $ \epsilon=60$ (bottom left), and $ \epsilon=80$ (bottom right), with $\rho=2.5$.}\label{raw3b}
\end{figure}

Thus,by denoting as $\boldsymbol\Phi(\textbf{x})=\{\Phi_j(\textbf{x})\}_{j=1}^M$ a set of nonlinear transformations from the original input space to the feature space, the corresponding variational problem leads now, in place of Eq. (\ref{sol1}), to the expression:
\be
\mathbf{w}_o=\sum_{k=1}^{N_T} \lambda_k d_k \boldsymbol\Phi(\textbf{x}_k) \label{sol2} \quad .
\ee
In our implementation of the SVM algorithm, we regarded the set $G$ as one of the two classes, whereas the other class, formerly introduced in Sec. \ref{sec:sec1} and denoted by $nG$, was thought of as given by the union $nG = D \cup I$.
We thus relied on the analysis of the histograms of the hazelnut nuclei, detailed in Sec. \ref{sec:sec1}. Therefore, we introduced two variables to identify each hazelnut: we set $\textbf{x}=(x_{mean},x_{max})$, where, for each histogram relative to an hazelnut nucleus, $x_{mean}$ and $x_{max}$ denote, respectively, the \textit{average} shade of gray and the shade of gray equipped with the highest probability.
Therefore, in the space spanned by the coordinates $x_{mean}$ and $x_{max}$, and parameterized by the values of $\epsilon$ and $\rho$, each hazelnut is represented by a single dot. The resulting distribution of dots, for different values of $\epsilon$ and  $\rho$, is illustrated in Figs. \ref{raw3a} and \ref{raw3b}, which evidence a clustering of points, for both the considered values of $\rho$, around the bisectrix of the plane. This is readily explained by considering that, when reducing $\epsilon$, the histograms attain a more and more symmetric shape. 

\begin{figure}[H]
\centering
\includegraphics[width=0.45\textwidth]{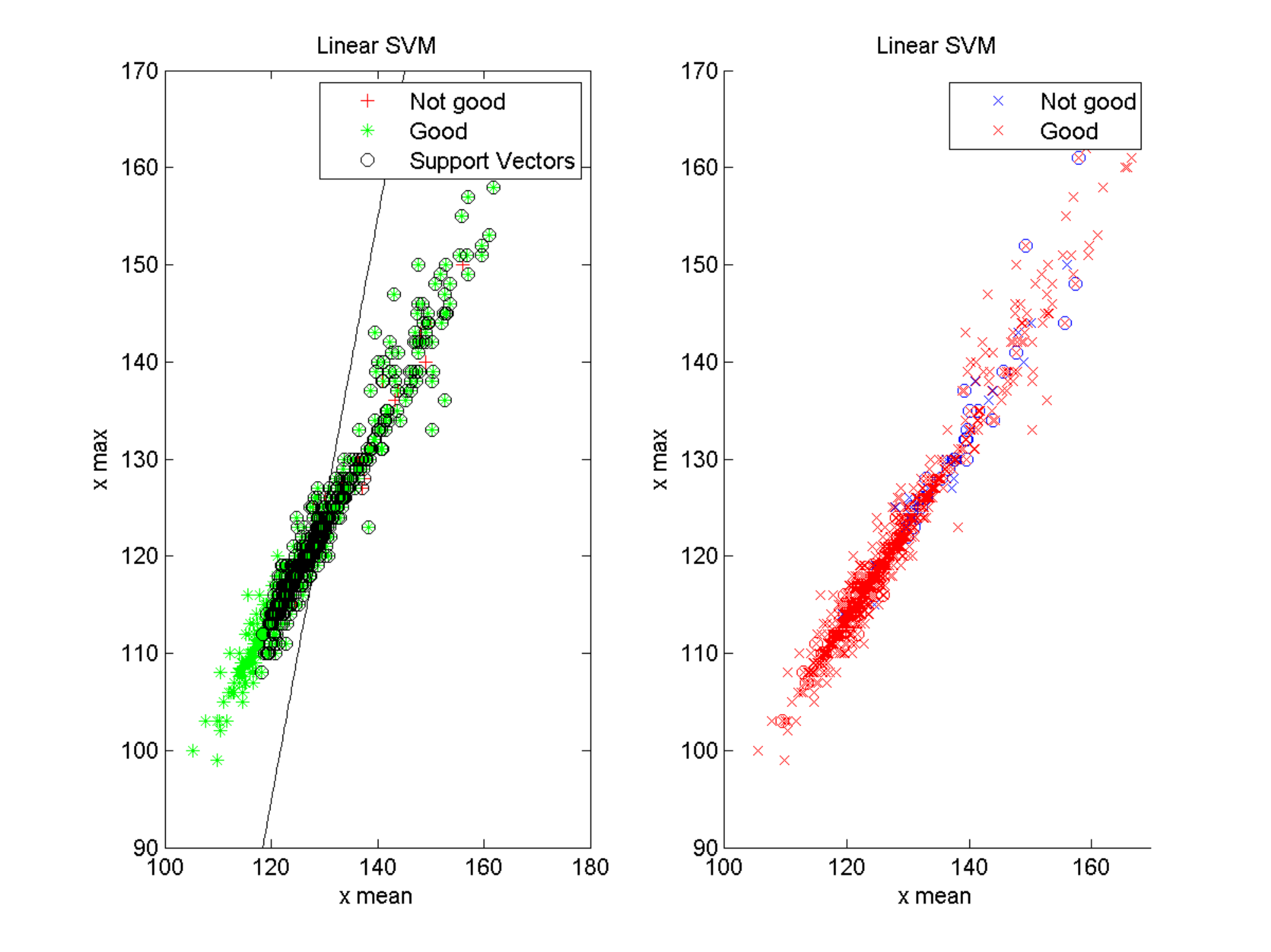}
\hspace{1mm}
\includegraphics[width=0.45\textwidth]{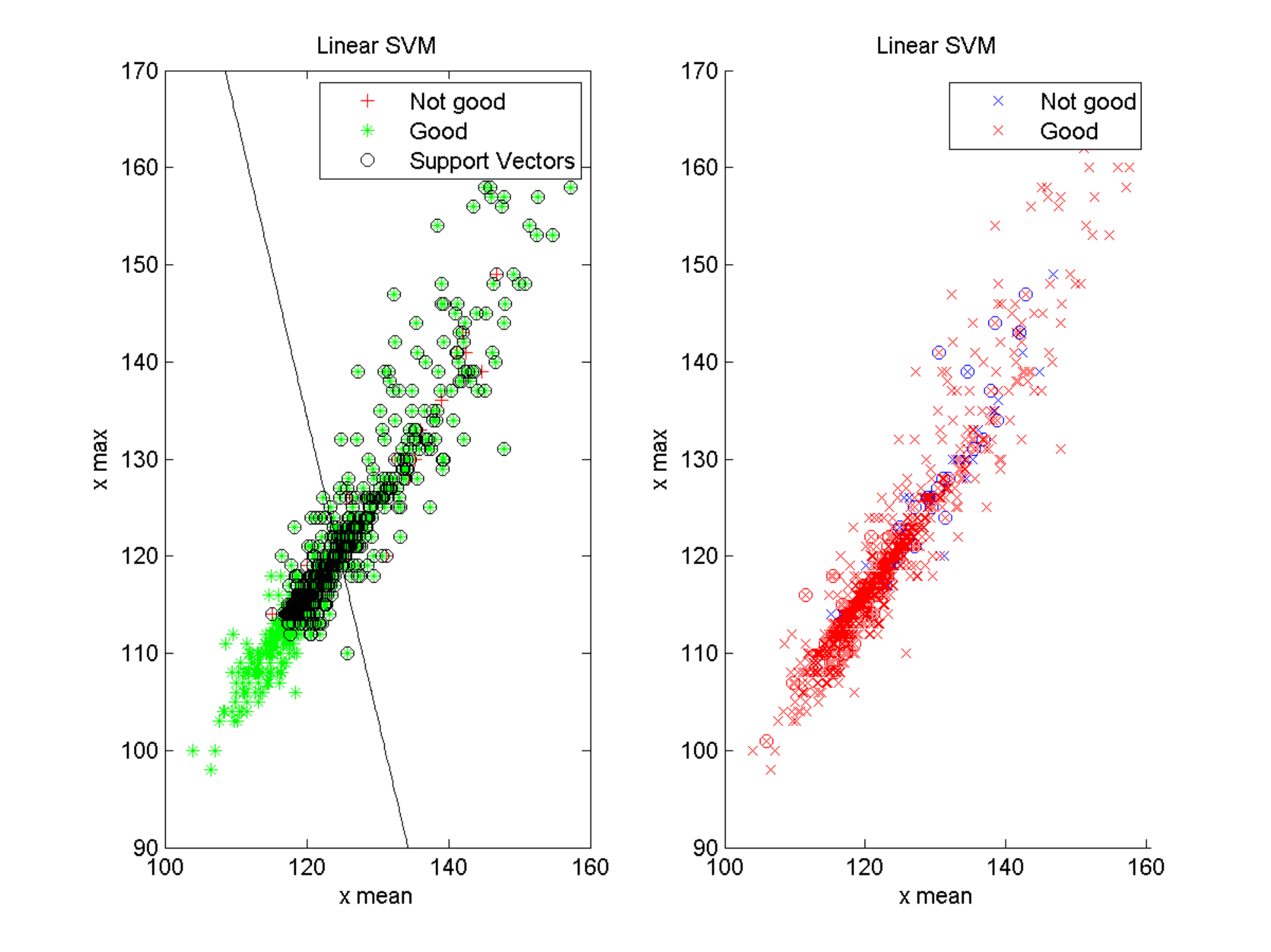}\\
\vspace{2mm}
\includegraphics[width=0.45\textwidth]{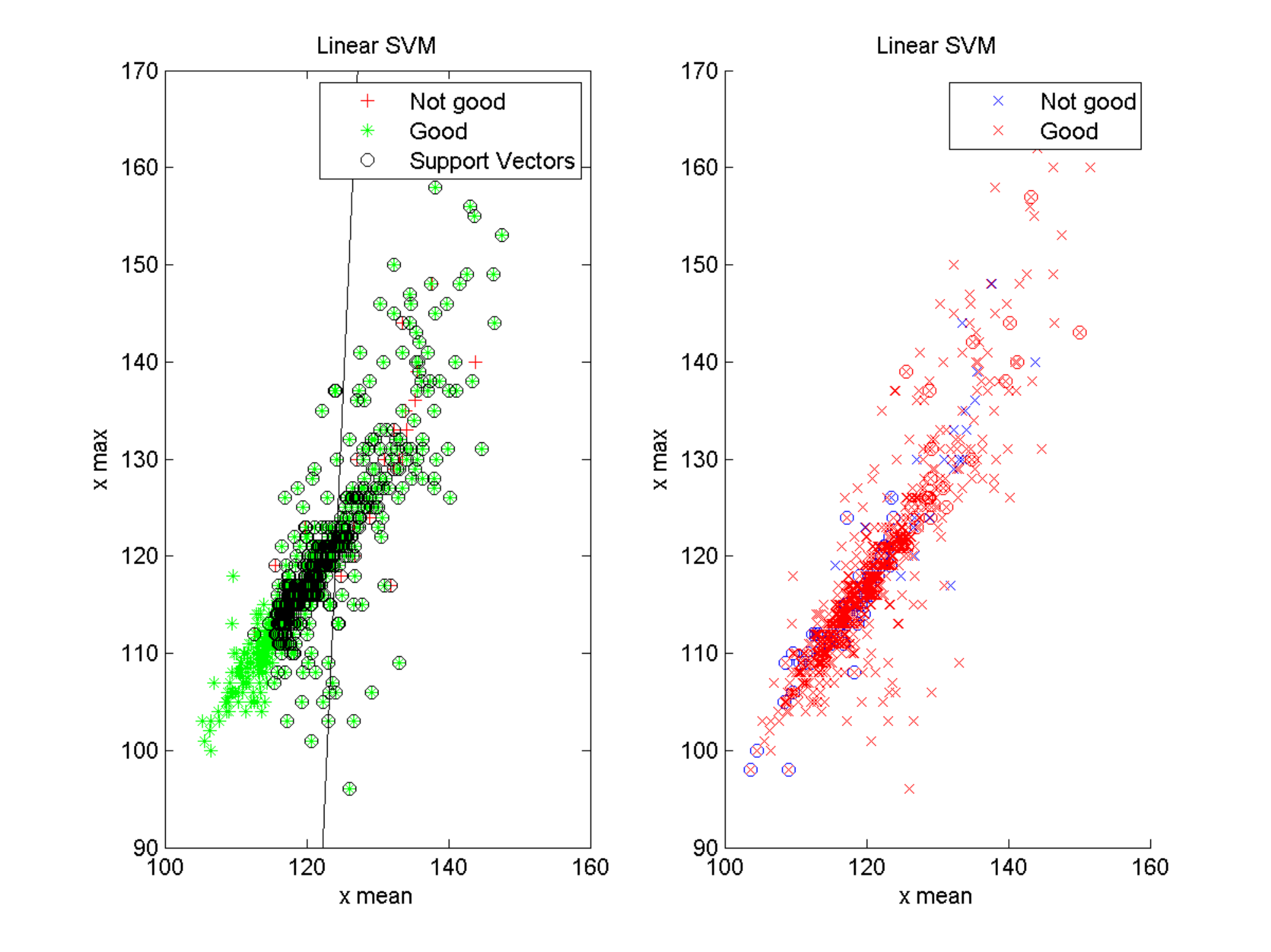}
\hspace{1mm}
\includegraphics[width=0.45\textwidth]{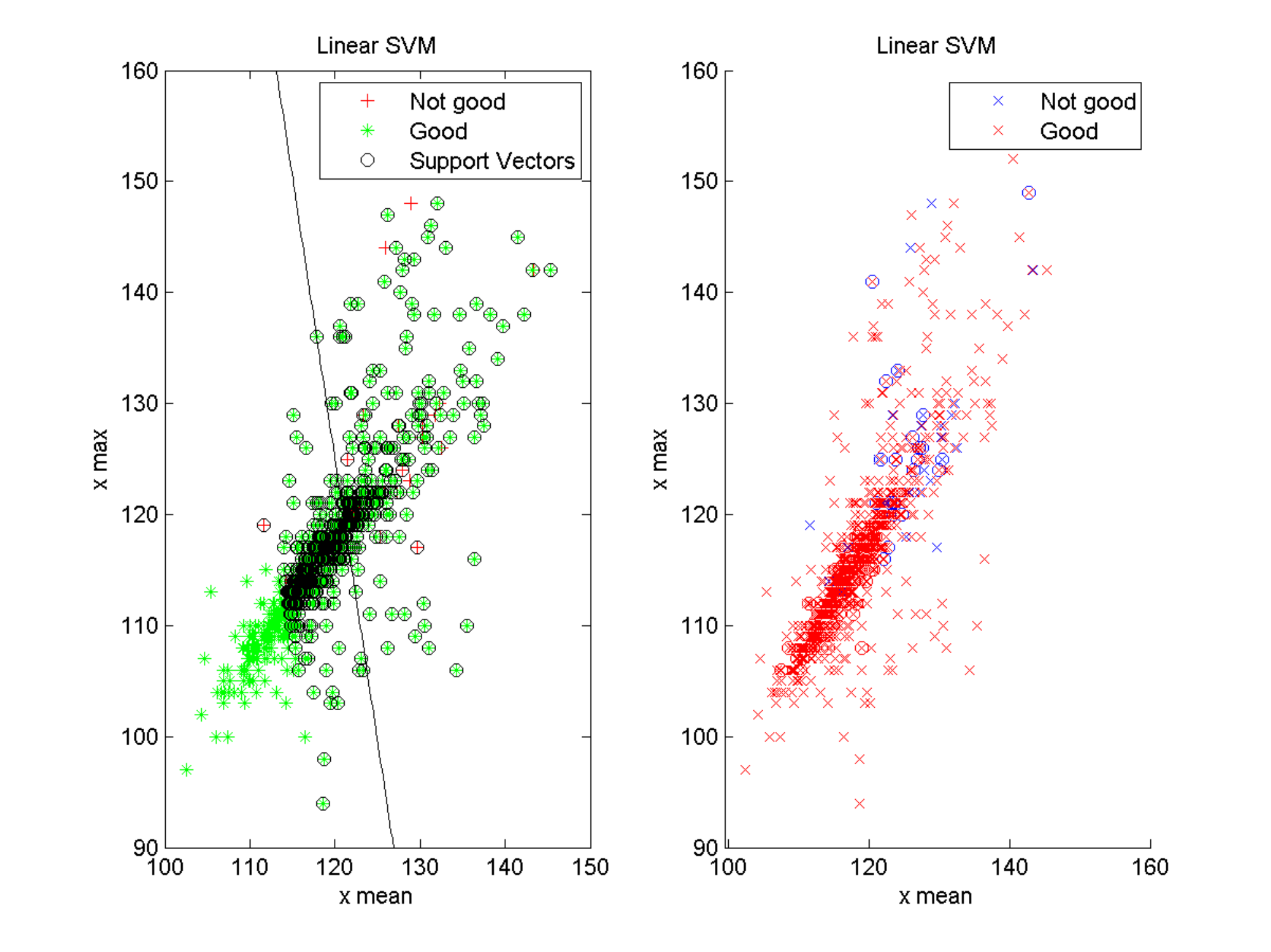}
\caption{Classification of the data through a linear SVM algorithm. Shown is the 2D space spanned by the values of the observables $x_{mean}$ (horizontal axis) and $x_{max}$ (vertical axis), for $\epsilon=20$ (top left), $\epsilon=40$ (top right), $\epsilon=60$ (bottom left), and $ \epsilon=80$ (bottom right), with $\rho=1.5$. In each left subfigure, shown are the training set of data (green and red crosses, denoting, respectively, the elements of the classes $G$ and $nG$), the support vectors (black circles) and the best separating hyperplane (black line). According to the SVM classification,the elements of the class $nG$ are expected to lie on the right of the boundary line. The right sub-figures, instead, display the 2D representation of all the available data (red and blue crosses, denoting, respectively, the elements of $G$ and those of $nG$) and the SVM output (red and blue circles).}\label{lin1}
\end{figure}

\begin{figure}[H]
\centering
\includegraphics[width=0.45\textwidth]{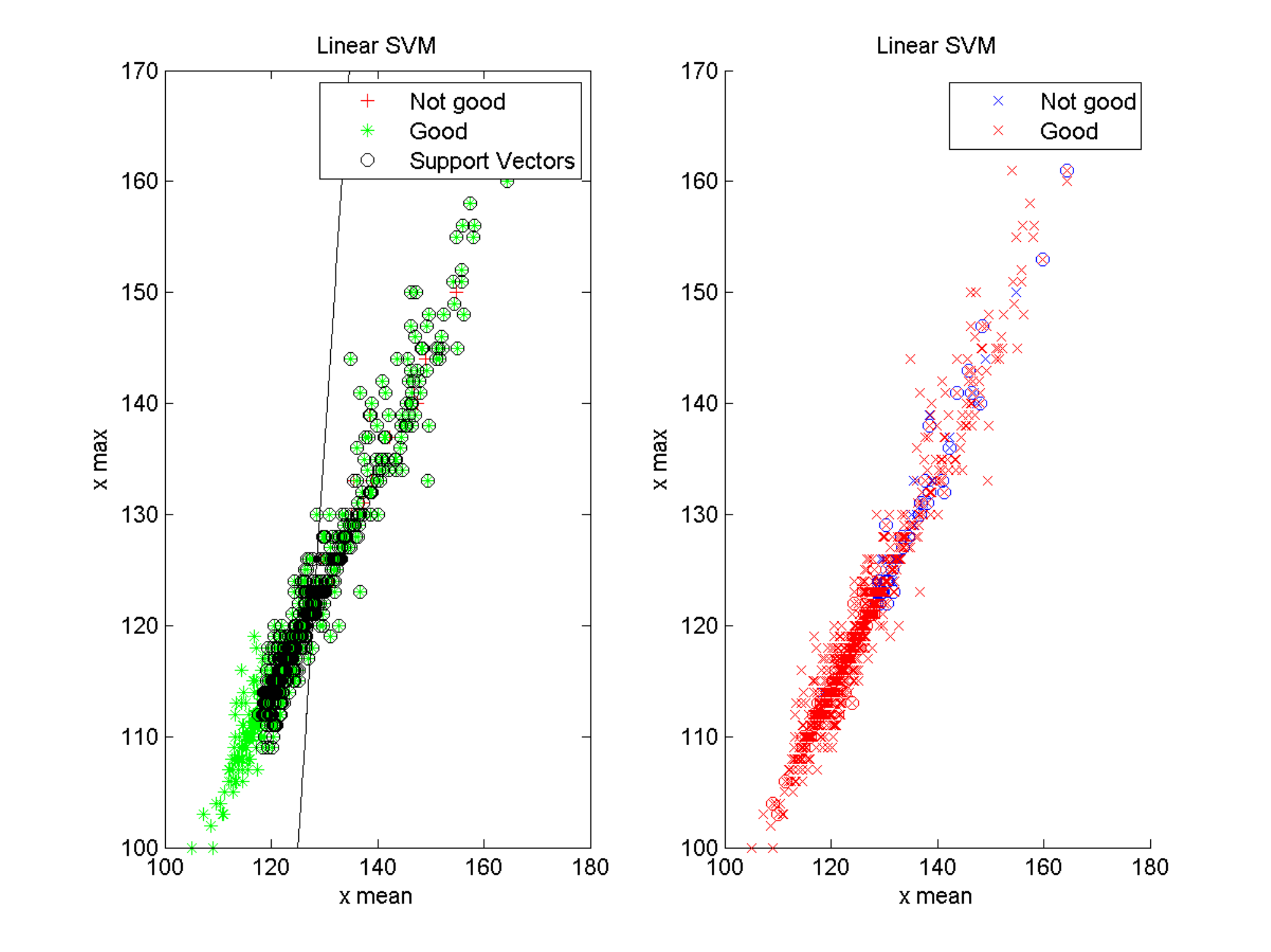}
\hspace{1mm}
\includegraphics[width=0.45\textwidth]{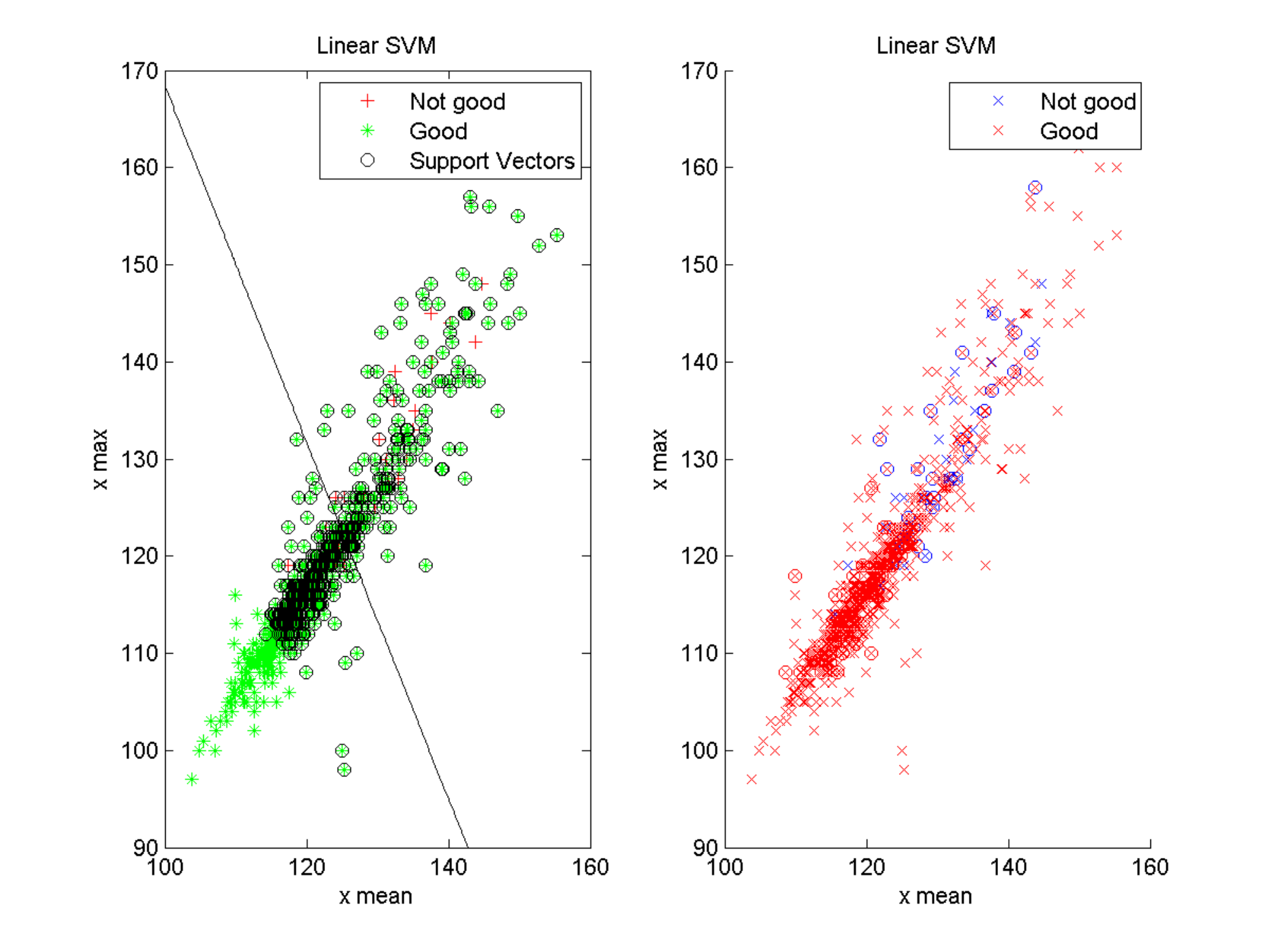}\\
\vspace{2mm}
\includegraphics[width=0.45\textwidth]{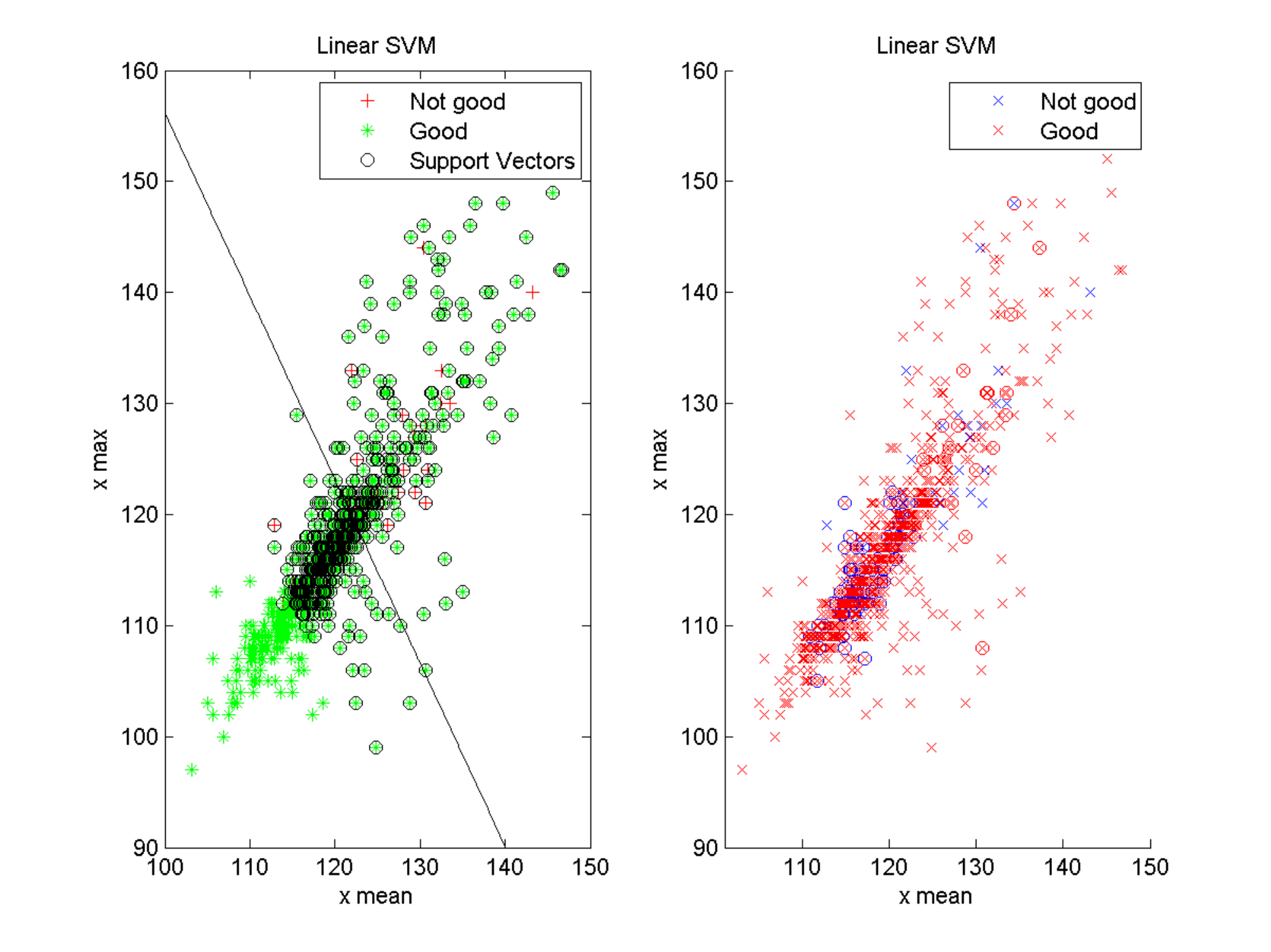}
\hspace{1mm}
\includegraphics[width=0.45\textwidth]{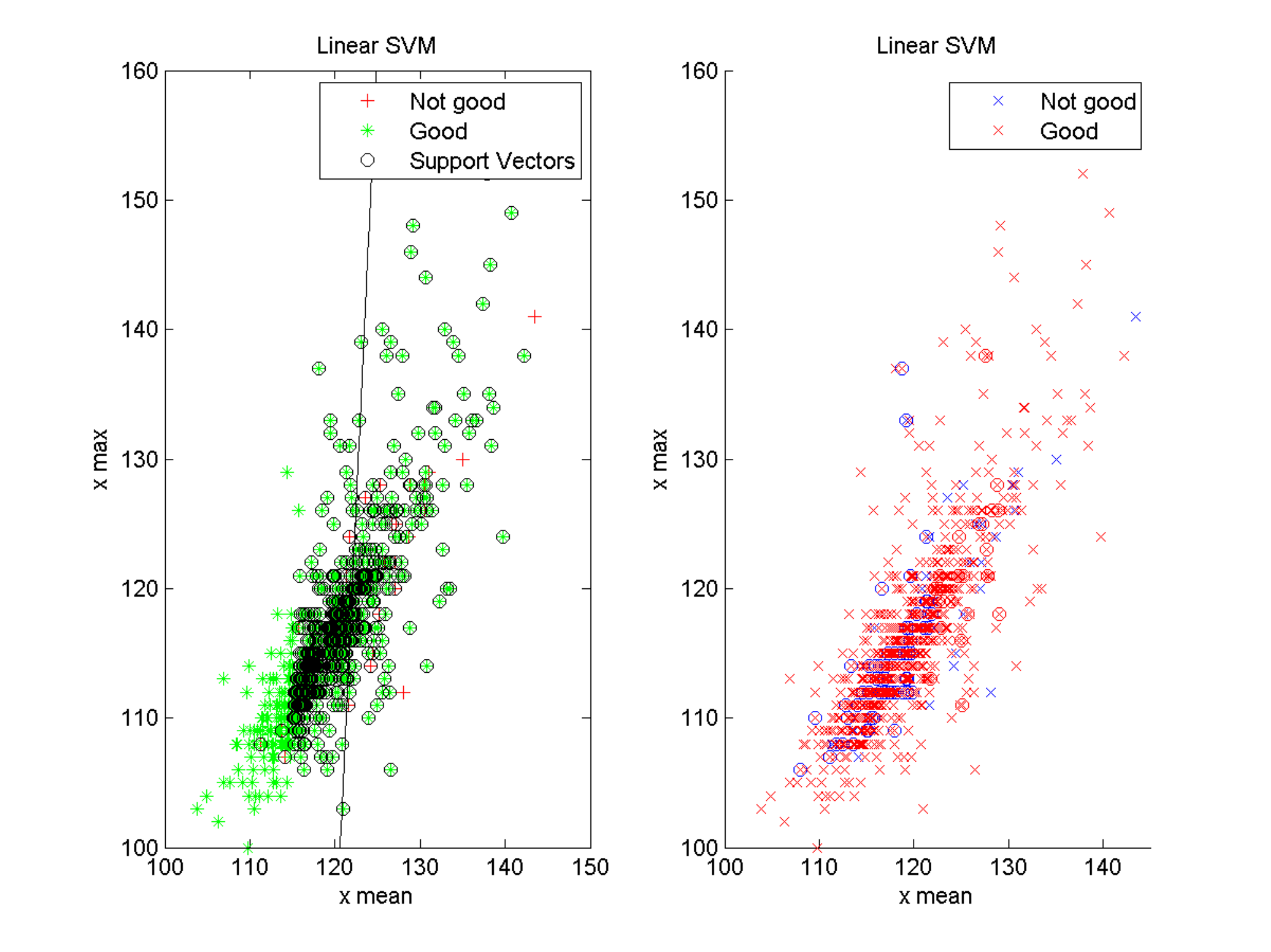}
\caption{Classification of the data with a linear SVM algorithm, as in Fig. \ref{lin1}, but with $\rho=2.5$.}\label{lin2}
\end{figure}

\begin{figure}[H]
\centering
\includegraphics[width=0.45\textwidth]{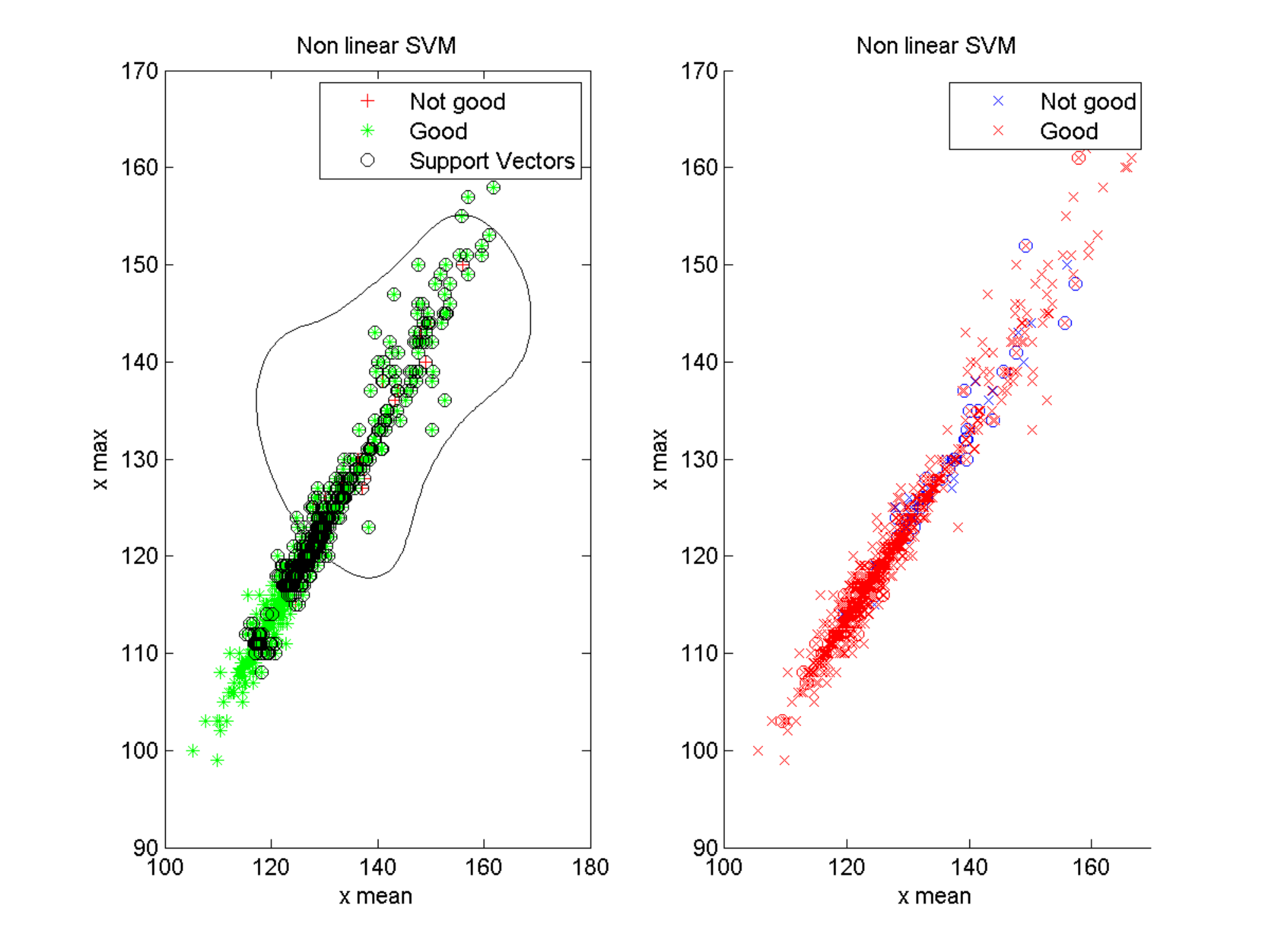}
\hspace{1mm}
\includegraphics[width=0.45\textwidth]{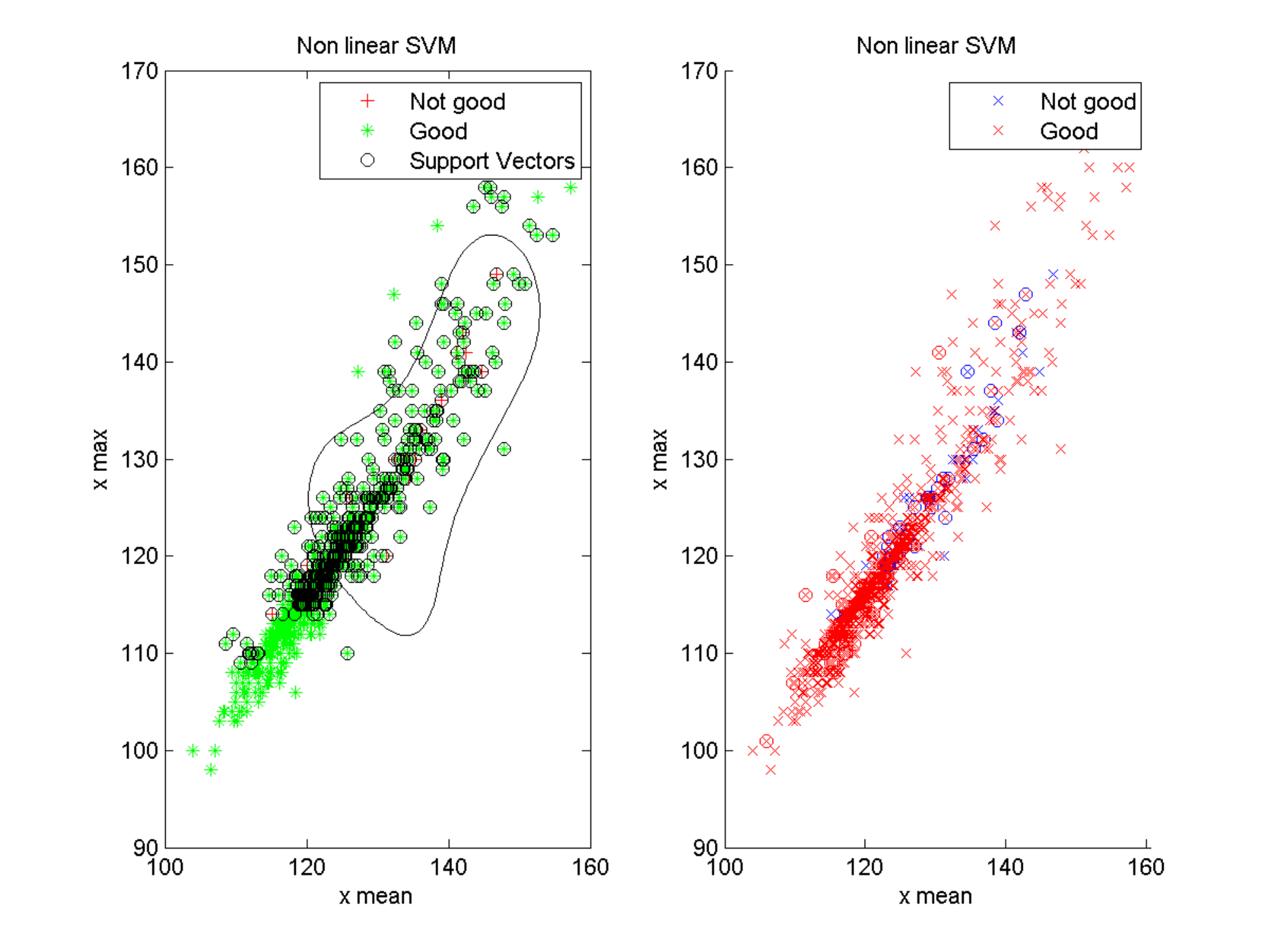}\\
\vspace{2mm}
\includegraphics[width=0.45\textwidth]{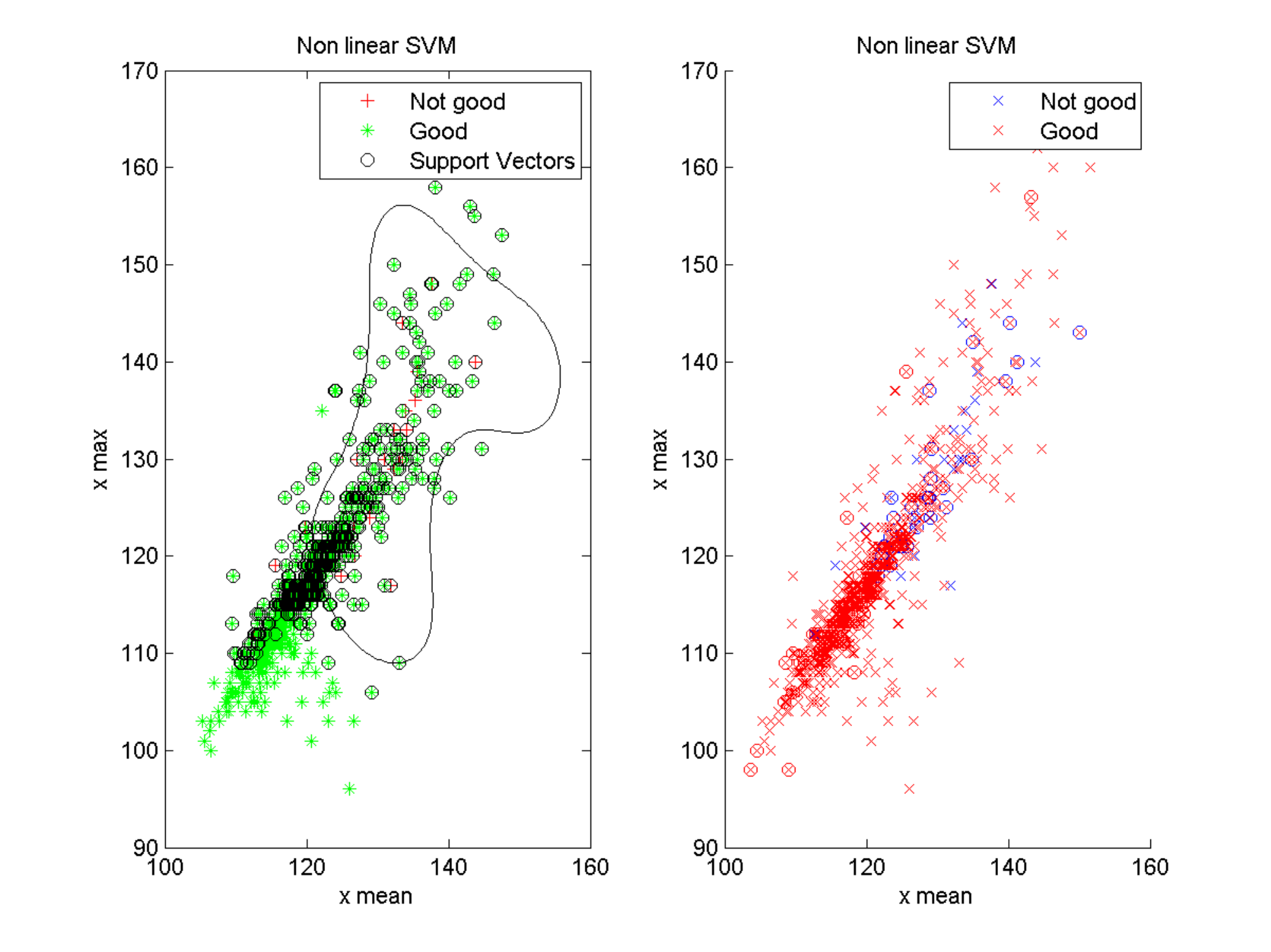}
\hspace{1mm}
\includegraphics[width=0.45\textwidth]{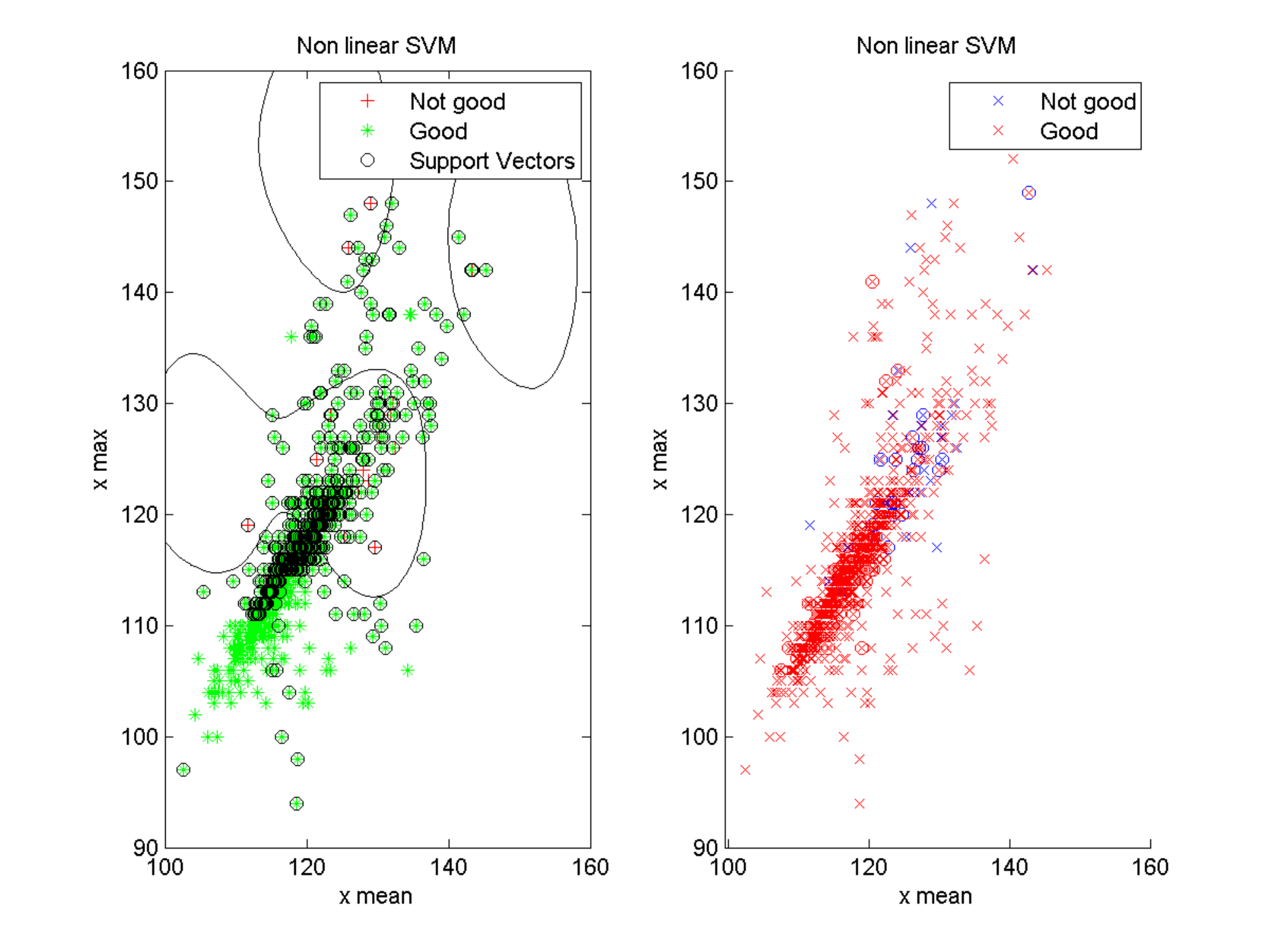}
\caption{Classification of the data through a nonlinear SVM algorithm (based on radial basis functions) for $\rho=1.5$ (cf. the caption of Fig. \ref{lin1}).}\label{nonlin1}
\end{figure}

\begin{figure}[H]
\centering
\includegraphics[width=0.45\textwidth]{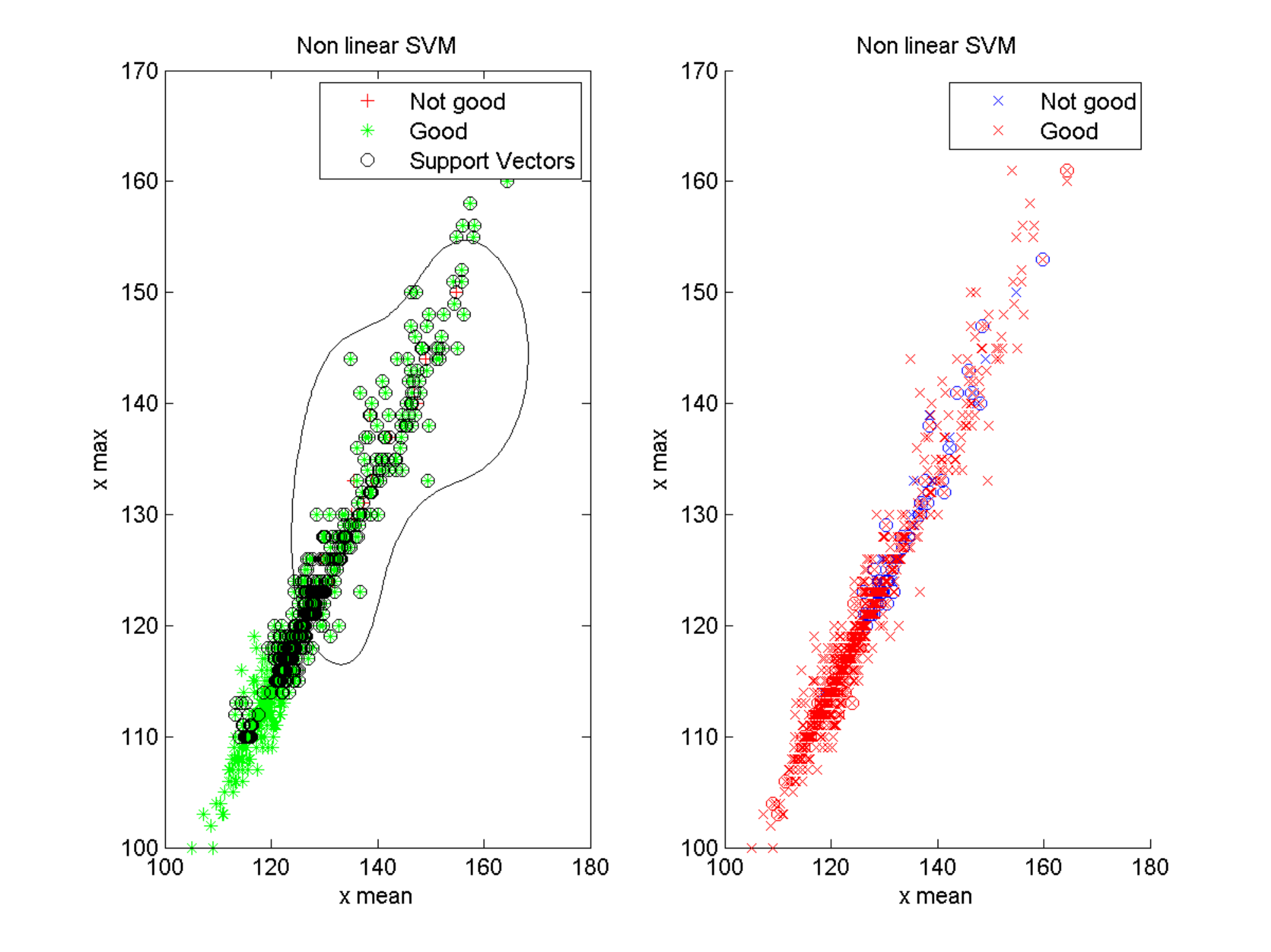}
\hspace{1mm}
\includegraphics[width=0.45\textwidth]{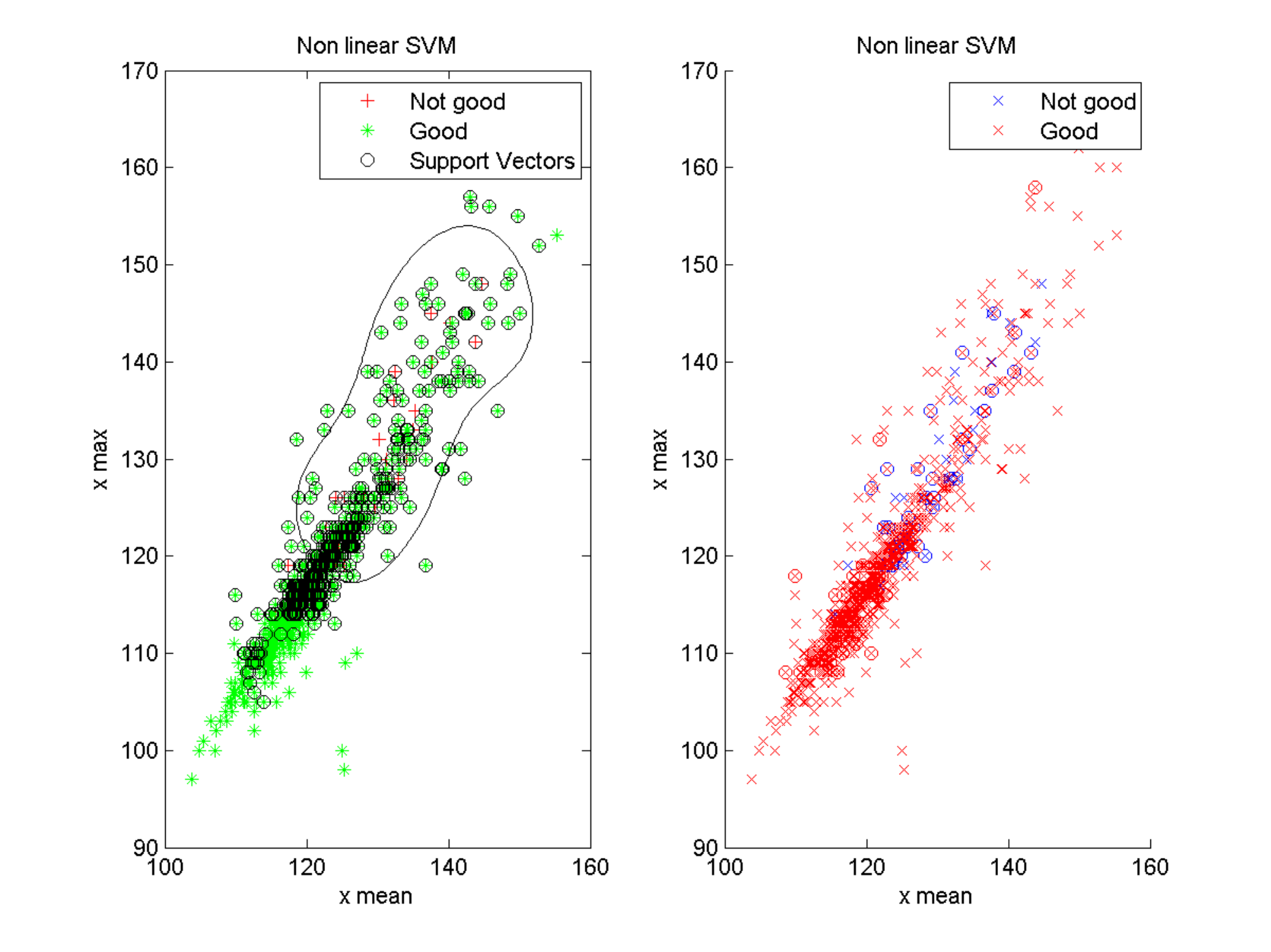}\\
\vspace{2mm}
\includegraphics[width=0.45\textwidth]{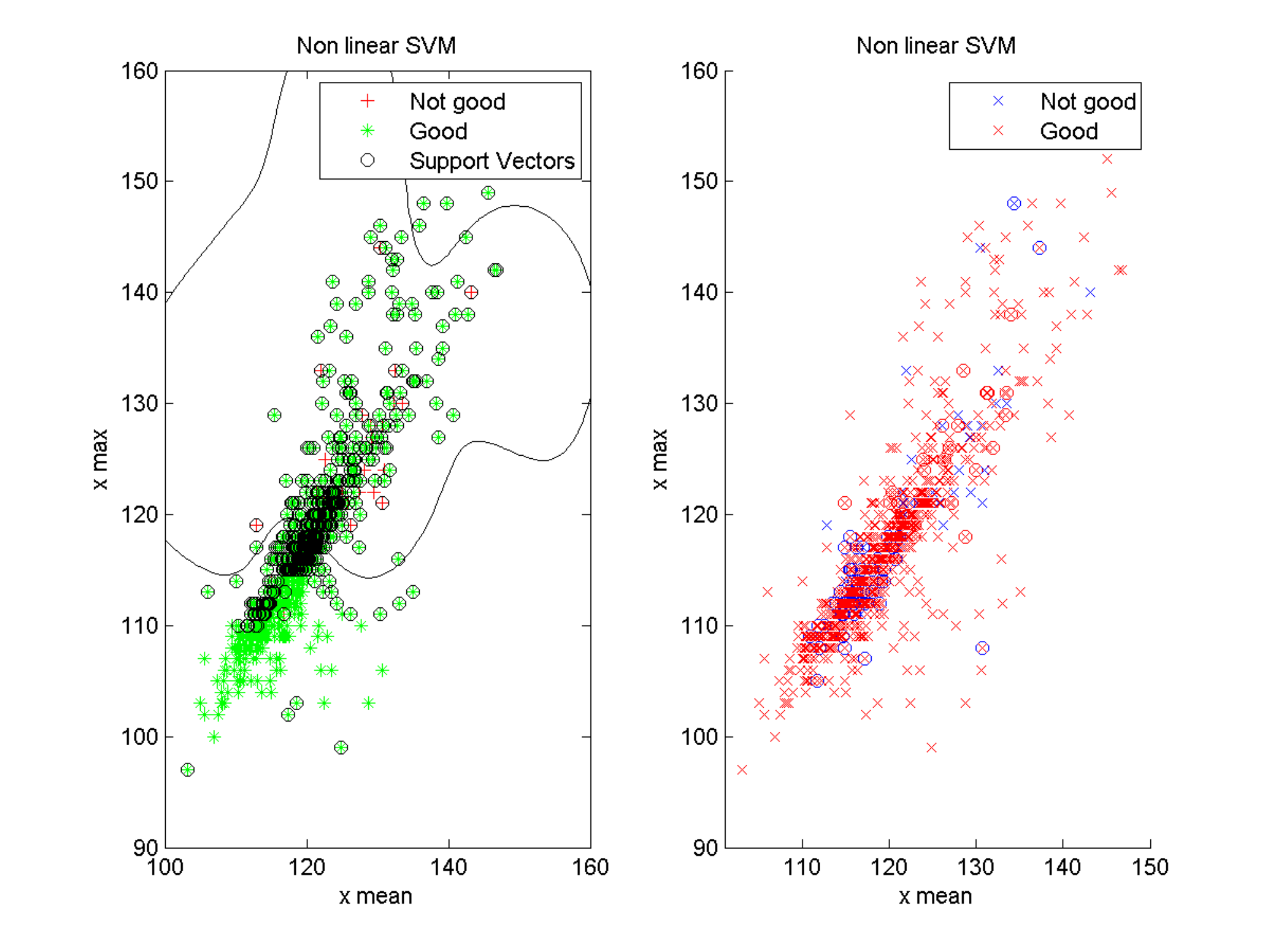}
\hspace{1mm}
\includegraphics[width=0.45\textwidth]{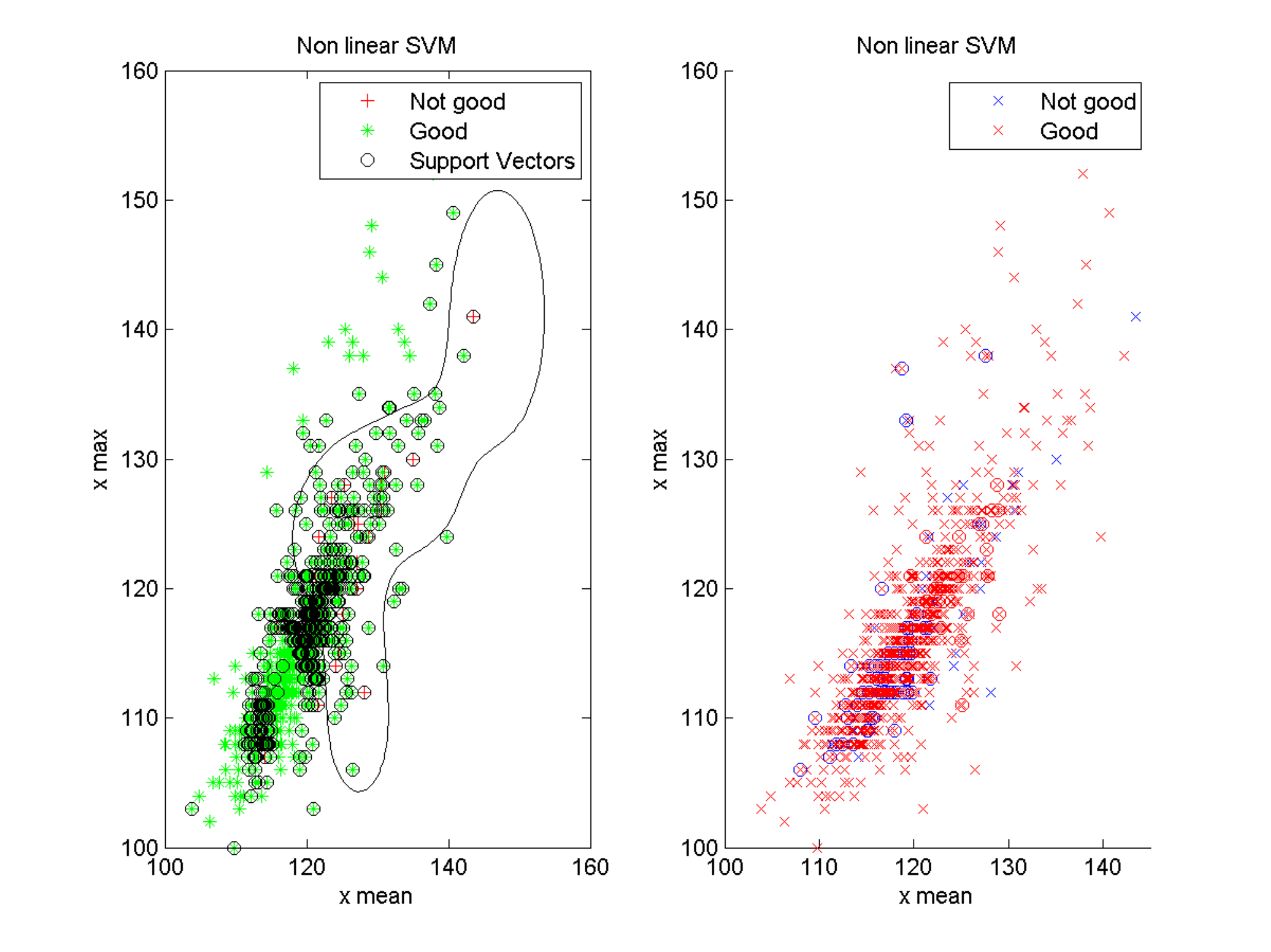}
\caption{Classification of the data with a nonlinear SVM algorithm (based on radial basis functions) for $\rho=2.5$ (cf. the caption of Fig. \ref{lin1}).}\label{nonlin2}
\end{figure}

Furthermore, an inspection of Figs. \ref{raw3a} and \ref{raw3b} reveals that the dots corresponding to the sets $D$ and $I$ are nested within the ensemble of points belonging to the set $G$: the classes $G$ and $nG$ are not amenable to be disentangled by a linear SVM regression, as also confirmed by the plots in Figs. \ref{lin1} and \ref{lin2}. 
In each of the two latter figures, the left plot shows the elements of the adopted (randomly selected) training set: green and red symbols identify the elements of the two classes $G$ and $nG$, while the black circles indicate the support vectors. The black line indicates the boundary (best separating hyperplane) detected by the SVM, which sensibly depends on the chosen training set. The right plot, instead, displays all the available data (red and blue crosses represent, respectively, the elements of the classes $G$ and $nG$), complemented by the SVM test set output (red and blue circles). 
The proper match between the colours of the circles and the crosses would indicate a successfully accomplished separation between the two classes, which, though, is not obtained with our data.
Furthermore, no remarkable improvement is obtained by attempting a classification of the data by means of a nonlinear SVM algorithm, based on radial basis functions \cite{Webb}, as shown in Figs. \ref{nonlin1} and \ref{nonlin2}. 
The results of this Section, confirm, therefore, the predictions of the statistical analysis outlined in Sec. \ref{sec:sec1}: the presence of a not linearly separable entanglement between points belonging to different classes can be thus traced back to the lack of a suitable statistical scales separation.\\
There is another relevant aspect, concerned with the implementation of the SVM algorithm, to be pointed out. \\
We remark, in fact, that each of the plots shown in Figs. \ref{lin1} and \ref{lin2} pertains to a specific training set of data $\mathcal{T}$.

\begin{figure}[H]
\centering
\includegraphics[width=0.65\textwidth]{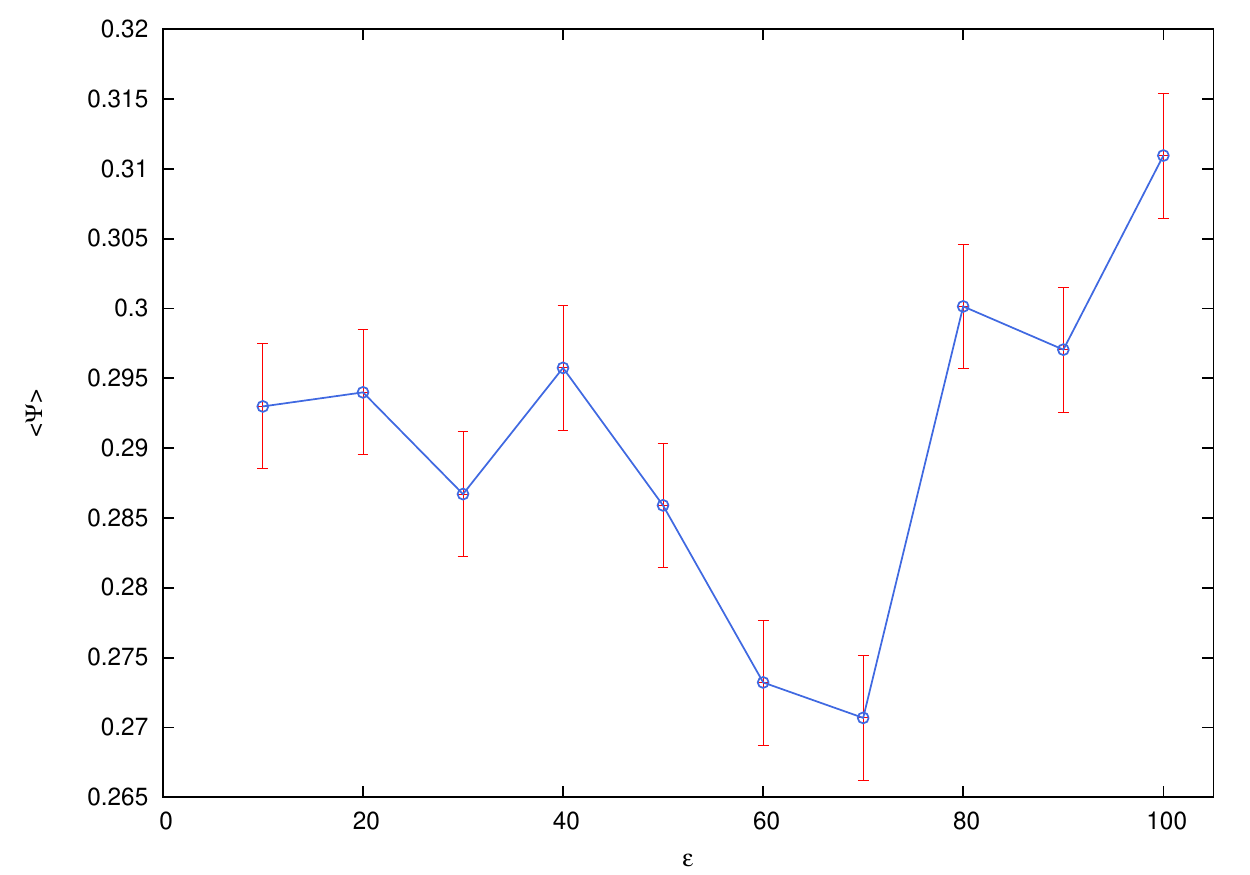}
\caption{Behavior of $\langle \Psi \rangle$, averaged over $N_c=500$ training samples, vs. $\epsilon$, with $\rho=1.5$, with error bars (in red).}\label{Psi}
\end{figure}

We can introduce, then, the quantity $\Psi_\ell(\mathcal{T}_\ell;\epsilon,\rho)$, relative to the specific training set $\mathcal{T}_\ell$, and defined as the ratio of the number of hazelnuts, belonging to the class $G$ and mistakenly classified as belonging to the class $nG$, to the total number of hazelnuts in the database, given by $N_G+N_{nG}$. The function $\Psi_\ell$ is an indicator of the performance of the SVM algorithm, and sensibly depends on the structure of the training sample considered in the simulation. 
Thus, while the behaviour of $\Psi_\ell$, pertaining to single training samples, yields no indication about the onset of an optimal scale $\epsilon^*$, the average $\langle \Psi \rangle$, given by
\be
\langle \Psi \rangle(\epsilon,\rho) = \frac{1}{N_c}\sum_{\ell=1}^{N_c}\Psi_\ell(\mathcal{T}_\ell;\epsilon,\rho) \quad , \nonumber
\ee
and computed over a sufficiently large number $N_c$ of training samples, attains a minimum precisely at $\epsilon^*=70$, cf. Fig. \ref{Psi}. The latter value of $\epsilon$ corresponds, in fact, to the scale of resolution maximizing the two statistical scales introduced in Sec. \ref{sec:sec1}, cf. Fig. \ref{ratio3}. The plot in Fig. \ref{Psi} confirms, hence, that the onset of an optimum scale $\epsilon^*$ can be numerically evinced also by means of SVM algorithms, provided that performance of the SVM is \textit{averaged} over a sufficiently large number of training samples.

\section{Conclusions}
\label{sec:conc}

In this work we performed a statistical analysis on the histograms of a set of hazelnut images, with the aim of obtaining a preliminary estimate of the performance of a machine learning algorithm based on statistical variables. We shed light, in Sec. \ref{sec:sec1}, on the relevance of two statistical scales, which need to be widely separated to accomplish a successful pattern recognition. The intrinsic lack of such scale separation in our data was also evidenced by the numerical results reported in Sec. \ref{sec:sec2}, revealing that no exhaustive classification can be achieved through SVM algorithms.
Moreover, the analysis outlined in Sec. \ref{sec:sec1} also unveiled the onset of an optimal resolution $\epsilon^*$, which is expected to optimize the pattern recognition. This observation was also corroborated by the results discussed in Sec. \ref{sec:sec2}, where the same value $\epsilon^*$, maximizing the performance of the SVM algorithm, is recovered by averaging over a sufficiently large number of training samples.
Our results, thus, strengthen the overall perspective that a preliminary estimate of the intrinsic statistical scales of the data constitute a decisive step in the field of pattern recognition and, moreover, pave the way for the further implementation of statistical mechanical techniques aimed at the development of a generation of more refined neural networks algorithms.

\newpage

{\bf Acknowledgments}

\vskip 5pt

We would like to thank Ferrero and Soremartec for their long-standing support of our research activity. We also thank Dr. A. Boscolo and Dr. L. Placentino, for providing us with the set of x-ray images used in this work.
This study was funded by ITACA, a project financed by the European Union, the Italian Ministry of Economy and Finance and the Piedmont Region.

\vskip 10pt


\end{document}